\documentclass[prodmode,acmtsas]{acmsmall}

\usepackage[english]{babel}
\usepackage{graphicx}
\renewcommand{\rmdefault}{cmr}

\usepackage[T1]{fontenc}
\usepackage[latin9]{inputenc}
\usepackage[table]{xcolor} 
\usepackage{fixltx2e}

\usepackage{epstopdf}
\usepackage{color}
\usepackage[caption=false]{subfig}
\usepackage{amssymb,amsmath}
\usepackage{paralist}

\usepackage[ruled]{algorithm2e}

\SetAlFnt{\small}
\SetAlCapFnt{\small}

\acmVolume{9}
\acmNumber{4}
\acmArticle{39}
\acmYear{2010}
\acmMonth{3}

\date{}

\usepackage{etoolbox}
\newtoggle{fullvers}
 \toggletrue{fullvers}

\def\Adj{\mbox{Adj}}
\def\length{link-length}
\def\Length{Link-Length}
\def\pbgdistance{path-based distance}
\newcommand{\distance}[2]{\overrightarrow{d}_{\hspace*{-.5ex}path}(#1,#2)}

\def\B{{\mathbb B}}
\def\R{{\mathbb R}}

\def\Frd{Fr\'echet distance}

\def\bsmall{Berlin-small}
\def\blarge{Berlin-large}
\def\asmall{Athens-small}

\newcommand{\thmref}[1]{Theorem~\ref{thm-#1}}

\newcommand{\lemref}[1]{Lemma~\ref{lem-#1}}
\newcommand{\defref}[1]{Definition~\ref{def-#1}}

\newcommand{\figref}[1]{Figure~\ref{fig-#1}}
\newcommand{\tableref}[1]{Table~\ref{table-#1}}
\newcommand{\corref}[1]{Corollary~\ref{cor-#1}}

\newcommand{\secref}[1]{Section~\ref{sec-#1}}
\newcommand{\subsecref}[1]{Subsection~\ref{subsec-#1}}

\def\image{\mathrm{im}}

\begin{document}


\markboth{Ahmed et al.}{A Path-Based Distance for Street Map Comparison}

\title{A Path-Based Distance for Street Map Comparison}
\author{Mahmuda Ahmed
\affil{University of Texas at San Antonio}
Brittany Terese Fasy
\affil{Tulane University}
Kyle S.\ Hickmann
\affil{Tulane University}
Carola Wenk
\affil{Tulane University}
}

\begin{abstract}
  Comparing two geometric graphs embedded in space is important in the field of transportation network analysis. Given
  street maps of the same city collected from different sources, researchers often need to know how and where they
  differ.  However, the majority of current graph comparison algorithms are based on structural properties of graphs,
  such as their degree distribution or their local connectivity properties, and do not consider their spatial
  embedding. This ignores a key property of road networks since the similarity
of travel over two road networks is
  intimately tied to the specific spatial embedding. Likewise, many
current algorithms specific to street map comparison either do not provide
 quality guarantees or focus on spatial embeddings only.

   Motivated by road network comparison, we propose a new path-based distance
measure between two planar
   geometric graphs that is based on comparing sets of travel paths
generated over the graphs.  Surprisingly, we are able to show that using paths
of bounded link-length, we can capture global structural and spatial
differences between the graphs.

%
   We show how to utilize our distance measure as a  local signature in order to 
identify and visualize portions
   of high similarity in the maps. And finally,  we give an experimental 
evaluation of our distance measure and its local signature on street map data 
from Berlin, Germany and Athens, Greece. 
\end{abstract}




\acmformat{Mahmuda Ahmed, Brittany Terese Fasy, Kyle Hickmann, and Carola Wenk.
 A Path-Based Distance for Street Map Comparison.}

\begin{bottomstuff}
  This work is supported by the National Science Foundation, under grant CCF-1301911.

Author's addresses: M. Ahmed, Computer Science Department,
University of Texas at San Antonio; B. T. Fasy  {and} C. Wenk,
Computer Science Department, Tulane University;
K. Hickmann, Center for Computational Science, Tulane University
\end{bottomstuff}

\maketitle

\section{Introduction}

\begin{figure*}[tbhp]
\centering
\subfloat[TeleAtlas, Berlin-small.]
{\includegraphics[width=.5\textwidth]{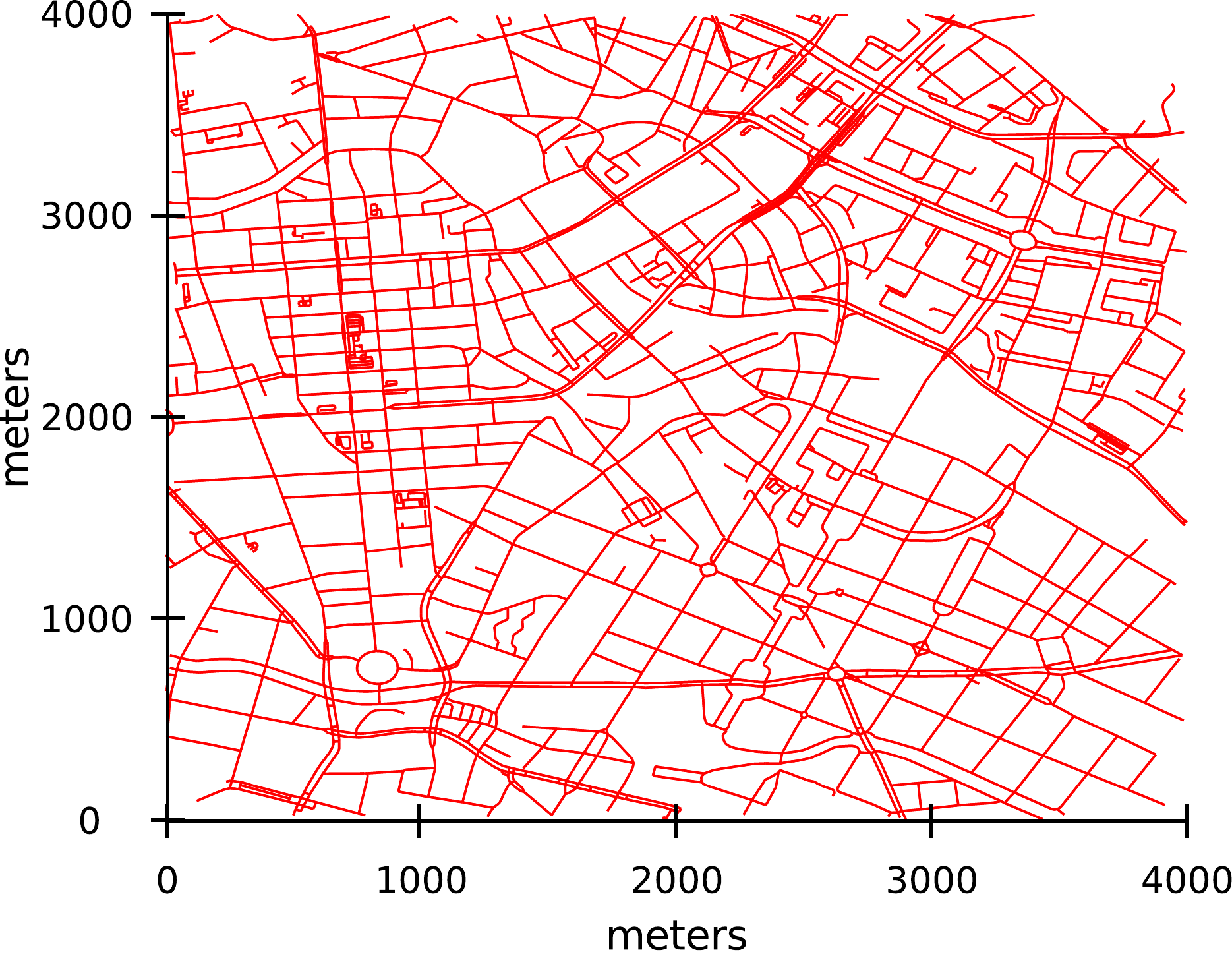}\label{
fig-berlin2t}}
\subfloat[OpenStreetMap, Berlin-small.]
{\includegraphics[width=.5\textwidth]{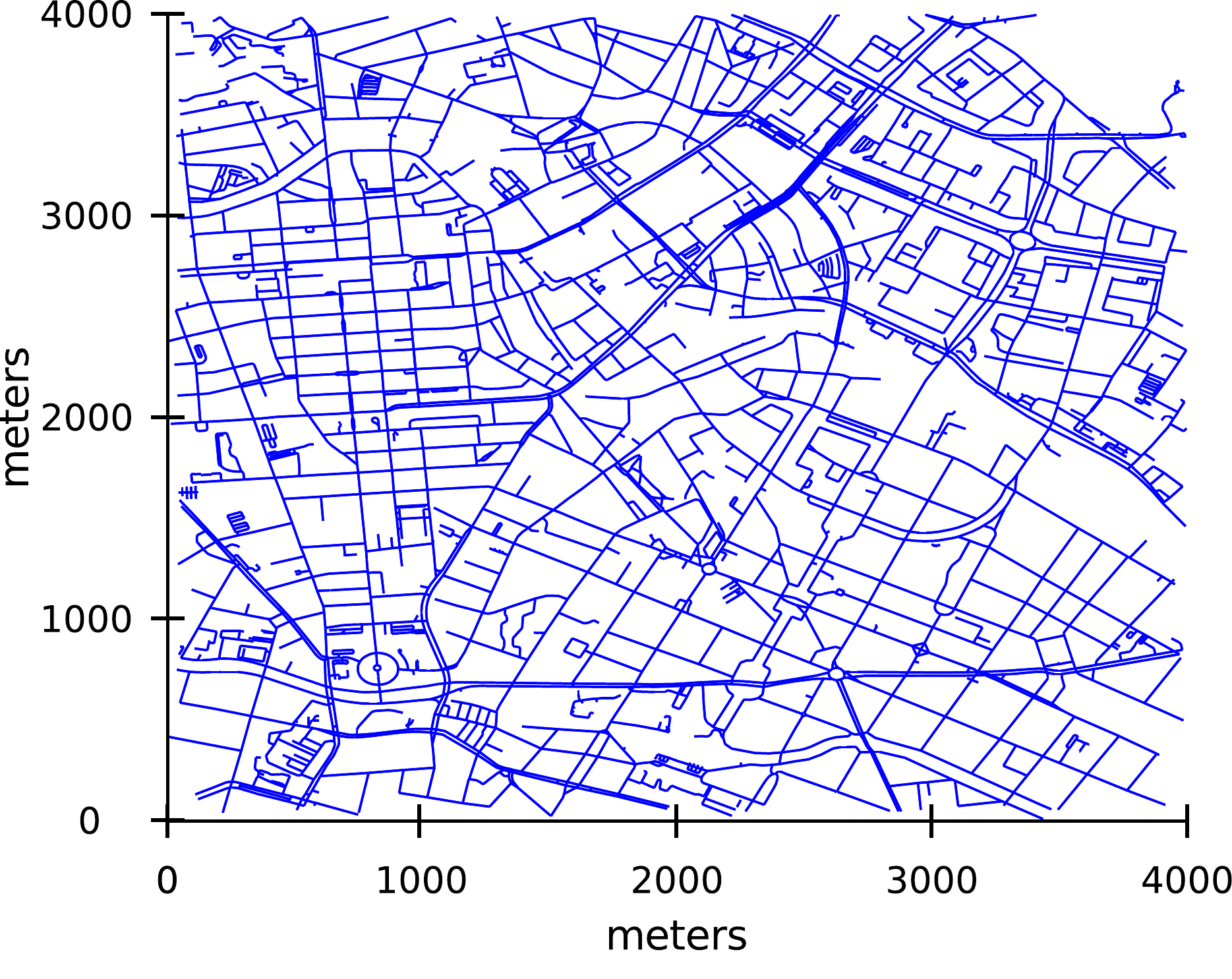}\label{
fig-berlin2o}}

\subfloat[TeleAtlas, Athens-small.]
{\includegraphics[width=.5\textwidth]{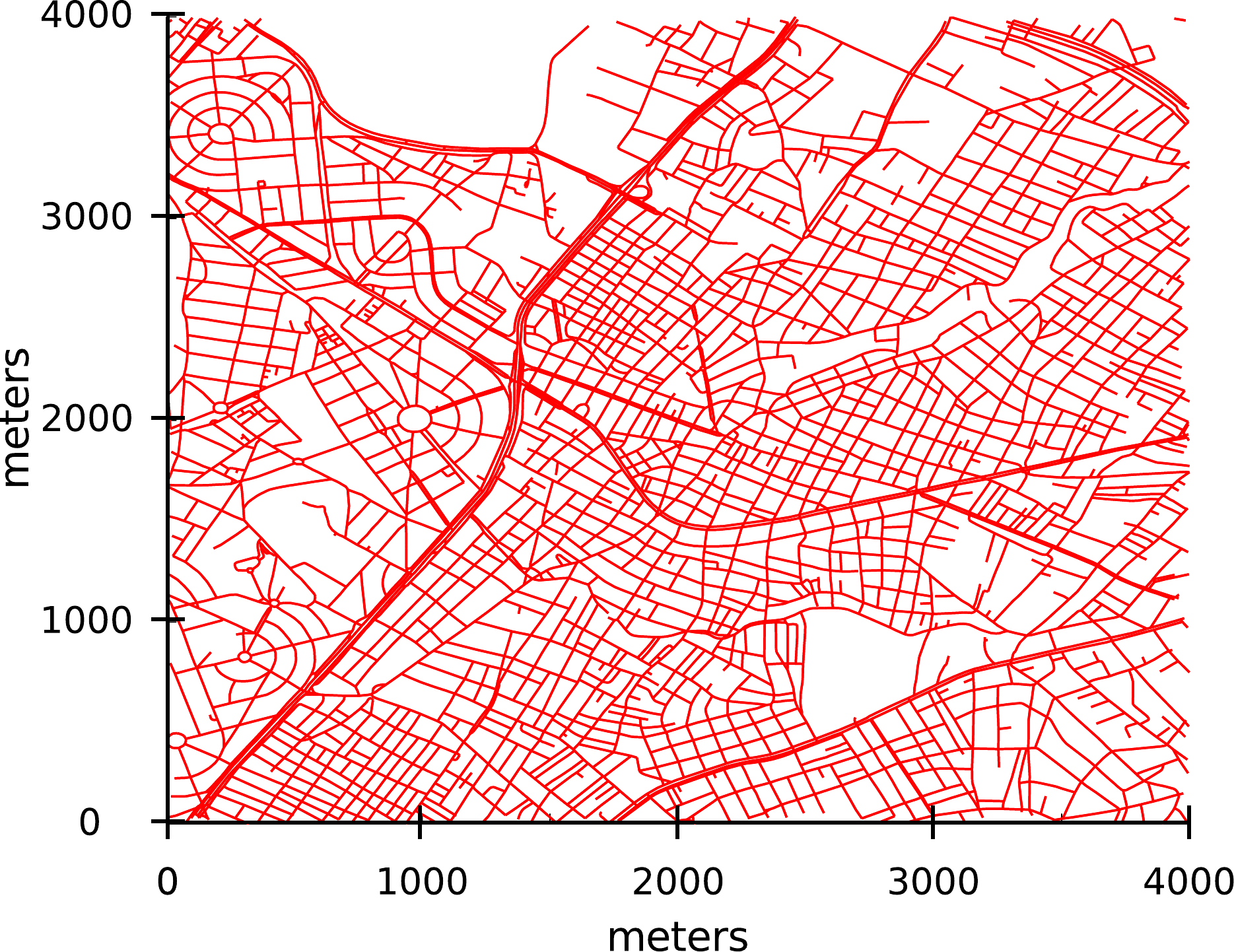}\label{
fig-athenst}}
\subfloat[OpenStreetMap, Athens-small.]
{\includegraphics[width=.5\textwidth]{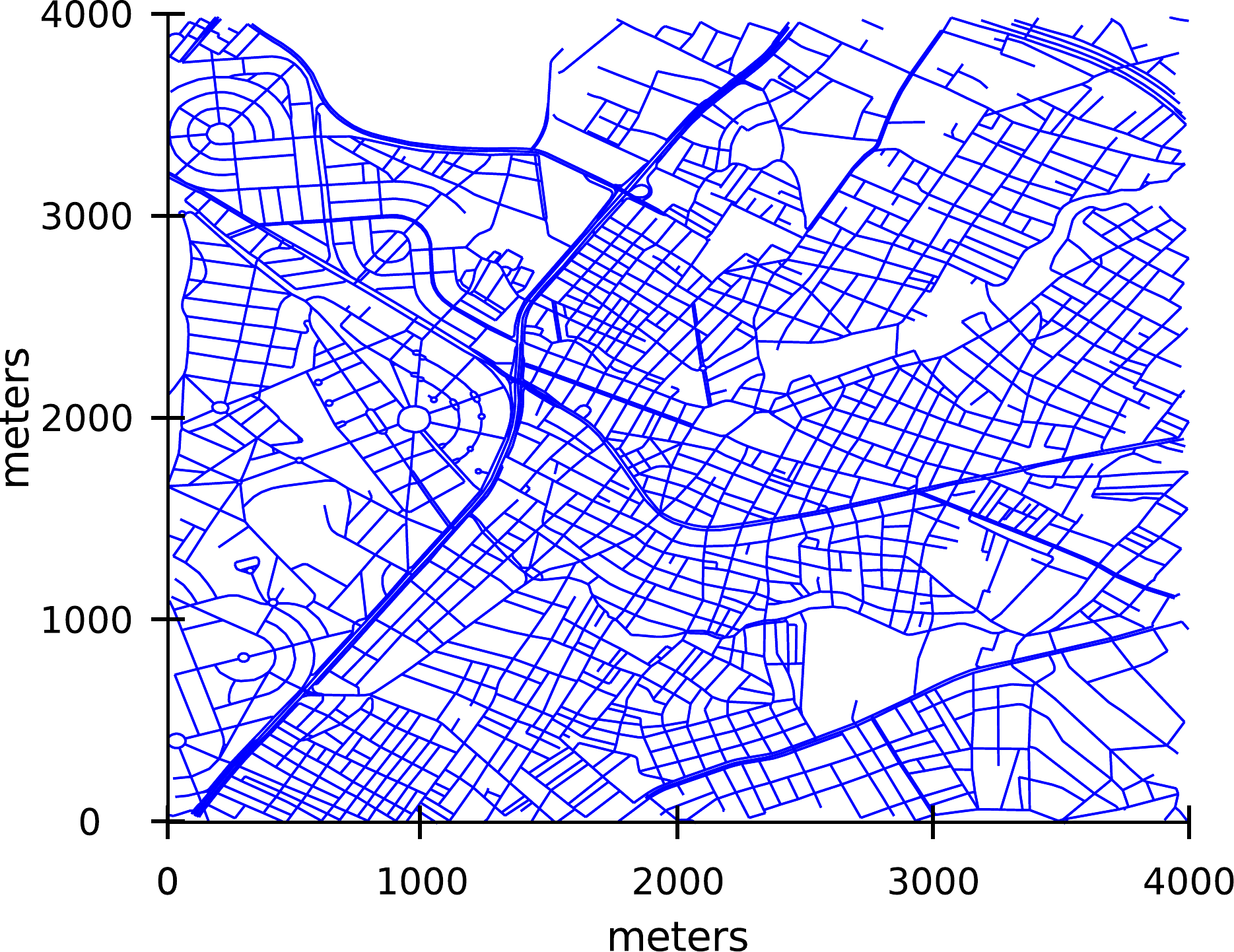}\label{
fig-athenso}}

\subfloat[TeleAtlas, Berlin-large.]
{\includegraphics[width=.5\textwidth]
{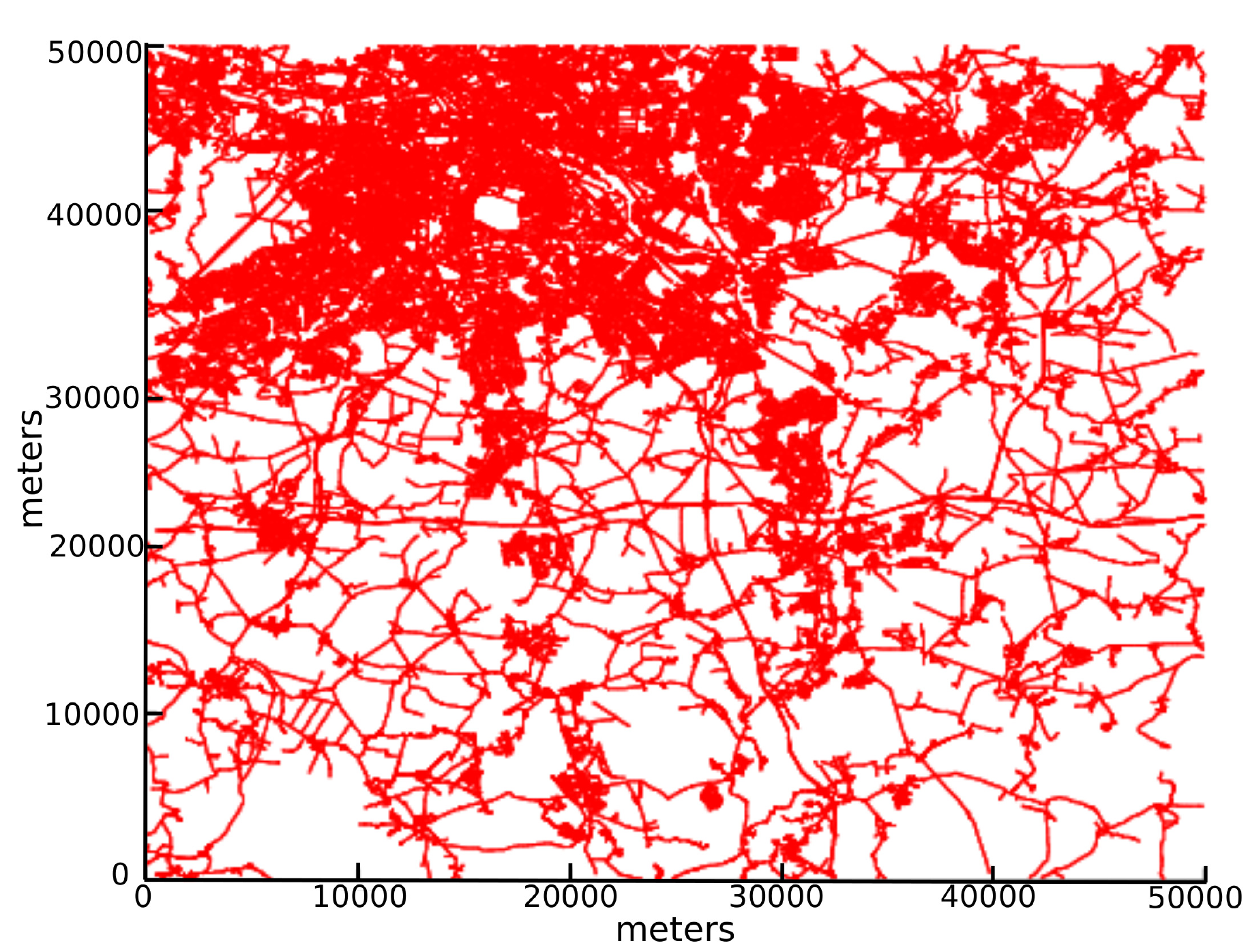}
\label{fig-blarget}}
\subfloat[OpenStreetMap, Berlin-large.]
{\includegraphics[width=.5\textwidth]{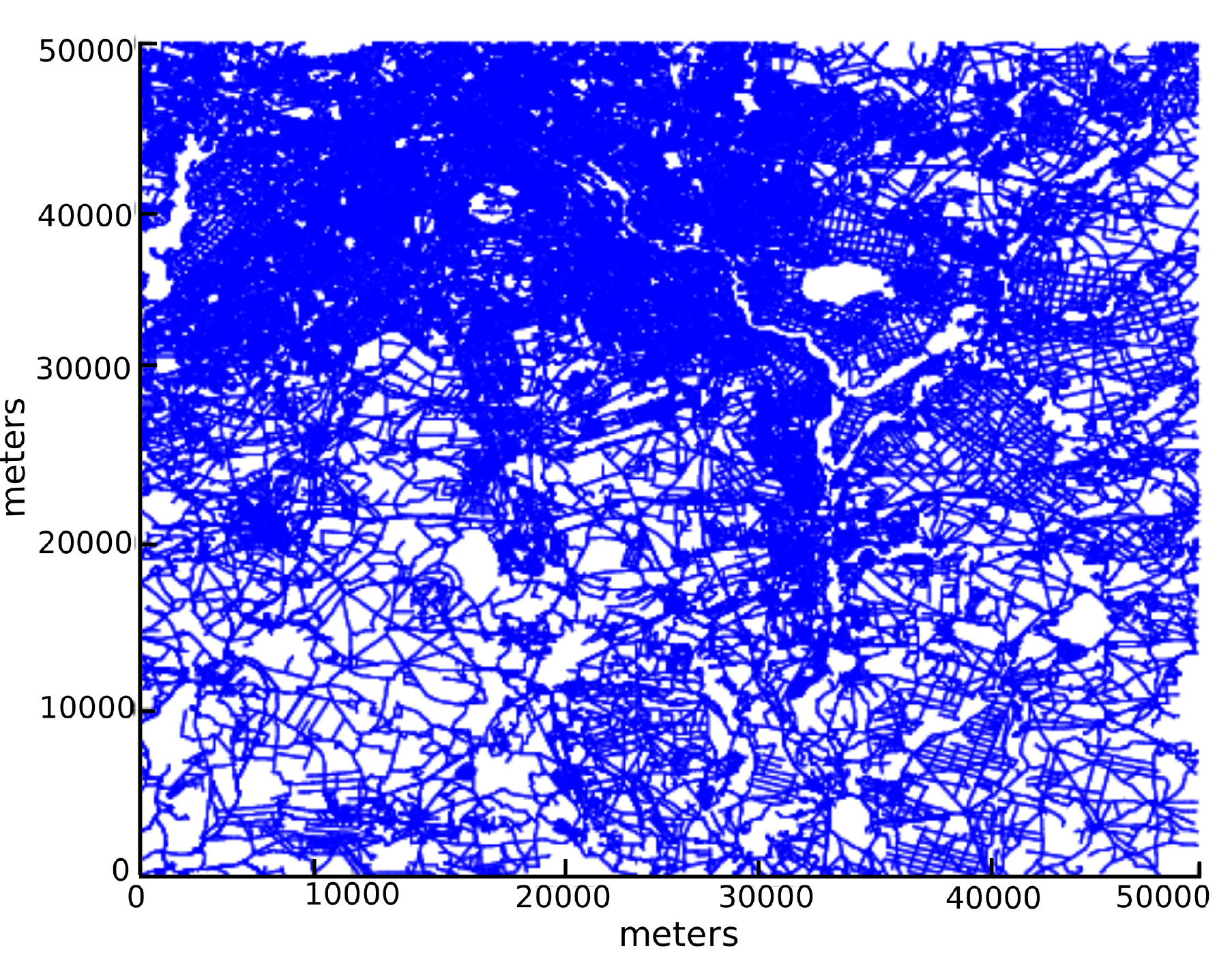}
\label{fig-blargeo}}
\caption{The TeleAtlas (TA) and OpenStreetMap (OSM) maps for the same
square sections of Berlin and Athens\footnotemark. For both Berlin datasets,
  OSM has more detail than TA and for \asmall, TA has more detail than OSM.
  Among these datasets, \asmall\ maps are the most similar to each other. }
\label{fig-dataset_example}
\end{figure*}

Comparing two embedded graphs is important in the field of transportation network analysis. Often, there exist more than
one record of a given transportation network; for example, multiple records can exist when a road network is
reconstructed from data.  In this case, we would want a method to evaluate the accuracy of the reconstruction against
the true map. Moreover, the ability to compare one road network map with a newer map allows one to quantitatively
determine the amount of change the transportation network has experienced. Given the street maps of the same city
collected from different sources, the goal of this paper is to understand how and where the road maps differ.
\figref{dataset_example} shows two sets of street maps for the same region of Berlin, Germany and Athens, Greece; while
many features are shared, there are large differences, and the goal is to quantify \mbox{such differences.}

The task of comparing street maps has received attention lately with
the emergence of algorithms to reconstruct
street maps from GPS trajectory data.  The OpenStreetMap
project\footnote{www.openstreetmap.org} provides street map data
 open to the public, and recently several automatic street map reconstruction
algorithms have been proposed in
\citeN{Aanjaneya:2011:MGR:1998196.1998203},
\citeN{csm_esa2012},
\citeN{Biagioni:2012:MIF:2424321.2424333},
\citeN{cghsRnrop10},
\citeN{DBLP:conf/nips/GeSBW11},
\citeN{Karagiorgou:2012:VTD:2424321.2424334}, and
\citeN{Liu:2012:MLS:2339530.2339637}.  
However, evaluating the
quality of the reconstructed networks remains a challenge. From a theoretical point of view, the problem is deceivingly
simple to state:
\begin{quote}
  {\bf Given two  embedded planar graphs, how
similar are they?}
\end{quote}
Stated this way, there seems to be an immediate connection to the NP-hard \emph{subgraph isomorphism} problem,which
requires a one-to-one mapping between edges and vertices of two graphs.  Given two graphs $G$ and $H$, it is {NP-hard}
to determine if there exists a {sub-graph} of~$G$ which is \textit{isomorphic}
to $H$.  There has been much work on the
subgraph isomorphism problem, and for very restricted classes of graphs it has been shown to be solvable in polynomial
time \cite{Eppstein:1995:SIP:313651.313830}.  The desired mapping for street map comparison, however, is not necessarily
one-to-one and requires spatial proximity; specifically, we desire a distance measure between two networks that
indicates when \emph{it feels the same to travel over the two networks.} That is, navigation on the two transportation
graphs works similarly.

Since we are assuming that the two networks being compared are embedded in the
plane, it is tempting to just treat the
networks as sets of points and use something well-known like the Hausdorff distance to evaluate similarity. However, one
could then allow networks with disconnected travel paths to be very similar,
even though driving routes on the two would
necessarily be very different. We are not aware of any algorithms with theoretical quality guarantees that explicitly
require travel paths on the two networks to be similar for the networks
themselves to be considered similar.

A second method of studying graph comparison is the \textit{graph edit distance}. This measure defines similarity
between $G$ and $H$, by quantifying how much $H$ must be changed so that it is 
isomorphic to $G$.   \citeN{DBLP:conf/wea/CheongGKSS09} defined 
geometric graph distance inspired by the graph edit
distance, applied to Chinese character recognition. But, unlike graph edit distance 
they restricted the operations to follow a specific sequence:
\begin{inparaenum}
\item edge deletions,
\item vertex deletions,
\item vertex translations,
\item vertex insertions, and
\item edge insertions.
\end{inparaenum}
The authors showed that their distance measure is NP-hard to compute, and they introduced a new
landmark distance that uses vertex distance signatures around landmarks and employs the \emph{earth mover's} distance.
The geometric graphs in their context, however, are embedded in two different coordinate systems.

Graph comparison lies at the core of many applied and theoretical research
avenues, and has been studied by both
theoretical and applied communities, see \citeN{cfsvTYGM04} for a
review.
But often, additional domain-specific information is used. For example, while
Chinese characters consist of graphs, the state-of-the-art in  Chinese character
recognition relies on additional knowledge, such as the hierarchical structure
of the characters or the individual strokes, and, sometimes, the stroke sequence
\cite{A.:2014:AMO:2676410.2629622,Shi:2003:OHC:964161.964163,%
Kim:1996:OCC:244032.244050}. 
In the context of street map comparison, \citeN{MondzechS11} and
\citeN{Karagiorgou:2012:VTD:2424321.2424334} have used shortest paths, 
independently computed in each graph
between randomly selected locations, to compare graphs. 
\citeN{Liu:2012:MLS:2339530.2339637} and 
\citeN{Biagioni:2012:MIF:2424321.2424333} have used bottleneck matching to 
compare point sets induced by the
two graphs. Another recent approach uses a technique from computational
topology to
compare local graph structures~\cite{afw2014lhdist}.  These approaches, however,
do not provide precise guarantees on how similar navigating over the two
networks will be
if the distance measures are small.

\addtocounter{footnote}{-1}
\footnotetext{The $x$
and $y$-coordinates are the offset (in meters)
from an arbitrary location, given in Universal Transverse Mercator (UTM)
coordinates.  That location is UTM Zone 33U, $390,000$ meters east,
$5,817,000$ meters
north in (a)-(b), UTM Zone 34S, $480,000$ meters east, $4,206,000$ meters north
in (c)-(d), and UTM Zone 33U, $375,000$ meters east, $5,775,000$ meters
north in (e)-(f).}

In judging the difference between two street maps, intuitively, one is concerned with the utility of the street map
graphs as a navigation tool. Therefore, we propose a distance measure on street
map graphs based on similarities of the
possible travel routes allowed by the networks. Here, a street map is
formally defined as a planar geometric graph $G$,
with vertices $V_G \subseteq \mathbb{R}^2$ and edges $E_G$ given by polygonal 
paths embedded in $\R^2$ connecting two  different vertices in $V_G$.  In this
framework, 
it is possible to have more than one edge connecting two vertices.
Under this definition the graph $G$ can be treated as a set of possible paths in 
$\mathbb{R}^2$, as opposed
to treating the graph as a set of points in $\mathbb{R}^2$.  Since paths of
travel are implicitly considered with our
measure, closeness in this distance represents similarity of navigation.

\paragraph{Our Contributions} We introduce a new distance measure for planar embedded geometric graphs $G$ and $H$,
which is based on covering $G$ and $H$ with sets of paths. The distance measure is the directed Hausdorff distance
between the path sets, where the \Frd\ is used to compute the distance between two paths.  We have emphasized use of the
\emph{directed} Hausdorff distance, since often in the map reconstruction problem, the reconstruction only represents a
subgraph of the larger transportation network. Thus, we are interested 
in measuring the second graph's closeness to
an appropriate subgraph of the true transport~network.

In \secref{distance}, we provide the theoretical guarantees of the path-based
distance.
Restricting our attention to paths with small link-length, we are
able to
capture the structural as well as the spatial properties of the graphs.  
Using
link-length one paths, we compute a generalization of the Hausdorff distance. 
Using link-length two paths, we capture intersections.  
Most surprisingly, using only link-length three paths, we are able to
approximate the distance between paths of arbitrary link-length in polynomial
time.
%
%

In \subsecref{signature}, we show how to utilize our distance measure as a local signature in order to identify and
visualize portions of high similarity, or of dissimilarity, between the maps. 
Such
local information is useful for detecting changing areas in road networks using historical map comparison and for
identifying types of street map formations that reconstruction algorithms may fail at detecting. Finally, in
\secref{experiments}, we give an experimental evaluation of our distance measure and its local signature on street map
data from Berlin, Germany and Athens, Greece. The code for computing our \pbgdistance\ is available on {\tt
  mapconstruction.org}.


\section{Street Map Graphs}
\label{sec-background}
We model a street map as a planar geometric graph, $G=(V_G,E_G)$, embedded in $\mathbb{R}^2$. We assume that each edge in $E_G$ is
represented as a polygonal curve and that no vertex in~$V$ has degree two. That
is, intersections in the street maps
become vertices and roadway segments with no intersections make up the edges.

\subsection{Comparing Street Maps}
\label{subsec-comparingStreetMaps}
\begin{figure}[t]
\centering
{\includegraphics[scale=0.6]{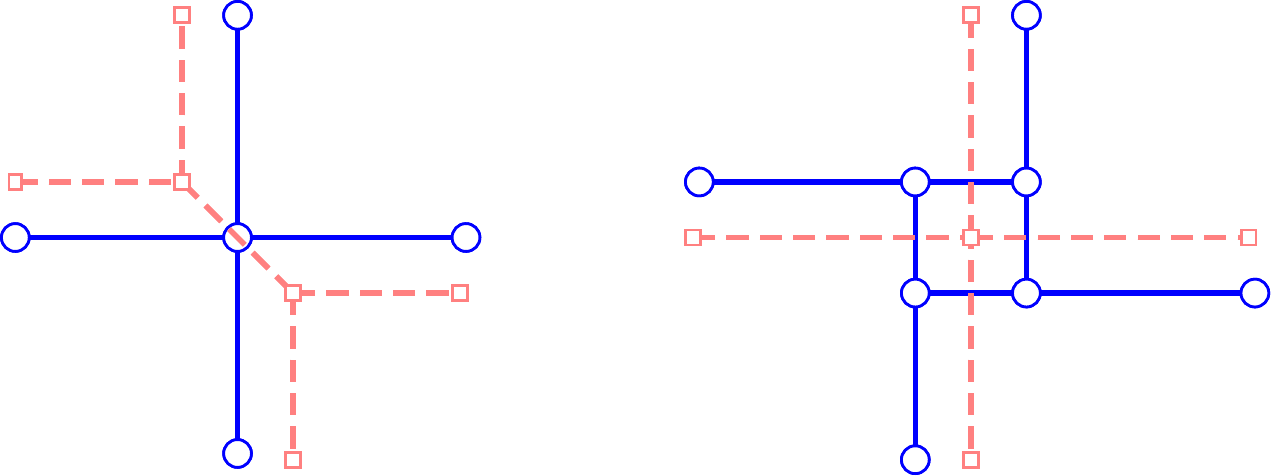}}
\caption{Two graphs $G$ (solid blue lines) and $H$ (dashed pink lines).  (a) One
vertex in $G$ represented as two
  vertices in $H$.  (b) Four vertices in $G$ represented as one vertex in $H$.}
\label{fig-ms}
\end{figure}

When designing a distance measure to compare two street map graphs, we would like to incorporate the following features:
\begin{enumerate}
\item{Spatial distance between corresponding vertices of the two graphs.}
\item{Similarity of the shapes of the edges.}
\item{Similar connectivity properties, i.e., similar navigation.}
\end{enumerate}
Difficulties can arise in evaluating whether or not the above three criteria have been accounted for properly. For
example, split and merge vertices (see \figref{ms}) may arise from different street map construction mechanisms. In both
subfigures, one can find a path on the blue graph that is \emph{close} to any arbitrary path on the dashed pink
graphs. Moreover, for every vertex on the pink graphs there is a \emph{close} vertex on the blue graph.
While designing the measure we wanted to make sure sure not to penalize too much for such cases but yet find these
differences.  In general, the current approaches to street map comparison fall into two categories. The first one treats
the graph as a set of points in the plane the second one treats the graph as a set of paths. Here, we briefly discuss
the ideas.

\begin{figure*}[t]
\centering
\subfloat[Four components.]
{\includegraphics[scale=0.85]{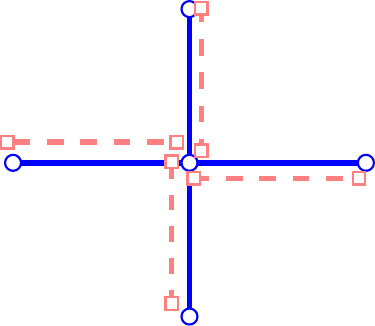}}
\hspace{.15in}
\subfloat[Missing vertex.]
{\includegraphics[scale=0.85]{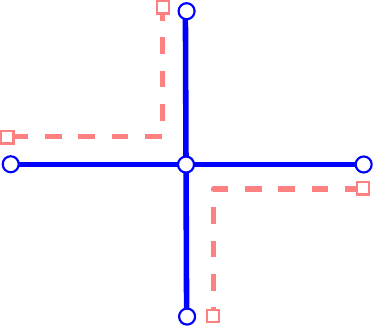}}
\hspace{.15in}
\subfloat[Split vertex.]
{\includegraphics[scale=0.85]{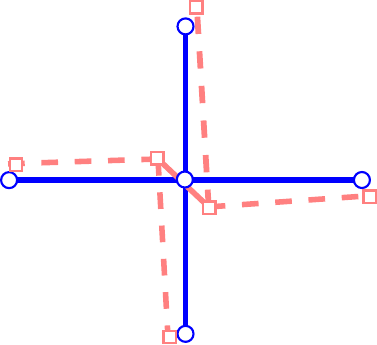}}
\caption{(a), (b), (c): Two graphs $G$ (solid blue lines) and $H$ (dashed pink lines). The \pbgdistance\ using $\Pi^1_G$
  is small despite large differences in combinatorial structures.}
\label{fig-length12}
\end{figure*}

\paragraph{Sets of Points}
This approach treats each graph as the set of the points that its vertices and edges cover in the plane. The main idea
is then to compute a distance measure between two point sets, such as the Hausdorff distance between the infinite complete set of points, or a one-to-one bottleneck
matching between a carefully selected finite subset of the points. The main drawback of using regular Hausdorff distance
is that no adjacency information is used, and the continuous structure of the graphs is largely ignored. Thus, in
\figref{length12}, the dashed pink graphs in (a), (b), and (c) would all be considered close to the solid blue graph
under this distance measure. However, their connectivity properties are all very 
different and the set of possible travel paths would be quite different.

The measure in \citeN{Biagioni:2012:MIF:2424321.2424333} compares the geometry 
and topology of the graphs by sampling its
edges. The main idea is as follows: starting from a random street location $p$ (the seed), walk in both directions,
choosing a sample point at regular intervals.  If an intersection is encountered, continue along every path possible
until a maximum distance from the seed is reached.  Repeating for the other graph using the closest point to $p$ as the
seed, two sets of locations are obtained.  These two point sets are compared by finding a maximal bottleneck matching
between them and counting the number of unmatched points in each set.
The sampling process is repeated for several seed locations, and a global tally of the matched and unmatched samples in
each graph is recorded.  In essence, the local neighborhoods of the seeds have been sampled, and these samples are
compared in the matching.
For the bottleneck matching, the sample points on one graph can be considered as \emph{marbles} and on the other graph
as \emph{holes}.  Intuitively, if a marble lands close to a hole it falls in, marbles that are too far from a hole
remain where they land, and holes with no marbles nearby remain empty. Each hole, however, can hold at most one marble.
Then the number of matched marbles (equal to the number of matched holes) is counted.

To produce a performance measure, \citeN{Biagioni:2012:MIF:2424321.2424333} use 
the well-known $F$-score, which they
compute as follows:
\begin{equation}
F\text{-score} = 2 * \frac{precision * recall}{precision+recall}
\end{equation}
where, $precision$ is defined to be $matched\_marbles / total\_marbles$ and $recall$ is defined to be $matched\_holes /
total\_holes$. In words, the precision measures the percentage of the marbles that are matched to holes and the recall
measures the percentage of the holes that are matched to marbles.  The $F$-score can range between zero and one, with a
score close to one indicating that nearly all holes and marbles are matched, and a score close to zero indicating that
very few marbles and holes are matched.
In \subsecref{comparejames}, we provide a comparison of our distance 
measures to this distance measure.

\paragraph{Sets of Paths}
The basic idea of this approach is to construct sets of paths to represent the two graphs, and then define a distance
measure based on distances between the paths.  Different distance measures can be used to compare paths, such as the
Hausdorff distance or the \Frd. Referring again to~\figref{length12}, we see that, in (a) and (b), there exist pairs of
connected vertices in the blue graph whose corresponding vertices are not connected in the pink dashed graph resulting
in a large path-based distance.  However, the dashed pink graph in (c) does have a \emph{close} path connecting the top
and bottom vertices so it would be closer in this distance measure. In this way using sets of paths to define a distance
measure preserves connectivity properties of the graph. The main challenge in defining a path-based distance measure is
then to select a set of paths from one graph such that the set as a whole preserves some structural properties of the
graph that can be utilized in an application. Such a set of paths must be small enough to check each path
computationally.
\citeN{MondzechS11} have introduced a heuristic measure by 
comparing shortest path lengths between pairs
of randomly selected points. 
\citeN{Karagiorgou:2012:VTD:2424321.2424334} have used a similar set of
paths, but used the discrete \Frd\ to compare routes.  In this paper we use a 
more general set of
paths.

\subsection{Paths}

In this paper, we use a path-based approach to compare two embedded geometric graphs. 
Let $a$ and $b$ be two points on any edge or vertex of $G$. 
A \emph{path} in $G$ between $a$ and $b$ is a, possibly non-simple, sequence of vertices in $G$ connecting $a$ to $b$ using valid adjacencies in the graph.  We consider such a path to be the image of a
continuous map $\alpha \colon [0,1] \to G$ such that $\alpha(0)=a$ and 
$\alpha(1)=b$. If a path starts and ends in vertices $u, v\in V_G$ we call it a 
\emph{vertex-path}, and we define its \emph{link-length} as the number of edges 
that comprise the path, and we may represent the path as sequence of vertices: 
$\alpha
= \langle uw_1w_2\cdots w_{k-1}v\rangle$
. Here we remind the reader that the
vertices of our graph do not have degree two. This distinction is made since
with street maps in mind, we consider only the
actual street intersections as graph vertices. We denote the set of all 
paths in $G$ by $\Pi_G$. We denote the set of
all vertex-paths of \length\ $k$ in $G$ as $\Pi^k_G$. Let 
$\hat{\Pi}_G=\cup_{k\geq 1}\Pi^k_G$ be the set of all vertex-paths, then we have 
that $\hat{\Pi}_G\varsubsetneq \Pi_G$.  Sometimes we may restrict our
attention to all paths of \length\ $k$ containing vertex $v \in V_G$ or an edge $e \in E_G$; we denote these restricted
sets of paths by $\Pi_{v}^k$ and $\Pi_{e}^k$, respectively.  We denote 
the Euclidean norm by $\Vert . \Vert$.  Our
distance measure between two embedded graphs are based on the \Frd\ between two 
paths.
\begin{definition}[Fr\'echet Distance]
  For two planar curves $f, g: [0,1] \rightarrow \mathbb{R}^2$, the Fr\'echet distance $\delta_{F}$ between them is
  defined as
\begin{equation}
  \delta_{F}(f,g) = \inf_{\substack{\alpha:[0,1]\rightarrow [0,1]}}
  \max_{\substack{t\in\left[ 0,1\right] }}\Vert f(t)-
  g(\alpha(t))\Vert,
\end{equation}
where $\alpha$ ranges over all continuous, surjective, non-decreasing
reparameterizations.
\end{definition}

The \Frd\ is a well-suited distance measure for comparing curves, or paths, because it takes continuity and monotonicity
of the curves into account. The \Frd\ between two polygonal curves with $m$ and $n$ vertices, respectively, can be
computed in $O(mn\log mn)$ time \cite{altgodau}.  Furthermore, the Fr\'echet 
distance induces a correspondence between the curves:
\begin{corollary}[Induced Correspondence]\label{cor-induced-corresp}
 Let $\delta = \delta_{F}(f,g)$.  Then, there exists a continuous function
 $C \colon [0,1] \to [0,1]^2$ such that $C$ is non-decreasing in each 
coordinate and $||f(s)-g(t)|| \leq \delta$ for all $(s,t)$ in the image of 
$C$.  
We define the function $M \colon \image{f} \to \image{g}$ 
where $f(s) \mapsto g ( \min\{t | (s,t) \in \image{C} \} )$.  And we 
define the generalized inverse $M^{-1} \colon \image{g} \to \image{f}$ 
where $g(t) \mapsto f ( \min\{s | (s,t) \in \image{C} \} )$.
\end{corollary}
In words, $C$ provides a parameterization that realizes the Fr\'echet 
distance between the curves.
We refer to $M$ as the \emph{Fr\'echet-correspondence} from $f$ to $g$.
Also, we note that $C$ (and hence $M$) need not be unique; however, 
it will suffice to choose an arbitrary correspondence~$C$. 
We extend $M$ to a function $\overline{M}$ from sub-paths of $f$ to sub-paths of $g$ as 
follows: $\overline{M}(p)$ is the shortest connected sub-path of $g$ containing all 
points $M(x)$, where $x \in p$ and $p$ is a sub-path of $f$.

In the {\em map-matching} problem, we ask to find a path $h \in H$ that minimizes the distance to a given curve $g \in
G$.  In our setting, we wish to minimize the \Frd\ $\delta_{F}(g,h)$.  
We call this map-matching the Fr\'echet-matching and denote it by 
$\delta_F(g,H)$. This distance can be computed in $O(mn\log^2 mn)$ time 
\cite{aerwMPM03}, where $m$ is
the number of vertices in $g$ and $n$ is the total number of vertices and edges in $H$.  The directed distance that we
define in the next section is the maximum map-matching distance over all paths 
$g \in G$.


\section{Path-Based Distance}
\label{sec-distance}

In this section, we formally define a path-based distance between street map graphs. The general idea is to summarize
each graph with a set of paths, and then to compare these sets using the directed Hausdorff distance. The {\em directed
  Hausdorff distance} between two sets $A$ and $B$ is defined as $\overrightarrow{d}(A,B) = \max_{a \in A}\min_{b \in B}
d(a,b)$. Usually, $d(a,b)$ is assumed to be the Euclidean distance.
However, the sets we
are considering are the sets of all paths on two different road network maps. Therefore we require a distance between
two paths instead of the standard Euclidean distance. For this, we use the \Frd\ between paths.
\begin{definition}[Path-Based Distance]
\label{def-pbDistance}
  Let $G$ and $H$ be two planar geometric graphs, and let $\pi_G \subseteq \Pi_G$ and $\pi_H \subseteq \Pi_H$. The
  directed {\em \pbgdistance} between these path sets is defined as:
  \begin{equation}\label{eq:pathbaseddist}
    \distance{\pi_G}{\pi_H} = \max_{p_G \in
\pi_G}\min_{p_H \in
    \pi_H}\delta_F(p_G,p_H).
  \end{equation}
\end{definition}
The undirected version of the distance, $d_{path}(\pi_G,\pi_H)$ is defined to be the maximum of the two directional
distances $\distance{\pi_G}{\pi_H}$ and $\distance{\pi_H}{\pi_G}$, similar to the undirected Hausdorff distance.  Like
the Hausdorff distance, the path-based distance is not symmetric, i.e., $\distance{\pi_G}{\pi_H}\neq
\distance{\pi_H}{\pi_G}$.  This anti-symmetry is desirable in our setting.  For example, $G$ could be the reconstructed
road network from bus route data.
In this case, the bus routes only correspond to a subgraph of the complete road network so the directed distance is more
informative.

The question now is how to define path sets $\pi_G$ and $\pi_H$ that
yield a suitable distance measure between $G$ and $H$.  Recall from
above that $\Pi_G$ is the set of all paths in $G$, and $\widehat{\Pi}_G$
is the set of all paths in $G$ that start and end in a vertex.
Ideally, $\pi_G=\Pi_G$ and $\pi_H=\Pi_H$, in order to capture the most
structure from both graphs. Interestingly, whether paths in $G$ start
and end in a vertex or anywhere on an edge, does not affect the
path-based distance.
\begin{lemma}[Vertex-Paths]
Let $G,H$ be two graphs.  Then, the following equality holds:
\label{lem-pihatequal}
  $\distance{\Pi_G}{\Pi_H}=\distance{\widehat{\Pi}_G}{\Pi_H}$.
\end{lemma}
\begin{proof}
For every $p\in \Pi_G$, there exists a path
$\widehat{p}\in \widehat{\Pi}_G$ such that $p$ is a sub-path 
of $\widehat{p}$, 
we
have that
$\distance{\Pi_G}{\Pi_H}\leq \distance{\widehat{\Pi}_G}{\Pi_H}$. 
In addition, from the inclusion
$\widehat{\Pi}_G\varsubsetneq \Pi_G$ follows
$\distance{\widehat{\Pi}_G}{\Pi_H}\leq \distance{\Pi_G}{\Pi_H}$.
\end{proof}

But still, these complete path sets are infinite in size, so an exhaustive
comparison is out of the question. Fortunately, if we restrict ourselves to a polynomial number of paths in $\pi_G$,
while $\pi_H=\Pi_H$, then $\distance{\pi_G}{\Pi_H} = \max_{p_G \in \pi_G} \delta_F (p_G,H)$ can be computed in
polynomial time, by computing a polynomial number of map-matching
distances~\mbox{\cite{aerwMPM03}.}

In what follows, we analyze the \pbgdistance\ for different subsets of paths $\pi_G \subset \Pi_G$ and fix
$\pi_H=\Pi_H$. In particular, we will closely examine what properties of the graphs $G$ and $H$ must be similar for
$\distance{\cdot}{\cdot}$ to be small when considering only paths of fixed 
link-length in $G$. For brevity, we define
the following notation for path-based distances for commonly used path sets.

\begin{definition}[Path-Based Distance for Common Path Sets]
\label{def-pbd_commonPathSets}
The overall path-based distance for graphs $G$ and $H$ is defined as $\Delta := 
\distance{\Pi_{G}}{\Pi_H}$. Let $v\in
V_G$ be a vertex, $e\in E_G$ be an edge, and let $k\geq 1$ be an integer. We 
define $\Delta_v :=
\distance{\Pi_{v}}{\Pi_H}, \Delta_e := \distance{\Pi_{e}}{\Pi_H}$, and 
$\Delta_{k} :=
\distance{\Pi_{G}^k}{\Pi_H}$. And we define $\Delta_{k,v} := 
\distance{\Pi_{v}^k}{\Pi_H}$ and $\Delta_{k,e} :=
\distance{\Pi_{e}^k}{\Pi_H}$.
\end{definition}

If two end vertices of an edge $e \in E_G$ are $v_1$ and $v_2$ then 
$\Delta_{k,e} \leq
\min\left(\Delta_{k,v_1},\Delta_{k,v_2}\right)$ as $\Pi_{e}^k \subseteq 
\Pi_{{v_1}}^k \cap \Pi_{{v_2}}^k$.
Observe that each path of \length\ two~$\langle uvw \rangle$ can be extended to 
a path of \length\ three by adding the second
edge backward $\langle uvwv \rangle$.  Extending this observation, we obtain 
the following: 
\begin{lemma}[Monotonicity]
 $\Delta_{k}$, $\Delta_{k,v}$ and $\Delta_{k,e}$ are non-decreasing in $k$.
\end{lemma}
\begin{proof}
 Let $p \in \Pi_G^k$.  We can then write $p=\langle v_1, v_2, \ldots, 
v_{k-1}, v_k \rangle$ for some set of vertices $v_1,\ldots, v_k \in G$.  Then, 
$p' = \langle v_1, v_2, \ldots, v_{k-1}, v_k, v_{k-1} \rangle$ is a vertex-path 
in $\Pi_G^{k+1}$.  Since $\distance{\{p\}}{\Pi_H} = \distance{\{p'\}}{\Pi_H}$ 
and $\{ p' | p \in \Pi_G^k\} \subseteq \Pi_G^{k+1}$, we can conclude that  
$\Delta_{k}$ is non-decreasing in~$k$.
 The proof is analogous for $p \in \Pi_{v}^k$ or $p \in\Pi_{e}^k $.
\end{proof}

Hence, it will be sufficient to compute $\lim_{k \to \infty} 
\distance{\Pi^k_G}{\Pi_H}$ in order to compute the
path-based distance between street map graphs $G$ and $H$. Here, we write $k \to 
\infty$ even though, for a finite graph
there is a maximum possible size for $k$ which, 
%
however, is computationally infeasible to use.

\subsection{Paths of \Length\ One}
\label{subsec-lengthone}
First, we consider the case where $\pi_H=\Pi_H$ is the set of all paths in $H$
and $\pi_G=\Pi^1_G=E_G$ consists of all paths
of \length\ one. In this case, one can only ensure that for
each edge $e\in E_G$ there is a similar
path $p_H$ in $H$ such that $\delta_F(e,p_H)\leq \distance{\Pi^1_G}{\Pi_H}$. As
$\Pi^1_G$ does not model the existence
of a vertex incident to two edges, a small $\distance{\Pi^1_G}{\Pi_H}$ does not
guarantee connections between edges of $H$. In
this case, different choices of $H$ with very different combinatorial structure
can have the same distance from $G$, as
shown in \figref{length12}.  Thus, the similarity of travel-paths is 
not captured using only link-length one paths.

\subsection{Paths of \Length\ Two}
Next, we consider the case where $\pi_H=\Pi_H$ and $\pi_G=\Pi^2_G$ consists of
all paths of \length\ two.  For this
case, we can define a correspondence between $V_G$ and $V_H$ under reasonable
assumptions on $G$ that guarantee the
vertices of $G$ are sufficiently spread out from each other and are at least of
degree four. The key observation here is
that $\Pi^2_{{v}}$ preserves all adjacency transitions around a vertex $v$,
because by definition $\langle v_ivv_j \rangle \in \Pi^2_{{v}}$ for all $v_i,
v, v_j\in V_G$ with $v_i, v_j\in
\Adj(v)$.  Here, we used $\Adj(v)\subseteq V$ to denote the set of vertices
adjacent to $v \in V$.

Letting $e_i$ be the edge $vv_i$, we define:

\begin{definition}[Intersection Radius]\label{def-int-radius}
  Given edges $e_0, e_1, e_2, \ldots, e_n$ with common endpoint $v$, we will
define the \emph{intersection radius} at
  scale $d$, denoted by $r_d(v)$.  Let $B$ be a ball of radius $r$ centered at
$v$.  We can then choose a point $w_i \in
  e_i \cap \partial B$ uniquely, if it exists, by starting at $v$ and walking
along $e_i$ until we reach $\partial B$.
  Let $r_d(v)$ denote the minimum radius such that $w_i$ exists for all $e_i$
and $|| w_i - w_j || > 2d$ for all $i\neq j$.
  If such a radius does not exist, then $r_d(v)=\infty$.
  We call $r_d(v)$ the intersection radius at $v$, and we define~$r_d = \max_{v
\in V} r_d(v)$.
\end{definition}

\begin{figure*}[tbph]
\centering 
\subfloat[Intersection region.]
{\includegraphics[height=1.5in]{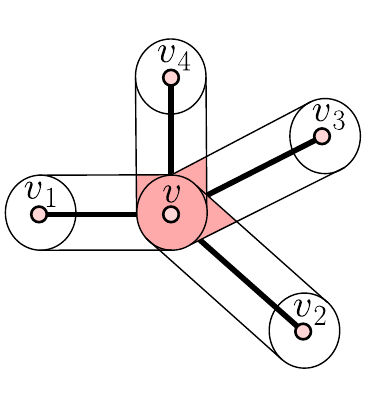}
\label{fig:verts-and-edges-vertregion}
}
\hspace{.15in}
\subfloat[Computing $||v-v'||$.]
{\includegraphics[height=1.5in]{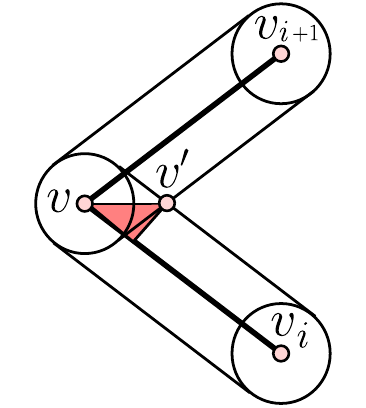}}
\hspace{.15in}
\subfloat[Vertex correspondence.]
{\includegraphics[height=1.2in]{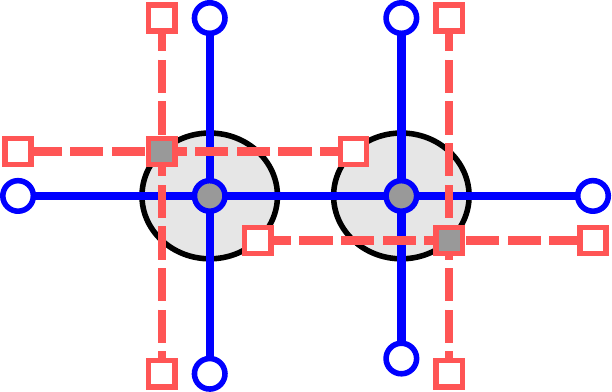}\label{
fig-vertcorr}}
\caption {(a) By definition, $\Pi^2_G$ contains all paths of \length\ two.  If a
vertex has at least four neighbors,
  then we can find a transverse intersection: $v_2vv_4$ and $v_1vv_3$.  (b) The
distance $d(v,v') =
  \Delta/\sin(\theta_v/2)$ is the hypotenuse of the right triangle shown in pink.
(c) Two graphs $G$ and $H$; edges in $G$
  are solid blue and edges in $H$ are dashed pink.}
\label{fig-length2}
\end{figure*}

\begin{figure}[tbph]
 \centering
 \includegraphics[height=1.5in]{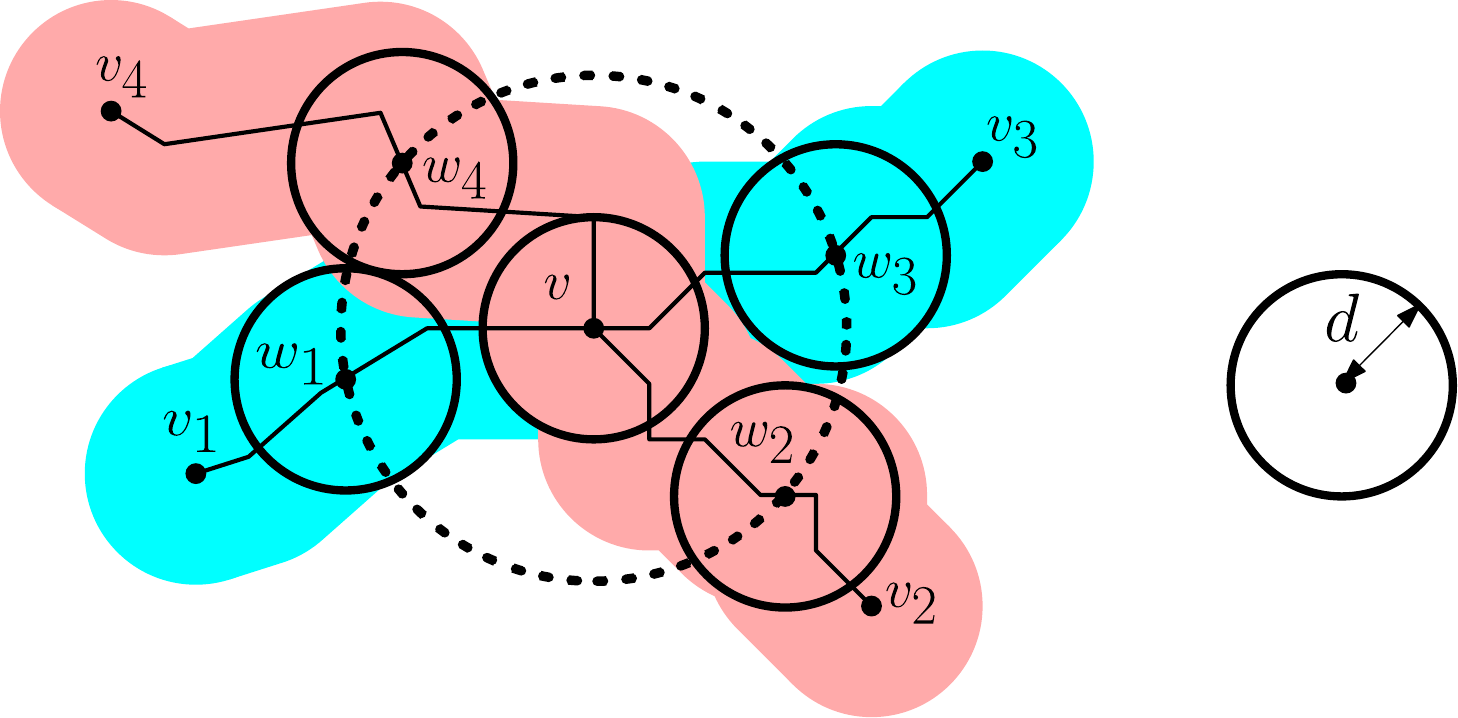}
 \caption{The intersection radius $r_d(v)$ is the smallest radius such that
each ball of radius $d$ centered at
   $\{w_i\}$ contains only one of the two paths.  In this example, $v$ is
$d$-separated since such a radius exists.
 }\label{fig-goodVertex-new}
\end{figure}

In the case that the edges are line segments, we denote with $\theta_v$ the 
smallest angle formed by
any two distinct edges at $v$, and we call this angle
the {\em minimum angle at vertex~$v$}.
In this case, the computation of $r_d(v)$ is straightforward:
\begin{lemma}[Straight Edge Intersection Radius]
\label{lem-theta}
  If all of the edges incident to~$v$ are line segments and if $r_d(v)$ is
defined, then $r_d(v) = d/\sin(\theta_v/2)$.
\end{lemma}
The distance $\Delta_{k,v}/\sin{(\theta_v/2)}$ is found by computing the
hypotenuse of the pink triangle in~\figref{length2}(b).

The theoretical guarantees that we give below work for \emph{well-separated
vertices} that have sufficiently high degree.
In particular, we are interested in the cases where $d = \Delta$,
$\Delta_v$, or $\Delta_{k,v}$ as defined at the end of the introduction to this
section. In this way, the specific
amount of vertex separation we require is dependent upon which subset of paths
our distance measure is being evaluated;
see~\figref{goodVertex-new}.

\begin{definition}[$d$-separated Vertex]
  A vertex $v \in G$ is $d$-separated if $r_d(v)$ is finite.
  \label{def-goodVertex}
\end{definition}

Having $d$-separated vertices of degree at least four implies that vertex-paths
crossing in~$G$ will
have corresponding crossing paths in $H$:

\begin{theorem}[Crossing Paths]
  Let $p_1$, $p_2$ be two vertex-paths in $G$ that intersect transversely at a
$d$-separated vertex $v$.  
If $p_1', p_2'$ are two paths in $H$ with $\delta_{F}(p_1,p_1'), 
\delta_{F}(p_2,p_2')< d$, then 
$p_1'$ and $p_2'$ must intersect.
%
\label{thm-crossings}
\end{theorem}
\begin{proof}
  Let $w_1$,$w_3$ (similarly, $w_2$,$w_4$) be the first intersections of $p_1$
($p_2$) with the ball $B$ of radius
  $r_d(v)$ centered at $v$, as shown in \figref{goodVertex-new}.  The path
connecting $w_1$ and $w_3$ has a
  corresponding path $\tilde{p}_1$ in $H$ within \Frd\ $d$.  Notice that this
path necessarily divides $B$ into two
  sets: one containing $w_2$ and one containing $w_4$.  Therefore, the path
$\tilde{p}_2$ in $H$ with \Frd\ at most $d$
  from the path connecting $w_2$ and $w_4$ must intersect~$\tilde{p}_1$.
\end{proof}

The previous theorem implies the following corollary that path-correspondences
between link-length two paths in $G$ to
paths in $H$ imply a guaranteed vertex correspondence between vertices in $G$ to
vertices in $H$.
\begin{corollary}[Vertex Correspondence]
  If a vertex $v \in G$ has degree at least four and is $\Delta_{k,v}$-separated
for some $k \geq 2$, then there exists a {\em
    corresponding} vertex $v'$ in $H$ such that $||v-v'|| \leq r_d(v) + d$.
\label{cor-length2}
\end{corollary}

Considering the example given in ~\figref{length2}(a), two paths that cross
transversely at $v$ in~$G$ will have a
corresponding vertex $v'$ in $H$ somewhere in the pink region containing $v$.


\subsection{Paths of \Length\ $k$ for $k \geq 3$}
\label{subsec-length3}
\nopagebreak

As we have seen in the last section, the \pbgdistance\ formed using the set
$\Pi^2_G$ of paths of \length\ two allows us
to define a particular vertex correspondence between $G$ and $H$ at vertices of
the road network that are not too
tightly clustered.  However, the distance $\distance{\Pi^2_G}{\Pi_H}$ can still 
be small for a connected
graph $G$ and a graph $H$ with multiple
connected components, as is the case in~\figref{length2}(c). In this section,
we 
analyze what
additional guarantees can be provided by
considering $\Pi^k_G$ for $k \geq 3$.


We prove that
$\distance{\Pi_G}{\Pi_H}$ can be approximated by $\distance{\Pi^3_G}{\Pi_H}$ as
long as assumptions about how much
vertices in $G$ are clustered are met. This is accomplished by showing that if
all \length\ three paths in $G$ have a
\emph{close} path in $H$, then for any path in $G$,
there is a \emph{close} path in $H$ as
well. This yields a polynomial-time algorithm to approximate
$\distance{\Pi_G}{\Pi_H}$ when all vertices in $G$ are
well-separated, overcoming the infinite complexity of using the full set of
paths, $\Pi_G$, to define our path-based
distance.

The following theorem shows that path-correspondences between link-length three
paths in $G$ to paths in $H$ suffice to
guarantee correspondences for longer paths (of link-length~four).

\begin{theorem}[Link-Length Three Surgeries]\label{thm-length4}
  Let $p=v_0v_1v_2v_3v_4$ be a vertex-path of \length\ four in~$G$, such that each
vertex $v$ in $p$ is $\Delta_{3,v}$-separated and
  does not have degree three. Then $\delta_F(p,H)\leq 2r_d(v_2) + d$, where
$d=\Delta_{3,v_2}$.
\end{theorem}

\begin{proof}
The general idea of this proof is as follows:
We will find two paths in $G$, which intersect at $v_2$.  Then, we will find 
the corresponding paths in $H$ and stitch them together to find a path close 
to~$p$.

We examine vertex $v_2$, which has degree at least four.
Let $v_2^a$ and $v_2^b$ be two neighbors of~$v_2$ which are neither $v_1$ nor
$v_3$;
  see~\figref{length3up} for the two possible configurations.  Let $p_{ab}$ be 
the path $v_2^a v_2 v_2^b$ and
$p_{13}$ be the path $v_1 v_2 v_3$.
\begin{figure}[h!tb]
  \centering
  \subfloat[Case a: surgery on transverse paths.]
  {\includegraphics[width=.45\textwidth]{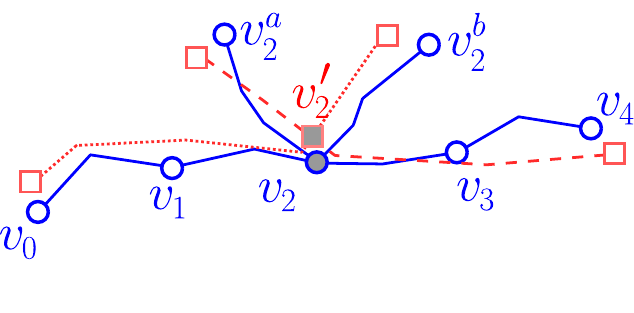}}
  \hspace{.25in}
  \subfloat[Case b: surgery on parallel paths.]
  {\includegraphics[width=.45\textwidth]{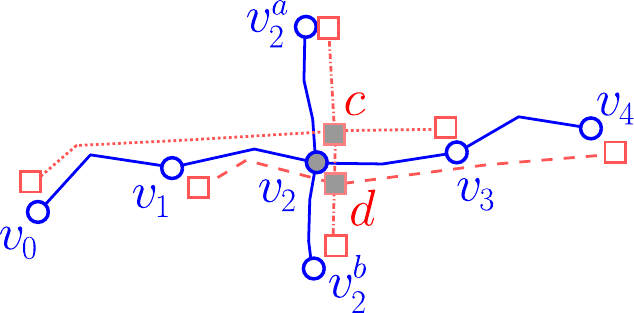}
  }
  \caption{We see two configurations of graphs $G$ (solid blue lines) and $H$
  (dashed pink lines).  In \thmref{length4},
    we show that for each path of link-length $\geq 4$, we can construct a
  corresponding path in $H$ by performing surgery
    on link-length three paths.  The pink paths shown are at most $\Delta_{3,v}$
  from the \length\ three paths.}
  \label{fig-length3up}
\end{figure}

  {\it Case a:} We first assume that $p_{ab}$ and $p_{13}$ form a non-transverse
intersection, as in
  Figure~\ref{fig-length3up}(a).  Consider the transverse paths
$p_{0b}=v_0v_1v_2v_2^b$ and $p_{a4}=v_2^av_2v_3v_4$. Let
  $p_{0b}'$, $p_{a4}'$ be the Fr\'echet-closest paths in $H$ to $p_{0b}$ and 
$p_{a4}$, respectively.  We observe that $\delta_F(p_{0b},p_{0b}')$ and
$\delta_F(p_{a4},p_{a4}')$ are at most
  $d=\Delta_{3,v_2}$, since $v_2 \in p_{0b} \cap p_{a4}$.  
  
  Let $M_a$ be a Fr\'echet-correspondence between $p_{a4}$ 
and $p_{a4}'$; 
likewise, let $M_b$ be a Fr\'echet-correspondence between
 $p_{0b}$ and $p_{0b}'$, as defined in \corref{induced-corresp}.  
Then, let
$\tilde{v}_2^{0b}=M_b(v_2)$ and $\tilde{v}_2^{a4}=M_a(v_2)$.
We know that $|v_2-\tilde{v}_2^{*}| \leq \Delta_{3,v_2}$ since $M_a$ and $M_b$ 
are Fr\'echet-correspondences.

\begin{figure}
\centering
{\includegraphics[height=1.2in]{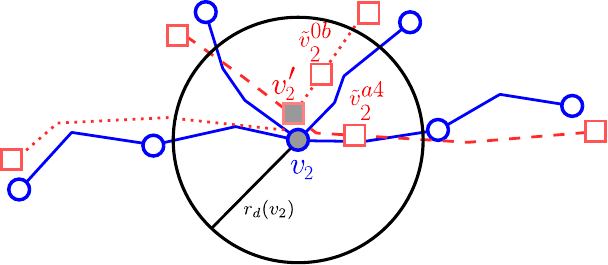}}
\caption{We focus on the intersection region of \figref{length3up}(a).  We draw 
a circle of radius $r_d(v_2)$ centered at
$v_2$.  The path $\rho_{02}'$ starts outside the circle 
and ends at $v_2'$.  It is obtained by shortening the path $p_{02}'$, which ends
at $\tilde{v}_2^{0b}$.  The path $\rho_{24}$ is obtained by extending the path 
$p_{24}$, moving the start vertex from $\tilde{v}_2^{a4}$ to $v'_2$.
}
\label{fig-length3up-case-b-proof}
\end{figure}
Next, we find an intersection of $p_{0b}'$ and $p_{a4}'$ close to 
$\tilde{v}_2^{0b}$ and $\tilde{v}_2^{a4}$. If 
$\tilde{v}_2^{0b} 
= \tilde{v}_2^{a4}$, then we have already found that intersection.  
Otherwise, assume $\tilde{v}_2^{0b} \neq 
\tilde{v}_2^{a4}$, as shown in \figref{length3up-case-b-proof}.   Let $\B$ be 
the 
ball of
radius $r_d(v_2)$ centered at~$v_2$.  Let $e_1$ be the edge $v_2v_1$, $e_3$ the 
edge $v_2v_3$, $e_2^a$ the edge $v_2v_2^a$, and $e_2^b$ the edge $v_2v_2^b$.  
Then, we can let $w_i$ (respectively, $w_i^j$) be the first intersection of 
the edge $e_i$ $(e_i^j)$
with $\partial \B$, as in \defref{int-radius}.
Furthermore, we can partition~$\partial \B$ into three sets: $A_{0b}:=$ points
within~$\Delta_{3,v_2}$ of~$\{ w_1,w_2^b \}$, $A_{a4}:=$ points
  within~$\Delta_{3,v_2}$ of $\{ w_2^a, w_3\}$, and the leftover points.  In
particular, each of the first two sets has exactly
  two connected components: $A_{0b} = A_{0b}^1 \sqcup A_{0b}^2$ and $A_{a4} =
A_{a4}^1 \sqcup A_{a4}^2.$ We can think of
  $A_{0b}^1$ and $A_{0b}^2$ as corresponding to $p_{0b}$ entering and leaving
$\B$; see
  \figref{verts-and-edges-partition}, where $A_{0b}$ is in cyan and $A_{a4}$ is
in red.
\begin{figure}
\centering
{\includegraphics[height=1.2in]{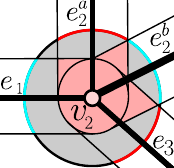}}
\caption{Since $v_2$ is $d$-separated for $d =
{\Delta_{3,v}}$ ,
  the boundary of the gray disc $\B$ can be decomposed into three parts, shown
in cyan (region $A_{0b}$), red (region $A_{a4}$), and black (neither).
}
\label{fig-verts-and-edges-partition}
\end{figure}

Consider $p_{0b}' \cap \B$, which could have multiple connected components.
Since $v$ is $d$-separated, there exists a
unique subpath of $p_{0b}'$ that enters on $A_{0b}^1$ and leaves on $A_{0b}^2$.
We call this subpath~$q_{0b}'$.
Similarly, there exists a unique subpath~$q_{a4}'$ of~$p_{a4}'$ that enters $\B$
on one of $A_{a4}^1$ or $A_{a4}^2$ and
leaves on the other.  Notice that $q_{0b}'$ and $q_{a4}'$ necessarily intersect
at least once by \thmref{crossings}.
Let $v_2'$ be one of those intersections.  We note that we can uniquely
choose an intersection by expanding the
subpaths around $\tilde{v}_2^{0b}$ and $\tilde{v}_2^{a4}$ until an intersection
is found.  Furthermore, the distance between $v_2$ and $v_2'$ is at 
most $r_d(v_2) + d$ by \corref{length2} 

We wish to perform surgery on the paths $p_{0b}'$ and $p_{a4}'$ in order to 
upper bound the \Frd\
between $p$ and $H$. 
Let $p'_{02} = \overline{M}_b(p_{02})$ and $p'_{24} = \overline{M}_a(p_{24})$; 
notice that both
$\delta_F(p_{02},p'_{02})$ and $\delta_F(p_{24},p'_{24})$ are at most 
$\Delta_{3,v_2}$.

Next, we
find a path $\rho_{02}'$ in $H$ Fr\'echet-close to
$p_{02} \subset p_{0b}$ that begins at $v_0'$ and ends at $v_2'$.  
In fact, we almost have that path already.  Informally,
the path that starts with $p'_{02}$ and
is either extended or shortened so that $v_2'$ is an endpoint.  We elaborate 
on the two scenarios (extending and shortening):
\begin{enumerate}
 \item[1. (Extending).]
First, suppose we need to extend $p_{02}'$, as is
the case when $v_2'$ is not on the path $p_{02}'$; see
\figref{length3up-case-b-proof}.
Here, we observe that $d_F(p_{0b},p_{0b}')
\leq d$ and $d_F(p_{02},\rho_{02}') \leq
r_d(v_2)+d$.

\item[2. (Shortening).]
Second, suppose we need to shorten $p_{02}'$.  Let $u=M_b^{-1}(v_2')$.  Observe 
$|v_2'-u| \leq d$ and
$|v_2'-v_2| \leq r_d(v_2)+d$, and for any $x$ on
the path between $v_2$ and~$u$, we have $|v_2'-x| \leq 2r_d(v_2) + d$.
\end{enumerate}

Thus, we have proven that
\begin{equation*}
\delta_F(p_{02},\rho'_{02})\leq 2r_d(v_2) + d
\end{equation*}
Using a similar argument, we can also obtain:
\begin{equation*}
\delta_F(p_{24},\rho'_{24})\leq 2r_d(v_2) + d
\end{equation*}
Hence, concatenating these two subpaths yields the path $p'$, which has \Frd\ at
most $2r_d(v_2) + d$ to~$p$.

{\it Case b:} We now assume that $p_{ab}$ and $p_{13}$ form a transverse
intersection, as illustrated in~\figref{length3up}(b).  We
observe that the path~$p$ consists of two paths $p_{03}=v_0v_1v_2v_3$ and
$p_{14}=v_1v_2v_3v_4$ of \length\ three, which
both have the subpath $p_{13} = v_1v_2v_3$ in common. Let $p_{03}'$, $p_{14}'$, 
$p_{ab}'$ be paths
in $H$ such that
$\delta_F(p_{03},p_{03}')$, $\delta_F(p_{14},p_{14}')$,
$\delta_F(p_{ab},p_{ab}')$ are at most~$\Delta_{3,v}$.  By an
argument analogous to the one above, we know that $p_{ab}'$ intersects both
$p_{03}'$ and $p_{14}'$. Denote these
intersection points with $c$ and $d$, respectively, as depicted in
\figref{length3up}(b).  (If there are multiple such
intersections, one can choose $c$ and $d$ arbitrarily among the valid choices).

Both $c$ and $d$ lie within distance $r_d(v_2)+d$ from $v_2$, by
\corref{length2}.  Furthermore, $c$ and~$d$ are
connected with a portion of~$p_{ab}'$ that lies completely within $\B$; denote
this subpath by~$\rho_{ab}'$.  Analogous
to {\it Case a} above, we can choose paths~$\rho_{02}$ ending at $c$ and
$\rho_{24}$ starting at~$d$ such that both
$\delta_F(p_{02},\rho_{02}')$ and $\delta_F(p_{24},\rho_{24}')$ are at most
$2r_d(v_2)+d$.  Concatenating the three
subpaths $\rho'_{02} \rho'_{ab} \rho'_{24}$ yields the path $p'$, which has
\Frd\ at most $2r_d(v_2)+d$ to~$p$.

Finally, we remark that this proof does not assume that the vertices and edges
in the paths are distinct, as long as the
assumptions stated in the theorem are satisfied.  In particular, it is possible
that $v_0=v_4$ or that the intersection
of two paths is a set of edges.
\end{proof}

The theorem below summarizes our main result for graphs with
$\Delta_3$-separated vertices that are not of degree three. If not all vertices
fulfill this property, then we can restrict our attention to a subgraph of $G$
containing only $\Delta_3$-separated vertices of sufficient degree.  In
the next section, we will show empirical evidence that requiring all vertices to
be \mbox{$\Delta_3$-separated} and not of degree three can be relaxed.
\begin{theorem}[Link-Length Three is Sufficient]
  If $G$ consists of only \mbox{$\Delta_3$-separated} vertices and no vertex of
degree
three, and if the distance between any two adjacent vertices is at least $2(r_d
+ d)$ for $d=\Delta_3$, then $\distance{\Pi_G}{\Pi_H} \leq 2r_d + d$.
\label{thm-final}
\end{theorem}

\begin{proof}
\begin{figure}
\centering
\includegraphics[height=1.5in]{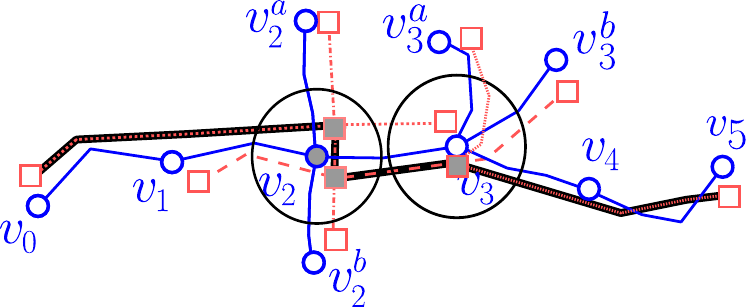}
\caption{Here we see three paths in $H$ (pink dashed lines) stitched together to
create a path close to a link-length
  five path in $G$ (blue solid lines).}
\label{fig-sufficient}
\end{figure}
Given a link-length $n$ vertex-path $p$ in $G$ for $n > 4$, we show how to find 
a path
in $H$ that is at most \Frd\ $2(r_d +
d)$ from $p$. 
Let $p$ be the path $v_0 v_1 \ldots v_{n-1} v_n$; for example,
see \figref{sufficient}.

Vertices $v_i$ for $i=1, 2, \ldots n-1$ must have degree at least four, since they are not end vertices and they are not
of degree three.  Therefore, for $i = 2,3, \ldots, n-2$, we can choose a vertex $u_i$ adjacent to $v_i$ so that there
exists at least one vertex between $u_i$ and $v_{i+1}$ in both a clockwise and a counter-clockwise ordering around
$v_{i}$. Likewise, we can also choose a vertex $w_i$ adjacent to $v_i$ so there exists at least one vertex between $w_i$
and $v_{i-1}$ in both a clockwise and counter-clockwise ordering around $v_i$ for $i = 2, 3, \ldots, n-2$. For example,
in \figref{sufficient}, $u_2 = v_1$, $w_2= v_3$, $u_3=v_3^a$ and $w_3=v_3^b$.
We define a set of paths that covers $p$.  The first path we consider is $p_1 = v_0v_1v_2 w_2$. The last path is
$p_{n-2} = u_{n-2}v_{n-2} v_{n-1} v_n$.  In between, for each edge $v_i v_{i+1}$, we add the path $p_i = u_i v_i v_{i+1}
w_{i+1}$.  Notice that each path corresponds to one edge in $p$, except the first and last paths, which each correspond
to two~edges.

Next, we Fr\'echet-match $p_i$ in $G$ to $p_i'$ in $H$ for $i=1,2,\ldots, n-2$, 
and perform surgeries on these paths in order to find a path $p'$ in $H$ that is
close to~$p$.  Notice that for each $i$, the Fr\'echet distance
between $p_i$ and $p'_i$ is at most $d$.  Let $M_i$ be the 
Fr\'echet-correspondence from $p_i$ to $p'_i$.

At each vertex $v_i$ for $i=2, 3, \ldots, n-2$, we find $v_i'$ and
perform surgeries between
$p_{i-1}'$ and $p_i'$, as described in
\thmref{length4} for $v_2$.  We notice that the induced correspondences between 
paths
after surgery are consistent with the
correspondences before surgery for the parts of $p$ outside of the balls of 
radius $r_d(v_i) + d$ centered at $v_i$.
Therefore, two adjacent vertices~$v_i$ and
$v_{i+1}$ separated by $2(r_d+d)$ can be combined by using the common 
correspondence on the part of the
edge $v_i v_{i+1}$ that is outside of the
balls of radius $r_d+d$ around $v_i$ and $v_{i+1}$.  Letting $p'$ denote the 
path in $H$ after surgery, we have $\delta_F(p,p') \leq 2 r_d + d$.
Together with \lemref{pihatequal}, this proves the claim.
\end{proof}

Let $m$ and $n$ be the the total number of vertices and edges in $G$ and $H$,
respectively.  Assume further that the edges consist of $O(1)$ line segments.
There are $|\Pi^3_G|\in O(m^3)$ paths of link-length three in $G$,
and their total complexity is $O(m^4)$. Using the map-matching algorithm of 
\citeN{aerwMPM03}, 
$\distance{\Pi^3_G}{\Pi_H}$ can be computed in $O(m^4 n\log^2 n)$ time, and
from \thmref{final} follows
that $\distance{\Pi_G}{\Pi_H}$ can be approximated in the same time. 

If all edges in $G$ are line segments, then the total complexity of all
link-length three paths in $G$ is $O(m^3)$, and Lemma~\ref{lem-theta} and 
\thmref{final} yield the following:

\begin{corollary}[Approximation]\label{cor-runtime}
  If all edges in $G$ are line segments, no vertex in $G$ is of degree three, and 
the distance between any two adjacent vertices is at least
$2\Delta_3(1+1/\sin(\theta/2))$, where $\theta$ is the smallest angle formed by
two incident edges. Then $\distance{\Pi_G}{\Pi_H} \leq
2\Delta_3(1+1/\sin(\theta/2))$, and this approximation can be computed in
 $O(m^3 n\log^2 n)$ time. 
\end{corollary}

In order to prove \thmref{crossings} (Crossing Paths) and 
\thmref{final} (Link-Length Three is Sufficient), we needed to use the 
assumption that all intersections are of degree four or more.  The reason we 
need to assume this is in order to obtain a vertex correspondence.  By using 
link-length two paths in $G$, we can find two transverse paths in $H$ and hence 
an~intersection.  

\begin{figure}
 \centering
  \subfloat[Missing Intersection]{\includegraphics[height=1.25in]{%
   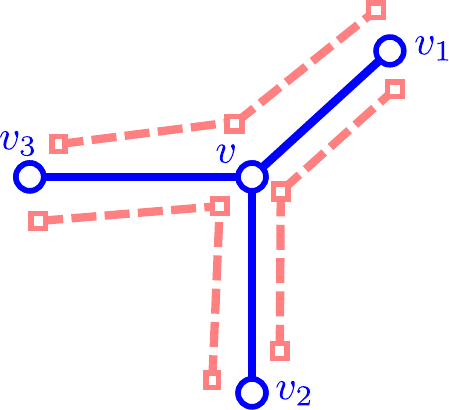}\label{fig-intersections-deg3}}\hspace{.2in}
 \subfloat[Graph $G$]{\includegraphics[width=.3\textwidth]{%
   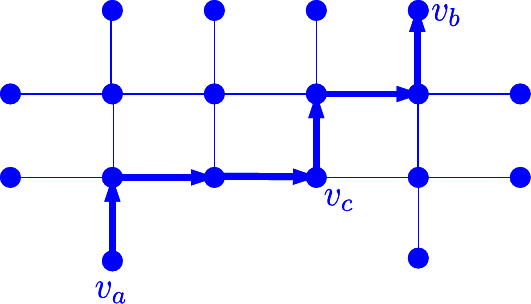}\label{fig-degree3-case-a}}
   \hspace{.2in}
 \subfloat[Graph $H$]{\includegraphics[width=.3\textwidth]{%
   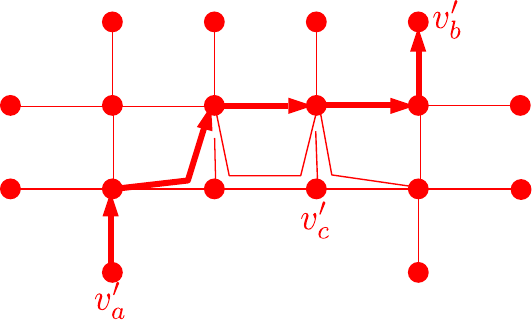}\label{fig-degree3-case-b}}
 \caption{
 A vertex correspondence can not be guaranteed  for a degree three vertex. (a)
The blue and the pink graphs have a small path-based distance, however no
vertex in the pink graph corresponds to $v$.
The graph $G$ in blue (b) and $H$ in
red (c) illustrate an example where $\distance{\Pi_{G}^3}{\Pi_H}$ is very small,
but the path-based distance can be
arbitrarily large. The closest correspondence of the bold path in $G$ is the
bold path in $H$.
}
\label{fig-degree3}
\end{figure}

When the vertices have degree three, then the link-length two 
paths can be close without an intersection occurring. 
\figref{intersections-deg3} is an example of such occurrence with one degree three vertex.  
Figures~\ref{fig-degree3-case-a} and~\ref{fig-degree3-case-b} show a contrived
example where $\distance{{\Pi^3}_G}{\Pi_H}$ is small but path-based distance can
be arbitrarily large. The closest correspondence of the bold path in $G$ is the
bold path in $H$ and their \Frd\ can not be bounded using link-length three
paths.

In \secref{experiments}, we observe that vertices of degree three are common in
road networks; see Tables~\ref{table-OSMstat}
and~\ref{table-TAstat}.  However, despite the datasets not meeting the
assumptions of our theorems, we still can use the path-based distance to
capture differences between the graphs.  In particular,
 we observe that there are two main types of 
dissimilarities between graphs from real city data: missing turns and missing
streets.
Both of these differences can be identified using (edge or vertex)
signatures; see \figref{lengthone} and \figref{linkbsmall}.  Next, we define
these signatures.

\subsection{Path-Based Signature}
\label{subsec-signature}

One benefit of the path-based distance measure is that there is a natural local signature that can be defined.  Given an
edge $e\in E_G$, we ask what is a tight upper bound for the Fr\'echet distance of paths going through that edge?  For a
connected graph, this would be a constant value, if we do not restrict the types of paths that we consider.  Instead, we
look at paths of a fixed link-length $k$.
For an edge $e\in E_G$ and for a given integer $k\geq 1$, the quantity 
$\Delta_{k,e}=\distance{\Pi_{e}^k}{\Pi_H}$
captures the distance of all link-length $k$ paths through $e$. In that sense, it represents a {\em local signature} for
the edge $e$ describing the local structural similarity of a sub-graph centered at $e$ to a subgraph in $H$. This local
signature can now be used to identify how well portions in $G$ correspond to portions in $H$. In
\subsecref{experimentsSignature}, we provide two approaches for using this
signature: First, we compute heat-maps that
map the value of the signature onto the graph edges, in order to visualize the degree of local similarity captured by
the signature. Second, we summarize the local signatures in a cumulative distribution function in order to capture a
summary of the local graph similarity.

Note that a signature could also be defined for each vertex $v$ by considering $\Delta_{k,v}$, however for visualization
purposes and for capturing the overall length of all edges in the graph, it appears more suitable to use $\Delta_{k,e}$.

\section{Experimental Results}
\label{sec-experiments}
In this section, we present our experimental results. We
implemented Java code to compute the \pbgdistance\ defined above. 
Besides using real street-maps from Berlin and Athens we used a set 
of perturbed graphs to analyze our distance measure; see \subsecref{controlled}.
Our code is available on {\tt mapconstruction.org}.

\subsection{Datasets and Runtimes}
We test our algorithm using map data from Berlin and Athens.  For Berlin, we
have maps from two sources:  TeleAtlas (TA) from 2007 and
OpenStreetMap (OSM) from April 2013.  We have both small (16 km$^2$) and large (2500
km$^2$) maps of Berlin. Similarly, for Athens, we have TA maps from 2007 and OSM maps
from 2010.  To compute the \pbgdistance, we selected several rectangular
regions with the same
coordinates from each map; \tableref{datasets} contains the UTM coordinates of
the
southwest-most and northeast-most corners of the rectangular
regions, and Tables \ref{table-OSMstat} and \ref{table-TAstat} contains some
statistics about the datasets.  From these tables, we see that the OSM Berlin
maps contain more vertices and edges than the TA maps; however, the OSM
Athens-small map contains slightly fewer vertices and edges than the TA map.
\begin{table}[tp]
\centering
\tbl{Data Set Regions.\label{table-datasets}}
{
\begin{tabular}{|l|r|r|r|r|r|}
\hline
Data set & xLow & xHigh & yLow & yHigh& Area\\
\hline
\asmall\ &$480,000$ m&$484,000$ m&$4,206,000$ m&$4,210,000$ m&$4$ km $\times$
$4$ km\\
\bsmall\ &$390,000$ m&$394,000$ m&$5,817,000$ m&$5,821,000$ m&$4$ km $\times$
$4$ km\\
\blarge\ &$375,000$ m&$425,000$ m&$5,775,000$ m&$5,825,000$ m&$50$ km $\times$
$50$ km\\
\hline
\end{tabular}}
\begin{tabnote}
  The data set regions are defined by the extreme southwest $(xLow,yLow)$ and
the extreme northeast $(xHigh,yHigh)$ UTM coordinates.
\end{tabnote}
\end{table}

\begin{table}[t]
\centering
\tbl{Statistics for OSM maps.\label{table-OSMstat}}{
\begin{tabular}{|l|r r  |r |r|}
\hline
&\multicolumn{2}{c|}{\# vertices}&\# edges& total length\\
&all&$degree \neq 3$&&\\
\hline
\asmall\ &2,318&1,026&3,758&323 km\\
\bsmall\ &2,166&841&3,051&267 km\\
\blarge\ &87,395&31,777&123,863&17,552 km\\
\hline
\end{tabular}}
\end{table}

\begin{table}[t]
\centering
\tbl{Statistics for TA maps.\label{table-TAstat}}{
\begin{tabular}{|l|r r  |r |r|}
\hline
&\multicolumn{2}{c|}{\# vertices}&\# edges& total length\\
&all&$degree \neq 3$&&\\
\hline
\asmall\ &2,770&1,210&4,343&339 km\\
\bsmall\ &1,507&634&2,303&227 km\\
\blarge\ &49,605&19,897&72,650&10,426 km\\
\hline
\end{tabular}}
\end{table}

\begin{table}[t!]
\tbl{Runtimes.\label{table-runtime}}
{
\begin{tabular}{|l|r|r r|r|r|}
\hline
Dataset&Link & Teleatlas & OSM to &  Execution & Machine\\
&Length & to OSM & Teleatlas &  Mode & Specification\\
\hline
& LinkOne &4.9 min&6.1 min& &\\
\asmall\ & LinkTwo &22.0 min&27.8 min& &Intel(R) Xeon(R)\\
& LinkThree &74.9 min&98.4 min& sequential &CPU E3-1270 v2\\
\cline{1-4}
 & LinkOne & 5.8 min & 5.6 min & &$@3.5GHz$ $8GB$ Ram\\
\bsmall\ & LinkTwo &24.4 min & 22.0 min &  & \\
& LinkThree &78.4 min & 69.0 min &  & \\
\hline
 &LinkOne& 4.0 h& 4.5 h& &\url{http://www.cbi.utsa.edu/}\\
 \blarge\ &LinkTwo& 15.0 h & 20.0 h &parallel &\url{hardware/cluster}\\
 &LinkThree& 49.0 h& 53.0 h& &\\
\hline
\end{tabular}
}
\end{table}

The path-based algorithm for computing the distance between two maps is scalable
and
has reasonable runtime; for the \bsmall\ dataset, it took $24.4$
minutes to
map all \length\ two paths of OSM to the TeleAtlas map.
The runtime and machine specification for corresponding experiments are
summarized in \tableref{runtime}.
Although this algorithm does not need to run in real-time, the computation can
be sped up by incorporating an efficient data structure for spatial search. As
the runtime is not our main focus in our current implementation, we perform
an exhaustive
search to find all points on the street-map which are close
to the start vertex of a path.

The algorithm can be trivially parallelized by decomposing the set of paths in
the first graph into
multiple sets and finding their (Fr\'echet-)closest paths in the
second graph independently. We implemented this parallel version of the
path-based distance computation.
For the \blarge\ dataset, there were almost $50,000$ paths to consider.  When
we executed the parallel computation, we used $25-250$ threads, resulting in
runtimes listed in \tableref{runtime}.
This simple
parallelization allowed for the path-based distance to be computed on the
\blarge\ map in only two days time, as opposed to taking weeks to compute.


\subsection{Computing the Path-Based Distance}
\label{subsec-goodVertices}
The results presented in \secref{distance} require that the graphs are
$d$-separated and that no vertex is of degree three. 
In the datasets that
we use in this section, $55-64\%$ of the vertices have degree three; see 
Tables~\ref{table-OSMstat} and~\ref{table-TAstat}.  If we
relax the degree three condition, then we can compare paths in one graph to
the (Fr\'echet-)closest path in the other graph.   In our experiments, we 
allow degree three vertices, at the cost of having a slightly less informative 
distance measure.
As discussed in \subsecref{length3}, even though there are contrived examples for which the approximation guarantees of our
theorems do not apply, we can still use the path-based distance to
capture differences between the graphs.

\begin{figure}[htb]
    \centering
    \subfloat[Spatial.]
    {\includegraphics[scale=0.6]{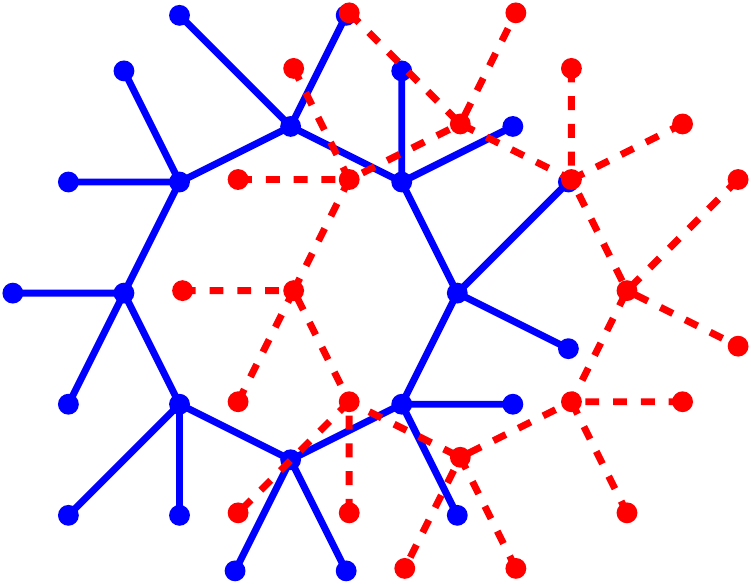}}
    \subfloat[Topological.]
    {\includegraphics[scale=0.6]{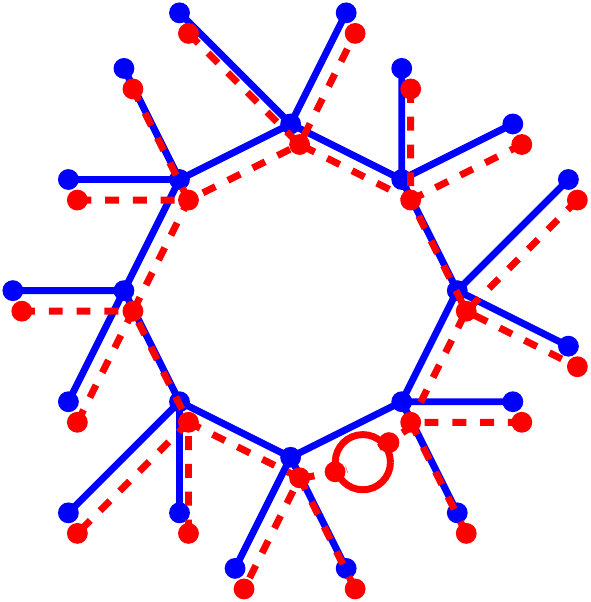}}
    \subfloat[Geometric.]
    {\includegraphics[scale=0.6]{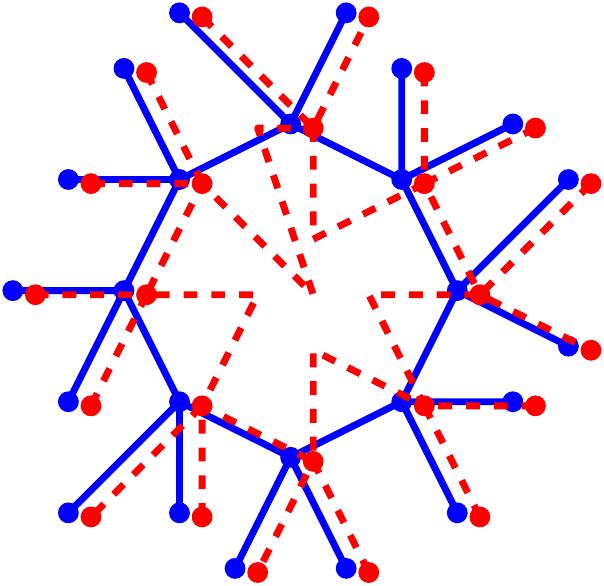}}
    \caption{Three typical examples where vertices that are not
    well-separated may arise: when one graph is a translation of the
    other, when one graph has more topological structure than the other, 
    and when
    seemingly corresponding edges (according to the topology of the graphs) have
    very different geometric embeddings.}
    \label{fig-badVertex}
\end{figure}
\begin{figure}[t]
  \centering 
  \subfloat[Mostly well-separated region.]{\includegraphics[height=1.5in]
  {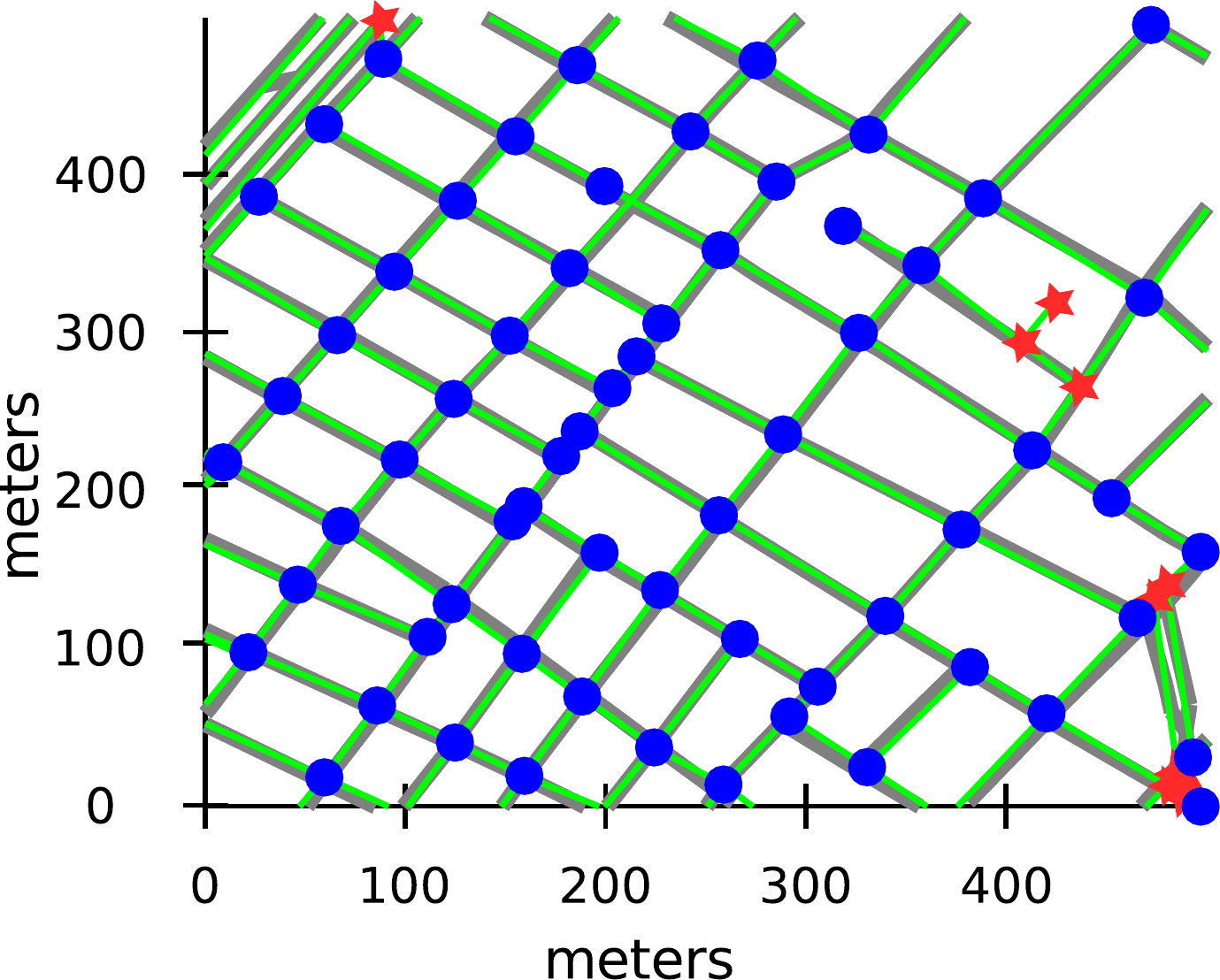}}
  \hspace{.2in}
  \subfloat[Mostly not well-separated region.]{\includegraphics[height=1.5in]
  {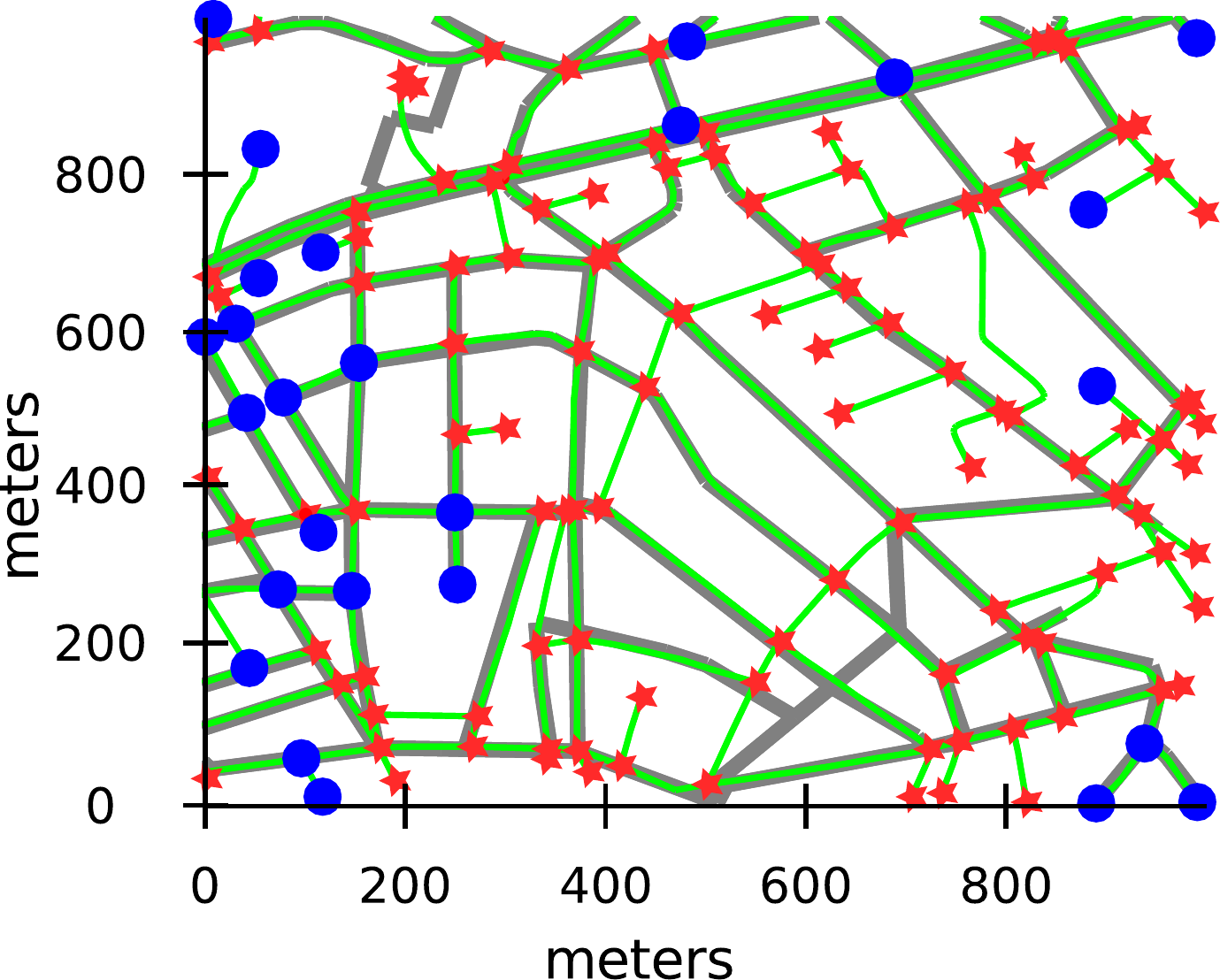} }
  \caption{We plot portions of the OSM and TA maps for
  the \asmall\ dataset\footnotemark.  The $\Delta_3$-separated
    vertices are displayed as blue circles and the other vertices as red stars.
  In (a), we see OSM
    (green) overlayed on TA (gray) where the road networks are very similar.  In
  (b), we see TA (green) overlayed on OSM
    (gray) where the TA map has more streets than the OSM one, resulting in a
  large number of vertices that are not $\Delta_3$-separated.
  }
  \label{fig-goodvsbad}
  \end{figure}

We look to Figures~\ref{fig-badVertex} and~\ref{fig-goodvsbad} to understand the
settings for which vertices are not well-separated.
\figref{badVertex} shows three
generic examples: poor separation due to spatial, topological, or geometric
differences in the graphs.
In practice, we see regions of well-separated vertices
and regions of vertices that are not well-separated; see \figref{goodvsbad}.
  \addtocounter{footnote}{-1}
  \footnotetext{The $x$
  and $y$-coordinates are the offset (in meters)
  from an arbitrary location, given in UTM
  coordinates.  That location is UTM Zone 34S, $481600$ meters east,
  $4208500$ meters
  north in (a) and UTM Zone 34S, $483000$ meters east, $4206000$ meters
  north in (b).}

When the vertices are not $d$-separated for sufficiently small $d$, then
discerning the topological structure of the individual maps -- let alone the
differences between them -- becomes difficult.
For our three datasets, we counted the number of $d$-separated vertices for
$d=\Delta_1, \Delta_2$ and $\Delta_3$; see
\tableref{goodbad}.  If we look at $\Delta_1$-separated vertices for the
\bsmall\ dataset, we see that $54\%$ ($1,159$ out
of $2,166$) vertices are well-separated.  Similarly, we found that $76\%$ of the
TeleAtlas vertices in \bsmall\ are
$\Delta_1$-separated.

\begin{table}[t]
 \centering
  \tbl{Statistics about well-separated vertices.\label{table-goodbad}}
  {
  \begin{tabular}{|l|c|r r|r r|}
  \hline
  & &\multicolumn{2}{c|}{\# vertices}&\multicolumn{2}{c|}{\#$d$-separated
vertices}\\
  Dataset&$d$&OSM&Teleatlas&OSM&Teleatlas\\
    \hline
  &$\Delta_{1}$& & &2,076 $\left(90\%\right)$&2,020 $\left(73\%\right)$\\
  \asmall\ &$\Delta_{2}$& 2,318 & 2,770
&1,803 $\left(78\%\right)$&1,616 $\left(58\%\right)$\\
  &$\Delta_{3}$&&&1,490 $\left(64\%\right)$&1,215 $\left(44\%\right)$\\
  \hline
           &$\Delta_{1}$& & &1,159 $\left(54\%\right)$&1,149
$\left(76\%\right)$\\
  \bsmall\ &$\Delta_{2}$& 2,166 & 1,507
&811 $\left(37\%\right)$&897 $\left(60\%\right)$\\
           &$\Delta_{3}$&  &&510 $\left(24\%\right)$&665 $\left(44\%\right)$\\
  \hline
   &$\Delta_{1}$&&&38,402 $\left(44\%\right)$&34,723 $\left(70\%\right)$\\
  \blarge\ &$\Delta_{2}$&87,395&49,605&27,445 $\left(31\%\right)$&32,344
$\left(65\%\right)$\\
   &$\Delta_{3}$&&&18,898 $\left(22\%\right)$&24,963 $\left(50\%\right)$\\
   \hline
\end{tabular}
  }
\end{table}

\addtocounter{footnote}{-1}
\begin{figure}[t]
\subfloat[$\Delta_1$]{\includegraphics[width=0.3\textwidth]
{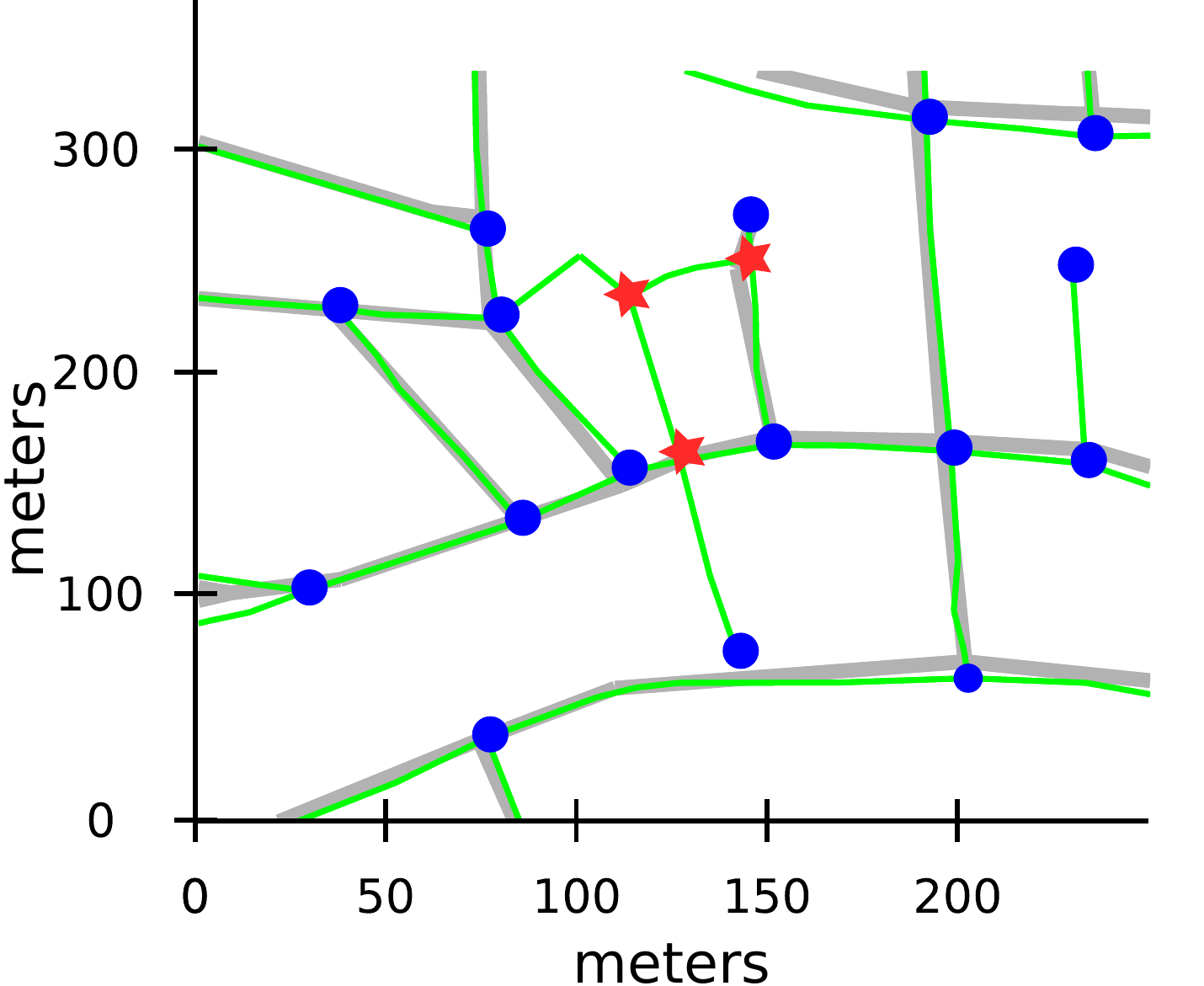}}
\hspace{0.1in}
\subfloat[$\Delta_2$]{\includegraphics[width=0.3\textwidth]
{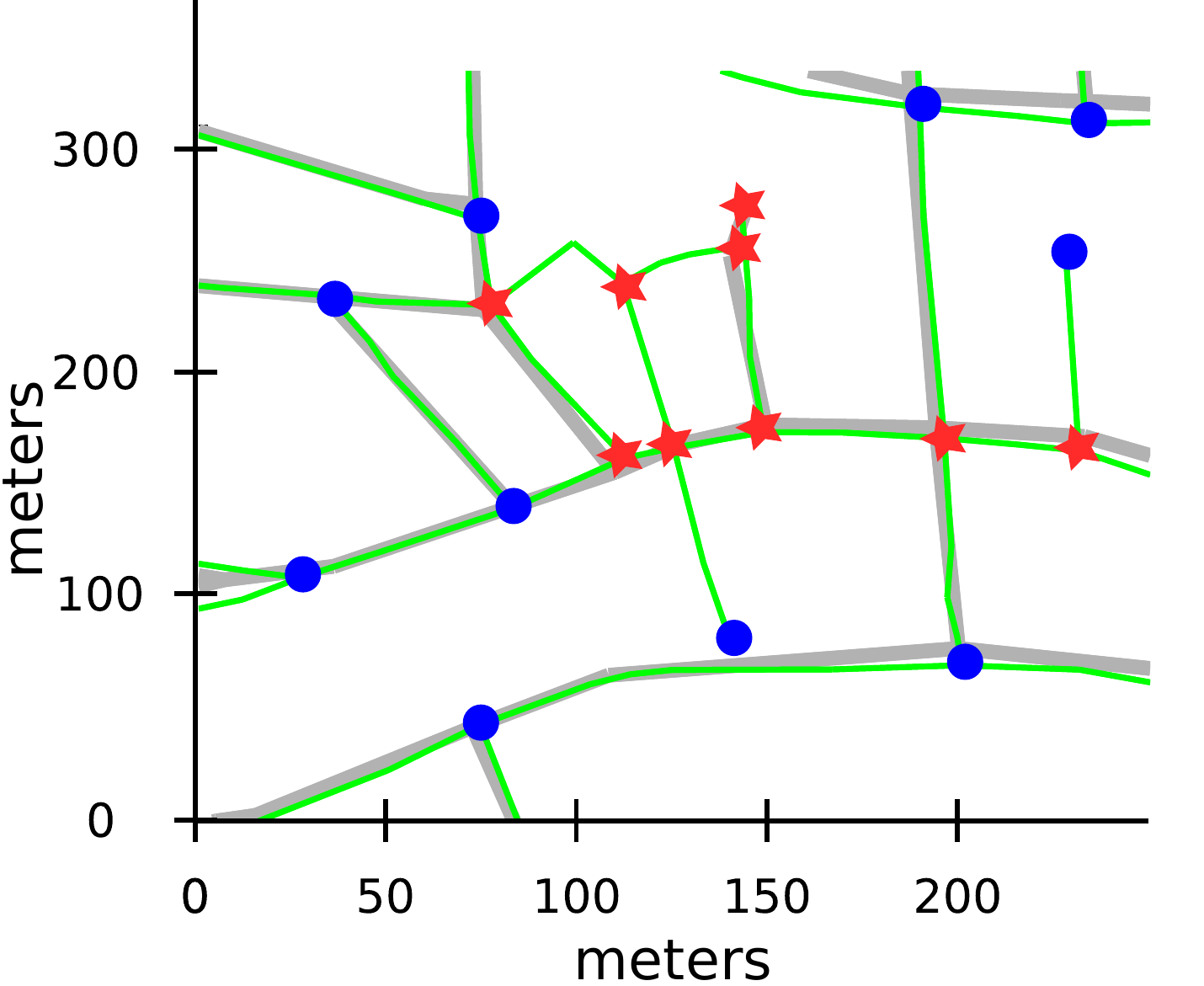}}
\hspace{0.1in}
\subfloat[$\Delta_3$]{\includegraphics[width=0.3\textwidth]
{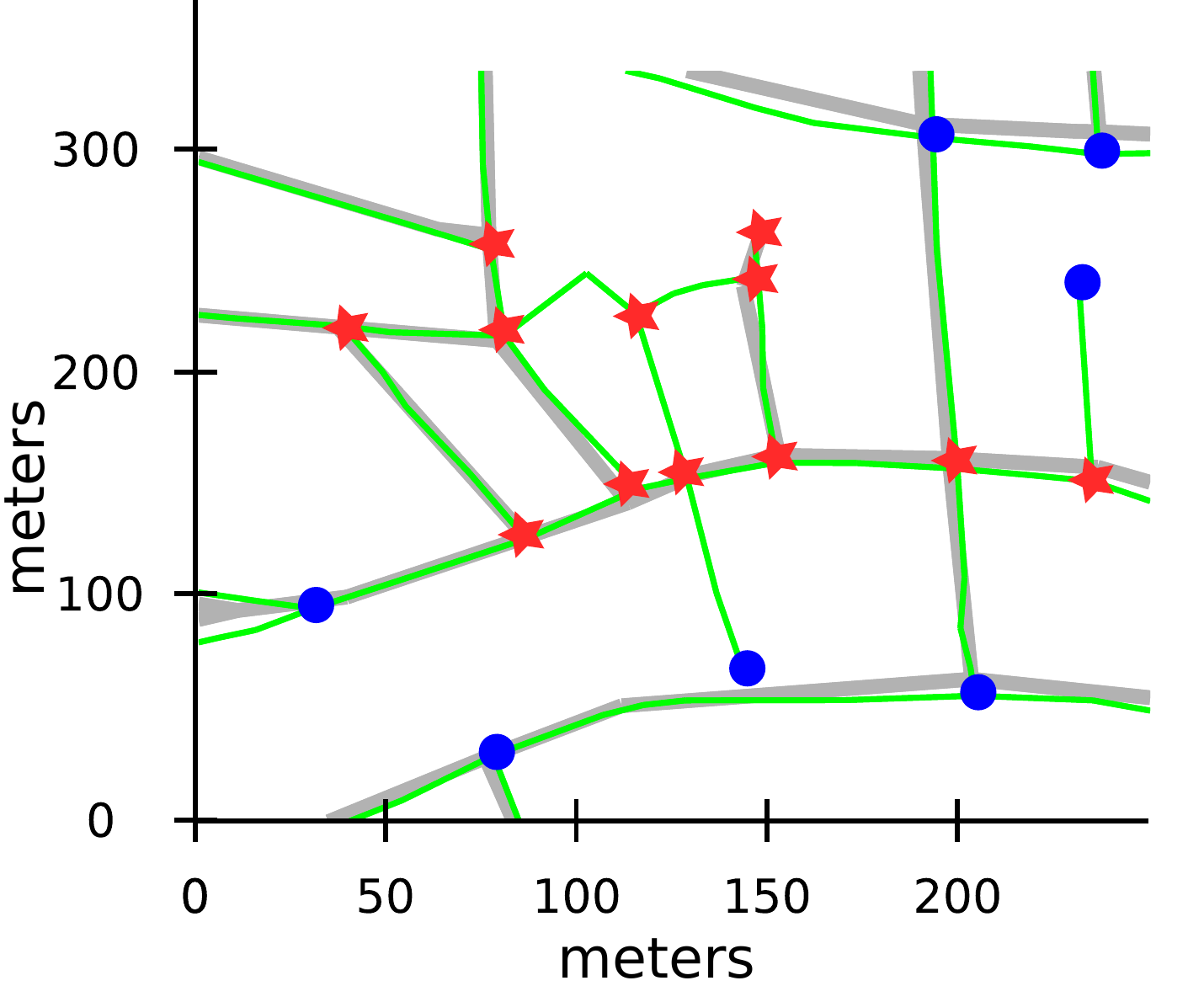}}
\caption{We demonstrate how a region of vertices that are not
$\Delta_k$-separated grows as $k$ increases.
Here, we see TA (in green) overlayed on OSM (in gray) for a section of the
\asmall\ dataset\footnotemark.  In blue circles, we mark the
$\Delta_k$-separated vertices, and in red stars, we mark the vertices that are
not $\Delta_k$-separated.}
\label{fig-growingbad}
\end{figure}
\footnotetext{The $x$
and $y$-coordinates are the offset (in meters)
from an arbitrary location, given in UTM
coordinates.  That location is UTM Zone 34S, $482450$ meters east,
$4206950$ meters north.}

Since, in our theorems,
$d$ depends on the directed
path-based distance between two sets of paths (one from each graph), that
implies that determining if a vertex is $d$-separated depends not only on the
graph containing the vertex, but also on the second graph to which we compare
it.
Recall from \defref{goodVertex} that a vertex $v\in G$ is $d$-separated if
$r_d$ is finite. As $d$ increases, the probability that $r_d$ is infinite also
increases.   As $\Delta_k$ is non-decreasing with respect to $k$, the number of vertices that
are not well-separated also grows with $k$, as demonstrated in
\figref{growingbad} for a section of the \asmall\ dataset.

In Table~\ref{tab-distance}, we show the values of the path-based distance for our datasets.
Since the path-based distance is defined as the maximum of map-matching 
distances, see Definition~\ref{def-pbDistance}, it is dominated by the most 
dissimilar sections in the maps. We therefore also record the $90^{th}$ 
percentile and the mean value of the set of map-matching distances, weighted by the  
overall length of the paths. These quantities may be more suitable in practice 
as these are less sensitive to outliers than the maximum. 
The distribution of map-matching distances can provide insights into differences
between the graphs; however, unless the
assumptions of the theorems of \secref{distance} are satisfied, it will not
guarantee vertex, edge, and intersection
correspondences for every vertex, edge, and intersection.

We see in Table~\ref{tab-distance} that for \asmall, the distance from OSM to TA is consistently smaller than the distance from TA to OSM, which suggests that the TA map contains more detail than the OSM map. This behavior is reversed for \bsmall\ and \blarge\, where the distance from TA to OSM is consistently smaller, which suggests that the OSM map contains more detail. 
When considering the mean values, the \asmall\ maps in particular appear to 
generally be in good correspondence, for one link, two link, and three link 
paths. For all data sets the effect of the different link-lengths is clearly 
noticeable: For the mean values highlighted in gray in Table~\ref{tab-distance}, 
when increasing the link-length by one, the distances increase by $9-16$ meters 
for the small data sets, and by $24-32$ meters for the large data set.
%
%
%
%

\begin{table}[t]
\centering
\tbl{Path-Based Distance.\label{tab-distance}}{
\begin{tabular}{|l| c| cc|c c| c c|}
\hline
&  & \multicolumn{2}{c|}{Path-based distance} 
&\multicolumn{2}{c|}{$90^{th}$-percentile} & \multicolumn{2}{c|}{Mean}\\
&& OSM& TA to &  OSM & TA to & OSM & TA to\\
Datasets& Path set   & to TA & OSM  & to TA & OSM & to TA & OSM\\
\hline
 &$\Pi_{G}^{1}$&150m&188m&11m&43m&\cellcolor{gray!25}8m&15m\\
 \asmall\ &$\Pi_{G}^{2}$&157m&251m&42m&71m&\cellcolor{gray!25}17m&27m\\
 &$\Pi_{G}^{3}$&166m&251m&71m&96m&\cellcolor{gray!25}30m&45m\\
 \hline
 &$\Pi_{G}^{1}$&203m&127m&70m&29m&25m&\cellcolor{gray!25}13m\\
\bsmall\ &$\Pi_{G}^{2}$&203m&130m&98m&61m&40m&\cellcolor{gray!25}24m\\
 &$\Pi_{G}^{3}$&210m&157m&121m&82m&63m&\cellcolor{gray!25}40m\\
\hline
 &$\Pi_{G}^{1}$&$> 1,600$m&1,150m&652m&57m&$>204$m&\cellcolor{gray!25}26m\\
\blarge\ 
&$\Pi_{G}^{2}$&$>1,600$m&1,450m&896m&116m&$>296$m&\cellcolor{gray!25}50m\\
 &$\Pi_{G}^{3}$&$> 1,600$m& 1,450m&1095m&174m&$>386$m&\cellcolor{gray!25}82m\\
\hline
\end{tabular}
}
\end{table}

When one map has more streets than the other, as is the case with the
Berlin-large map, the
path-based distance between them can be very large (as should be expected).
Since the Berlin-large map contains part of the city of Berlin
as well as parts of its southern suburbs (see Figures \ref{fig-blarget} and
\ref{fig-blargeo}), the local path-based distances observed by the edges and
the vertices are not randomly distributed; there are regions in which
the graphs are very similar, as well as regions where the graphs are dissimilar.
For this reason,
the value of the distance alone is difficult to interpret; therefore, we turn
to the local signatures to provide more insight.

\subsection{Using the Local Signature}
\label{subsec-experimentsSignature}

In this subsection, we study the local signature
$\Delta_{k,e}=\distance{\Pi_{e}^k}{\Pi_H}$ for a fixed $e\in E_G$ and
for a given integer $k\geq 1$. 
%
We first show how heat-maps can be used as a visualization tool for the local
similarity captured by the signature. This
can help identify similar regions in the map. In a different approach, we then
define a cumulative distribution function
in order to capture a summary of the local signature in terms of percentage of
the overall graph length.

\paragraph{Heat-Maps}
We compute heat-maps by coloring an edge lighter shades of yellow to darker
shades of red based on the value of that
signature (smaller to larger).  An edge $e \in G$ which is drawn in lighter
shade implies correspondences in $H$ between $e$
and a subgraph induced by its neighborhood (depending on $k$) which is very
similar spatially, structurally and topologically.
On the other hand, a darker red edge implies one or more possible
dissimilarities: missing edge, missing vertex,
spatial/structural differences, etc.

\begin{figure}[t!p!]
\centering 
\subfloat{\includegraphics[width=.45\textwidth]
{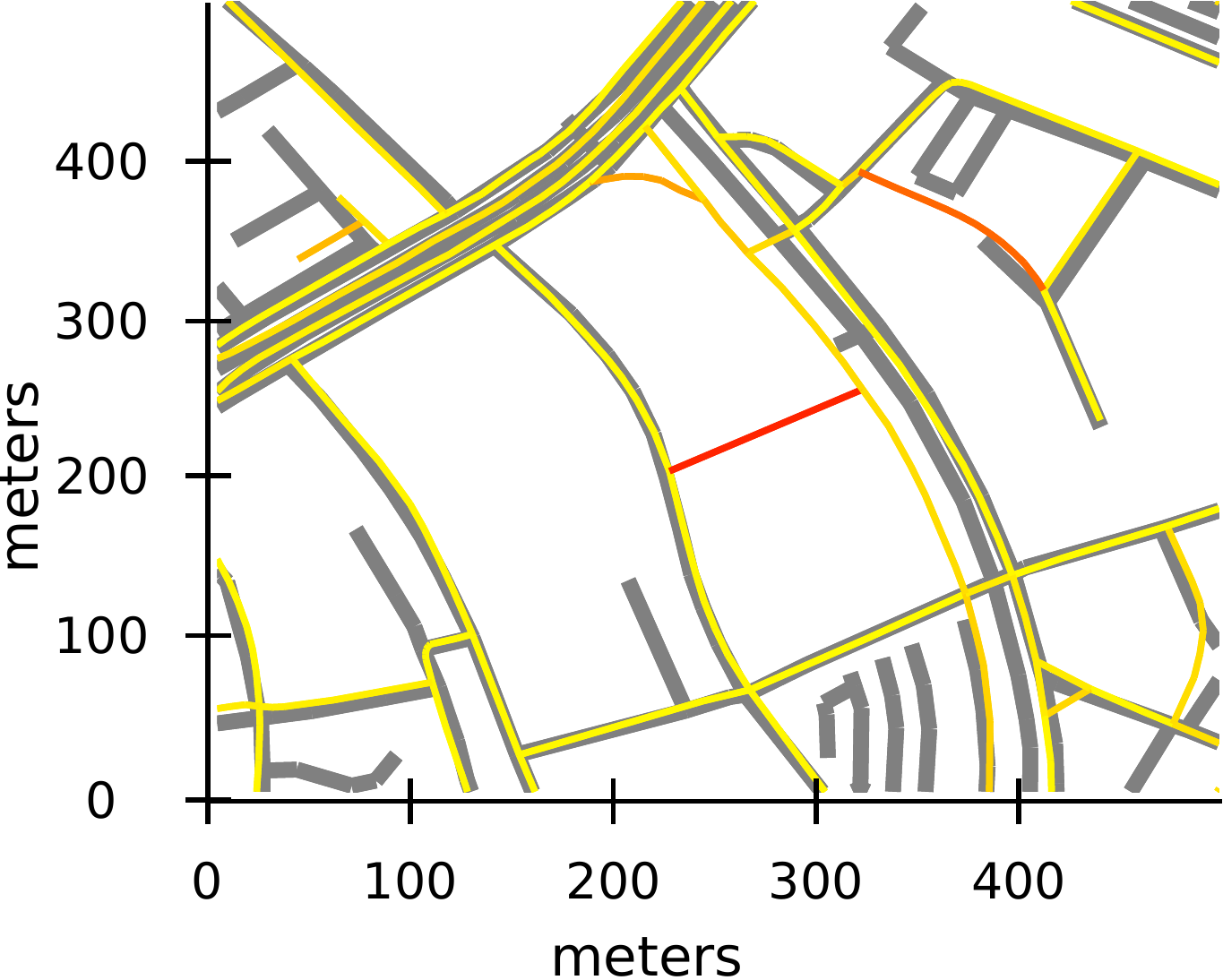}}\ \ \
\subfloat{\includegraphics[width=.45\textwidth]
{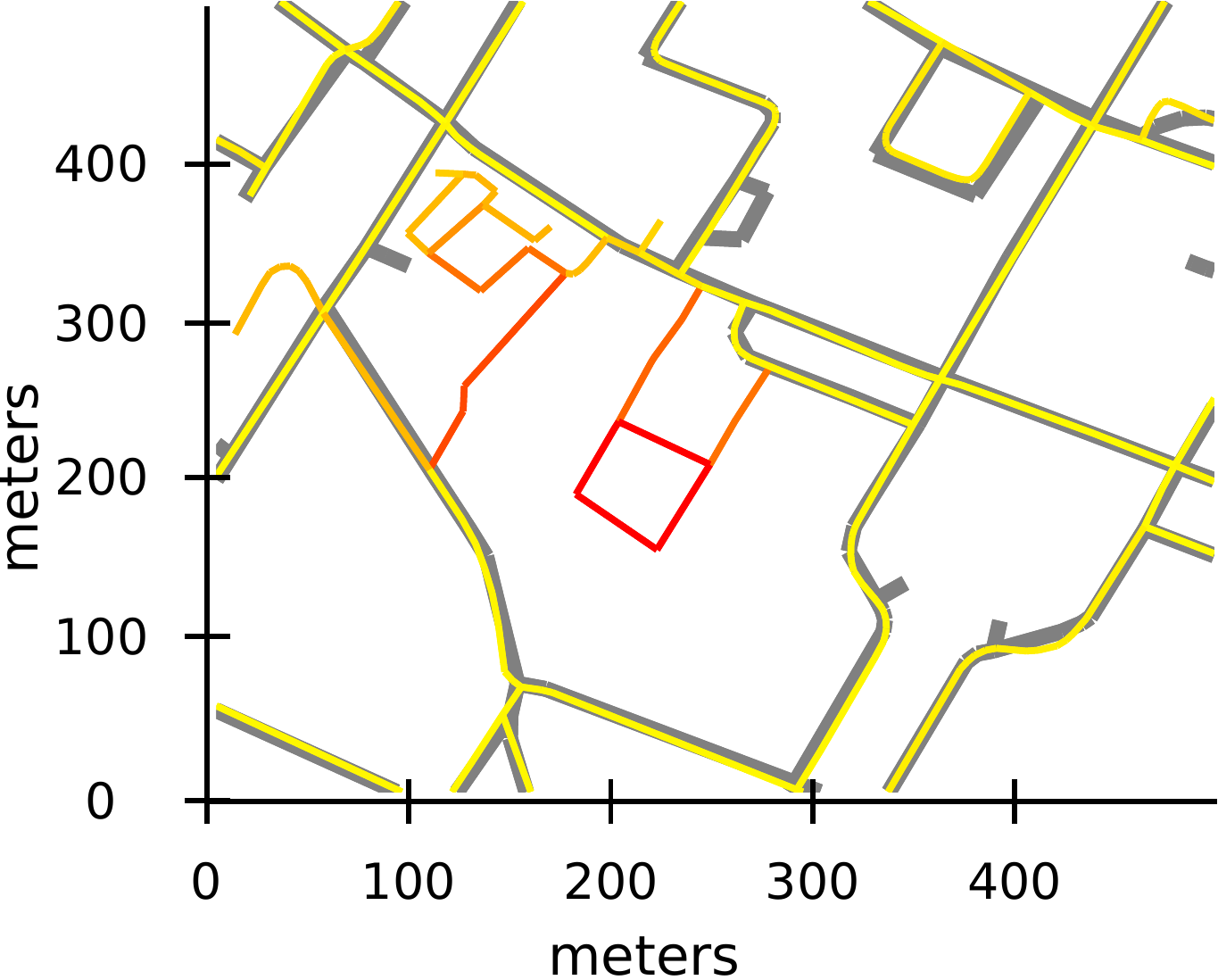}}
\caption{We plot the $\Delta_{1}$-signature of the TeleAtlas map, overlayed on
the OpenStreetMap (in
gray) for the \bsmall\ dataset\footnotemark.  
TA edges with a close OSM path have a low signature and
are shown in yellow.
TA edges without any corresponding path in OSM have a high signature and are 
shown in red.
}
\label{fig-lengthone}
\end{figure}

\begin{figure}[ptbh]
\centering 
\subfloat[$\Delta_{1}$-signature]
{\label{linkonebsmall1}\includegraphics[width=.45\textwidth]
{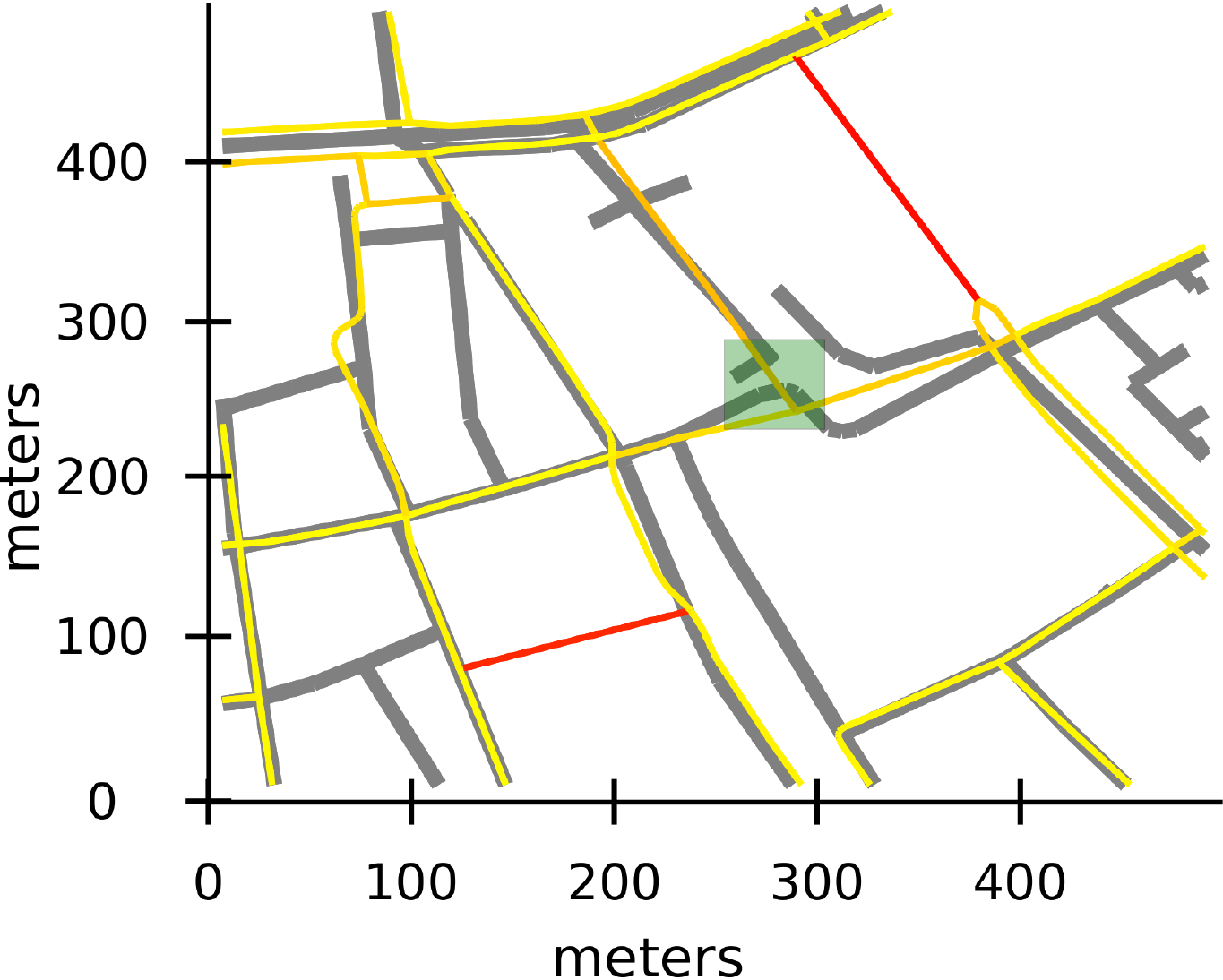}}\ \ \
\subfloat[$\Delta_{2}$-signature]
{\label{linktwobsmall1}\includegraphics[width=.45\textwidth]
{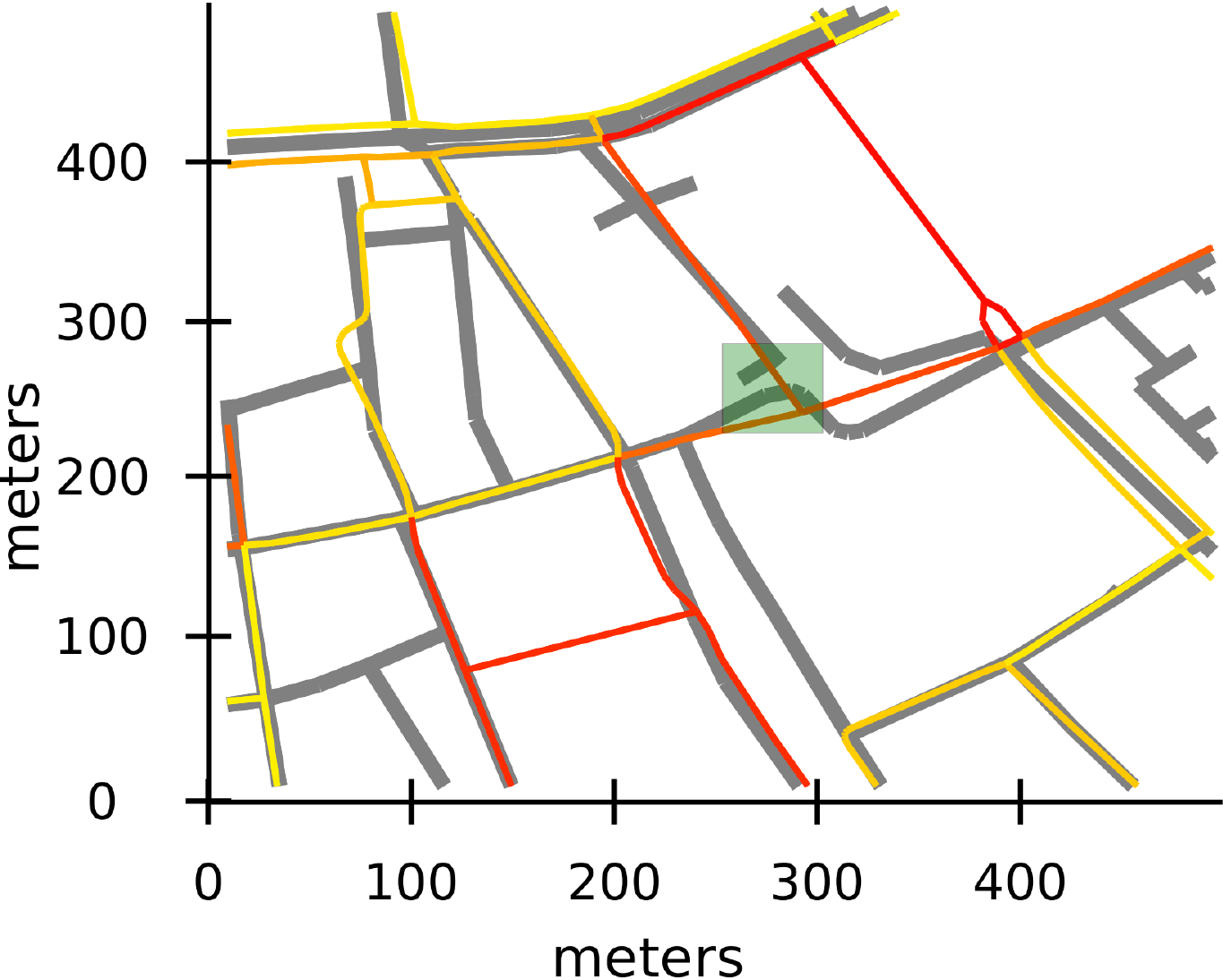}}

\subfloat[$\Delta_{1}$-signature]
{\label{linkonebsmall2}\includegraphics[width=.45\textwidth]
{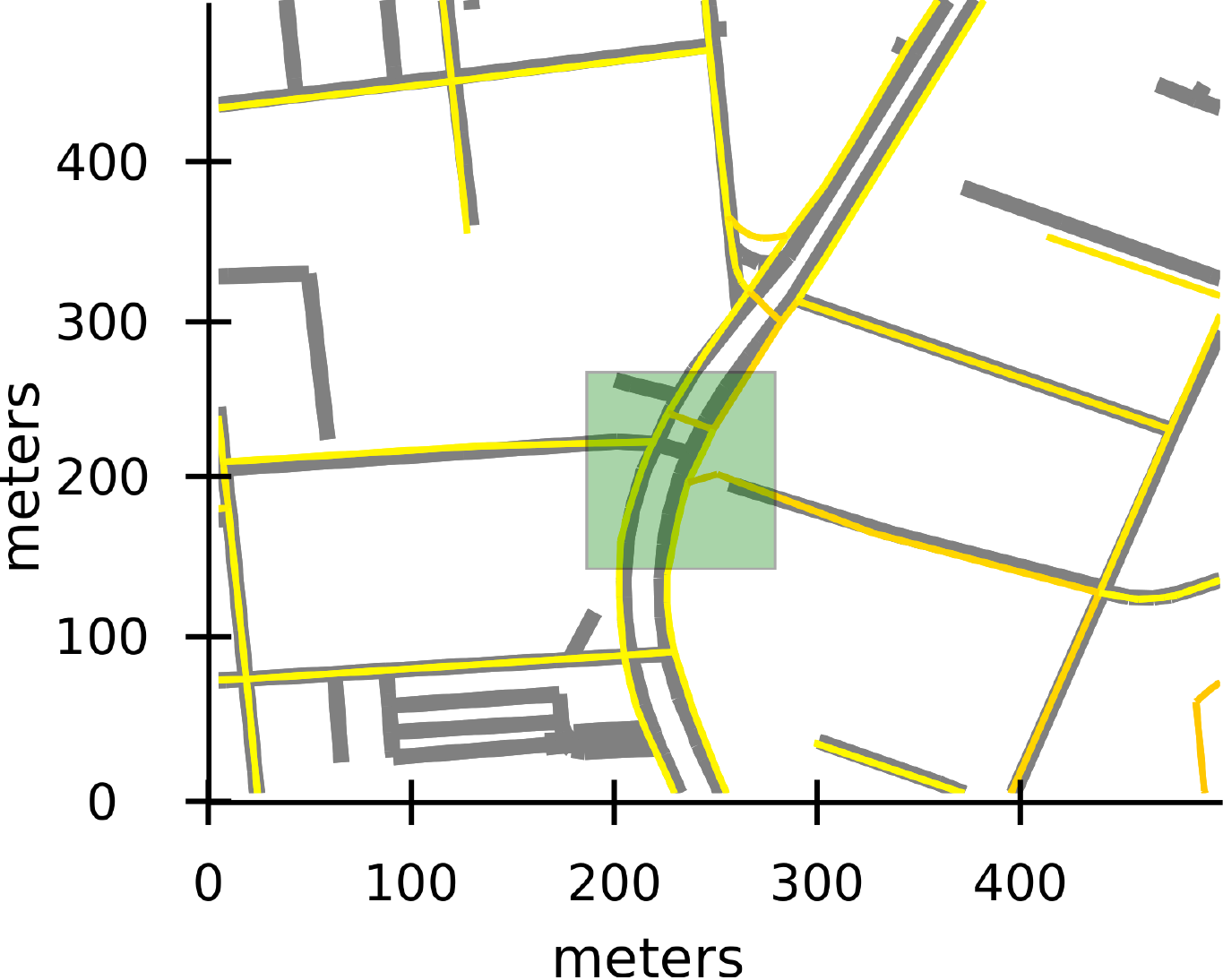}}\ \ \
\subfloat[$\Delta_{2}$-signature]
{\label{fig-linktwobsmall2}\includegraphics[width=.45\textwidth]
{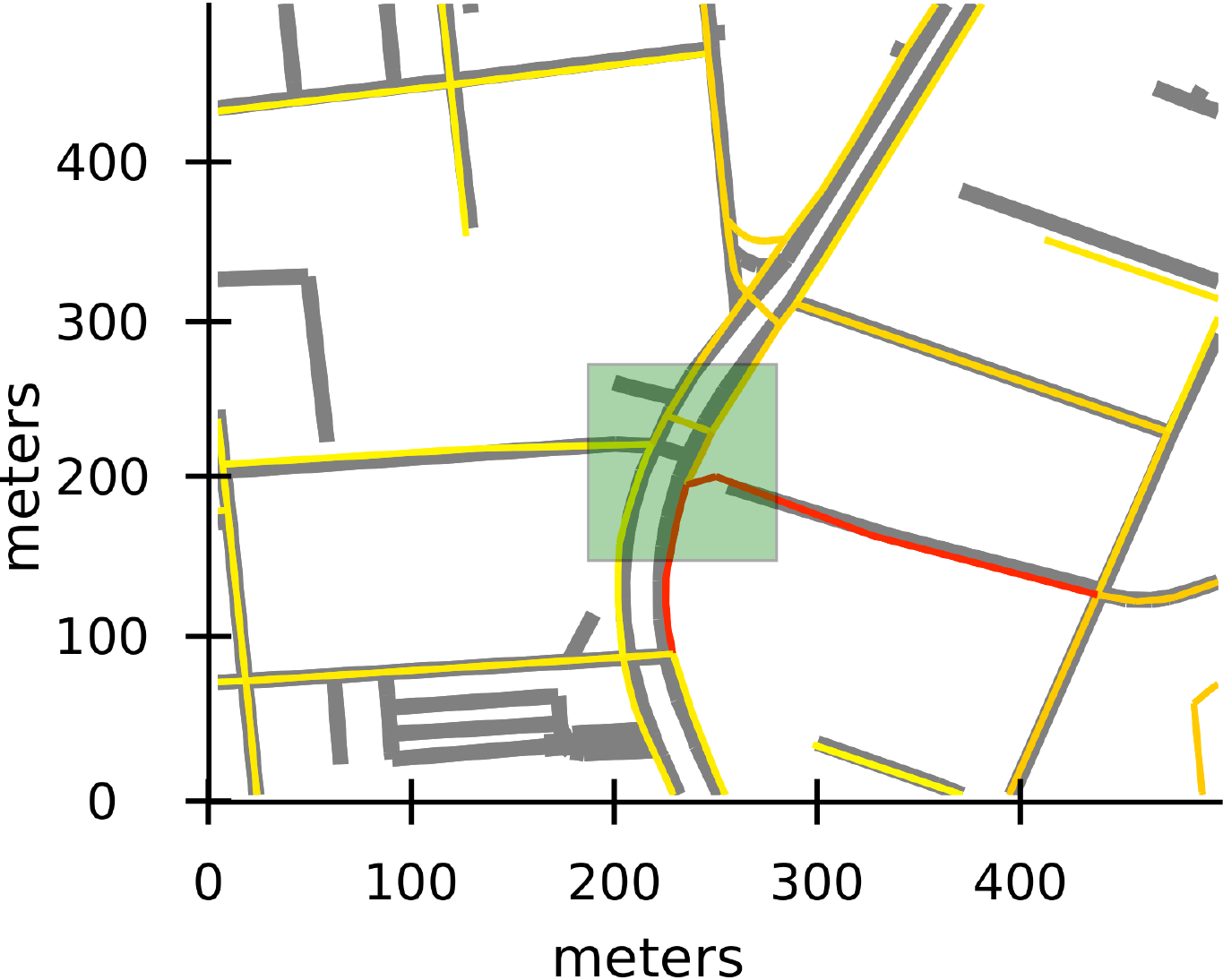}}
\caption{
We plot the path-based signature of the TeleAtlas map, overlayed on the
OpenStreetMap (in gray) for the \bsmall\ dataset\footnotemark. Note that
  $\Delta_{1}$ fails to capture missing vertices, but, as shown in the green boxes,
 $\Delta_{2}$ shows a large 
distance when the gray TA map has an intersection (in Figures (a) and (c)) which is missing
in the colored OSM map (in Figures (b) and (d)). }
\label{fig-linkbsmall}
\end{figure}

First, we demonstrate the special features \length\ one, \length\ two and 
\length\
three signatures can capture.  As was shown in
\subsecref{lengthone}, the value $\Delta_1$ can bound distances between
edges and their corresponding paths, but fails to identify when a
vertex (i.e.,  connection between two edges) is missing.  In 
Figures~\ref{fig-lengthone} and
\ref{fig-linkbsmall}, the heat-map of the 
TA map is overlayed on the OSM map (in gray) of \bsmall.  In
\figref{lengthone}, we can see that edges which are in the TA map but do not 
have corresponding edges in the OSM map have a large directed distance 
(indicated by the red color), when signatures
are computed using \length\ one paths.  
At the same time, the \length\ one signature fails to identify 
the  difference between two edges that
become close and two edges that intersect.  This difference, however, can be
identified
using \length\ two paths; see the regions inside the green boxes in
\figref{linkbsmall}.

\addtocounter{footnote}{-1}
\footnotetext{The $x$
and $y$-coordinates are the offset (in meters)
from an arbitrary location, given in UTM
coordinates.  That location is UTM Zone 33U, $392255$ meters east,
$5819700$ meters north and UTM Zone 33U, $392255$ meters east,
$5818300$ meters north respectively.}

\begin{figure}[tbph]
\centering 
\subfloat[$\Delta_{2}$-signature]
{\label{linktwotest}\includegraphics[width=.34\textwidth]
{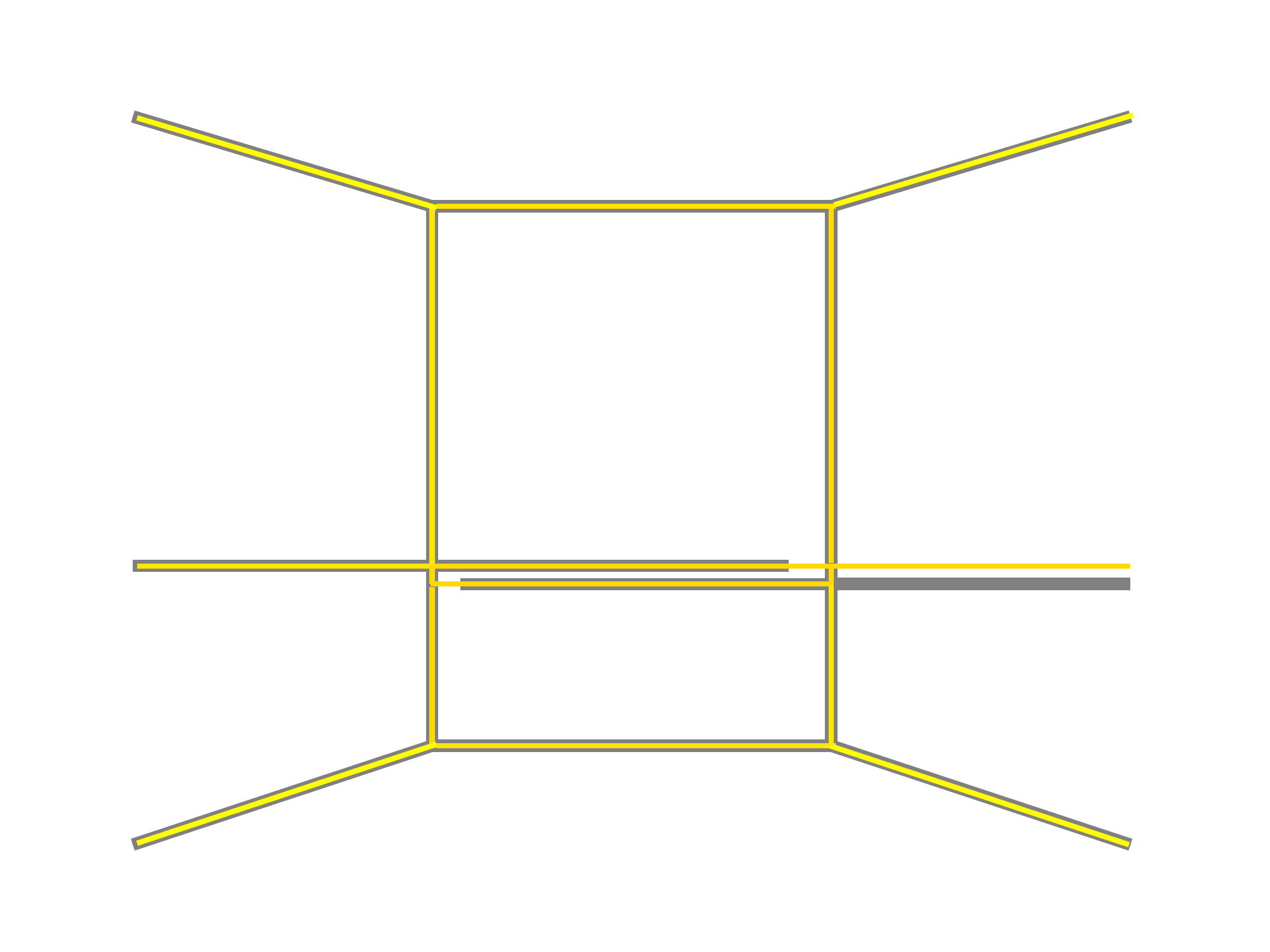}}
\subfloat[$\Delta_{3}$-signature]
{\label{linkthreetest}\includegraphics[width=.34\textwidth]
{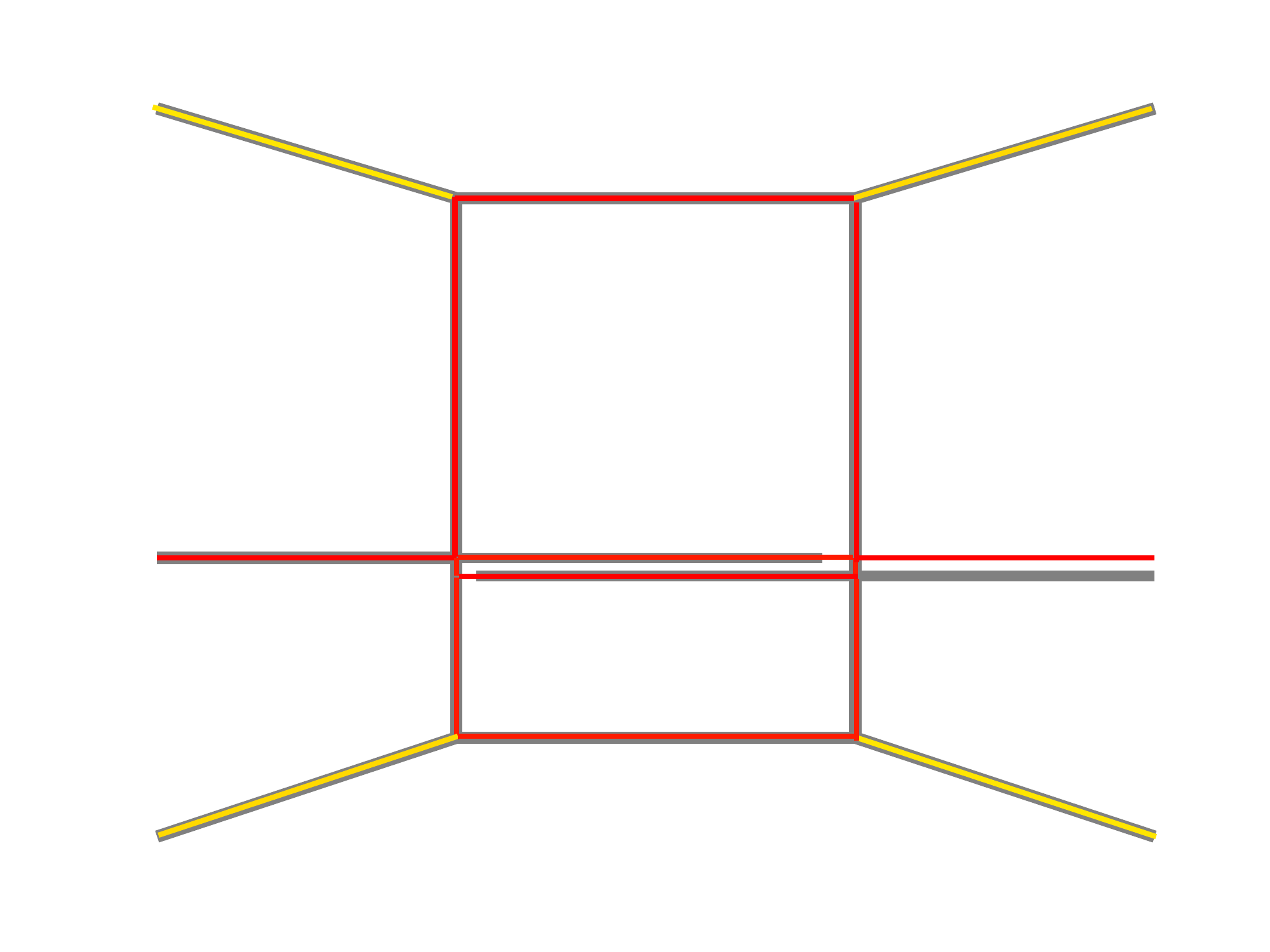}}
\caption{$\Delta_2$ can be arbitrarily small with $\Delta_3$
arbitrarily large.}
\label{fig-testlinkthree}
\end{figure}

Although \length\ two helps to find missing vertices, in some cases,
it fails to capture differences in more global notions of connectivity.
As one can see in the hypothetical example in
\figref{testlinkthree}, $\Delta_2$ can be arbitrarily small with $\Delta_3$
arbitrarily large.  \thmref{final} ensures that for $k>3$, the
value of $\Delta_k$ cannot become arbitrarily larger than $\Delta_3$.

\begin{figure}[tbph]
\centering 
\subfloat[Athens-small, TA mapped to OSM.]{\includegraphics[width=.45\textwidth]
{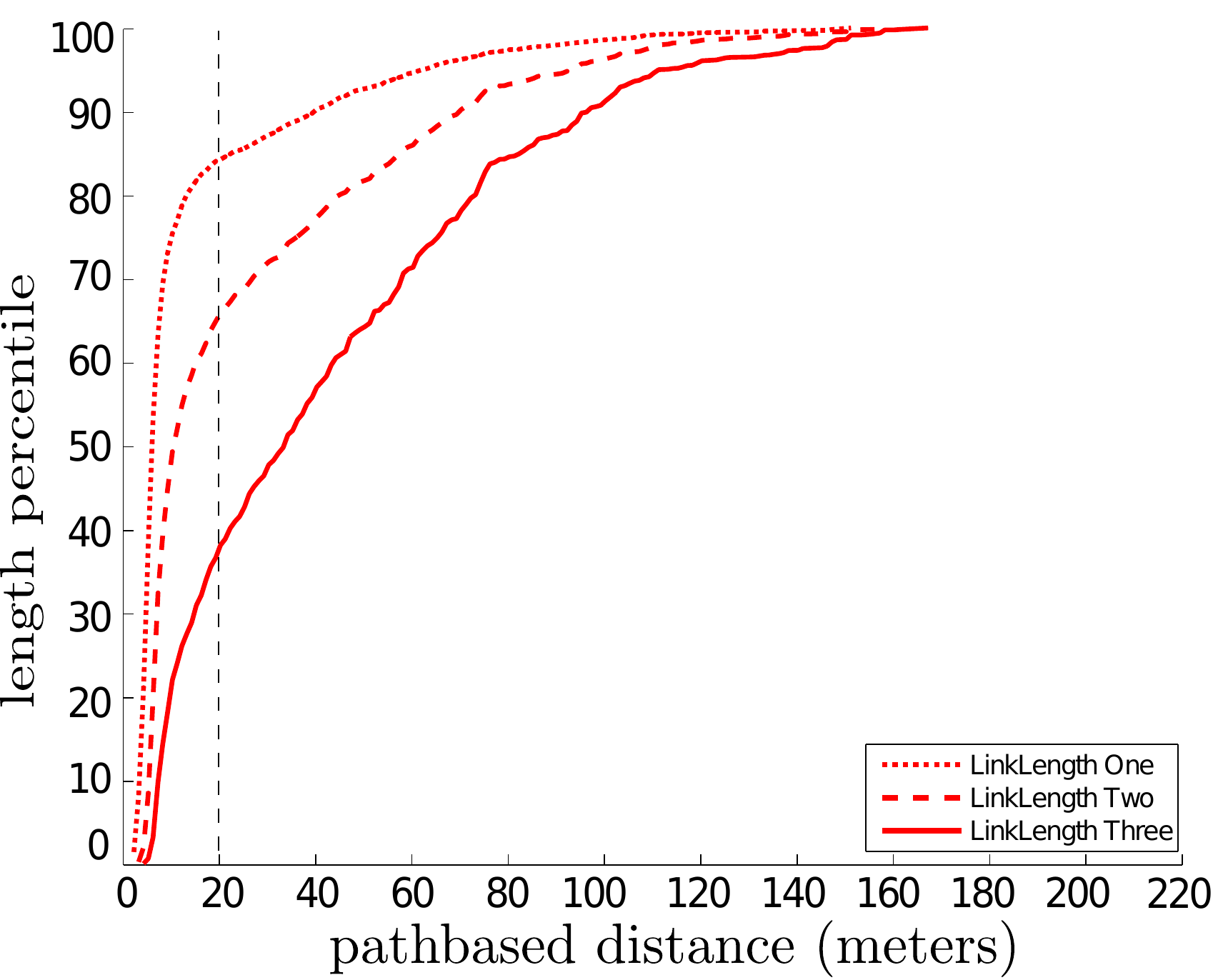}\label{fig-percentilePBathens_teleatlas
}}
\subfloat[Athens-small, OSM mapped to TA.]{\includegraphics[width=.45\textwidth]
{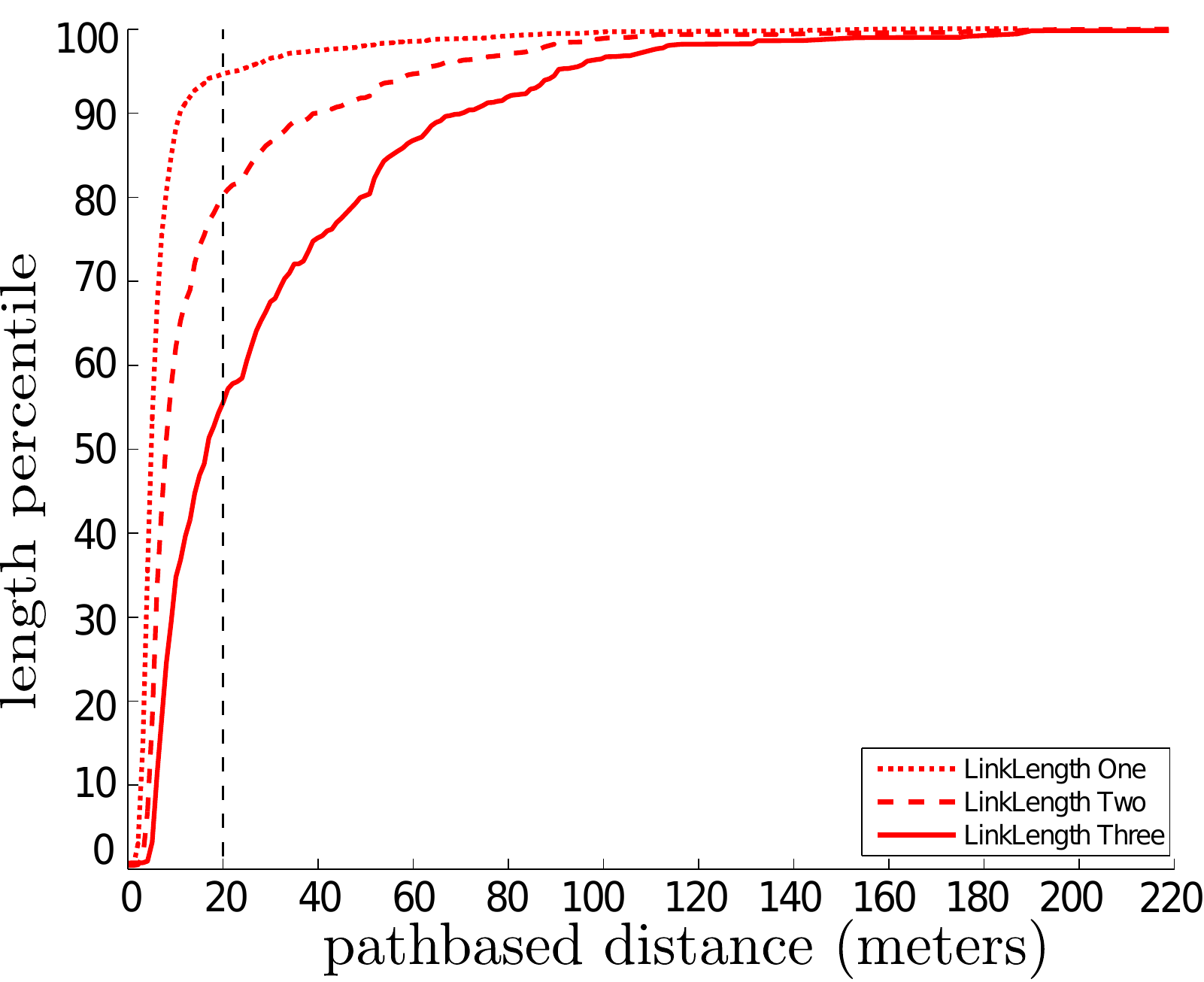}\label{fig-percentilePBathens_osm}}

\subfloat[Berlin-small, TA mapped to 
OSM.]{\label{fig-percentilePBberlin_teleatlas}\includegraphics[
width=.45\textwidth]
{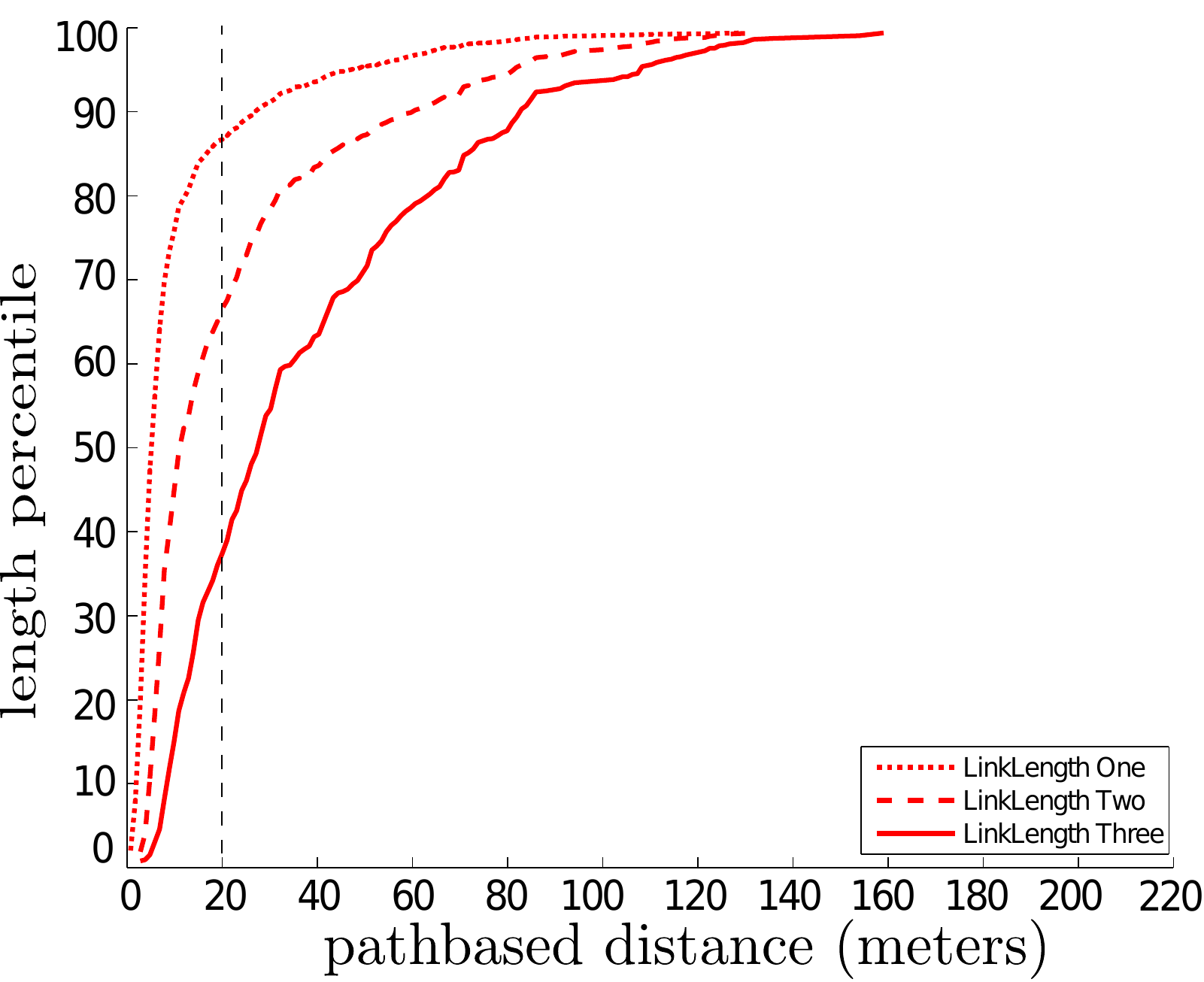}
}
\subfloat[Berlin-small, OSM mapped to TA.]{\includegraphics[width=.45\textwidth]
{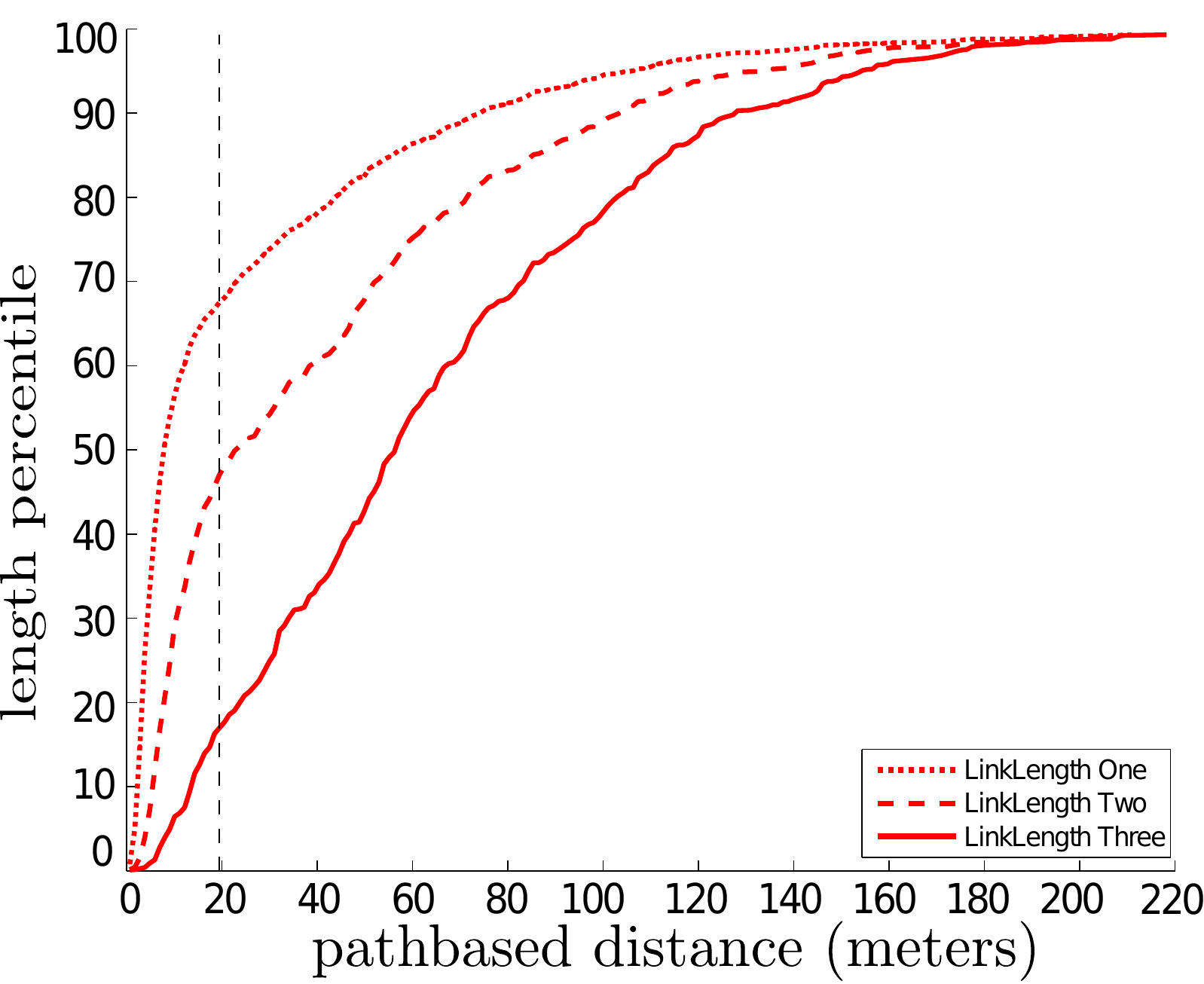}\label{fig-percentilePBberlin_osm}}

\subfloat[Berlin-large, TA mapped to OSM.]{\includegraphics[width=.45\textwidth]
{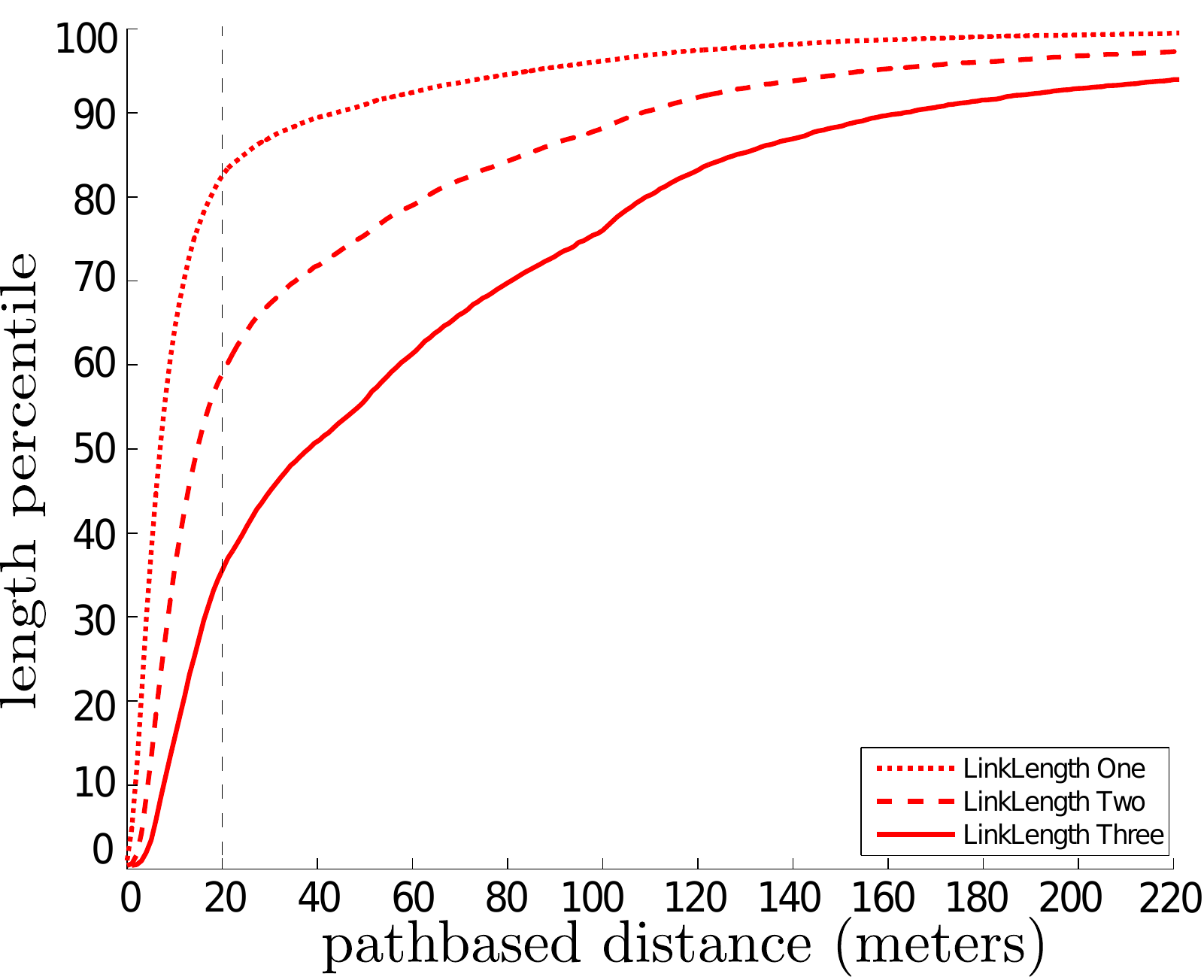}\label{
fig-percentilePBberlinl_teleatlas}}
\subfloat[Berlin-large, OSM mapped to TA.]{\includegraphics[width=.45\textwidth]
{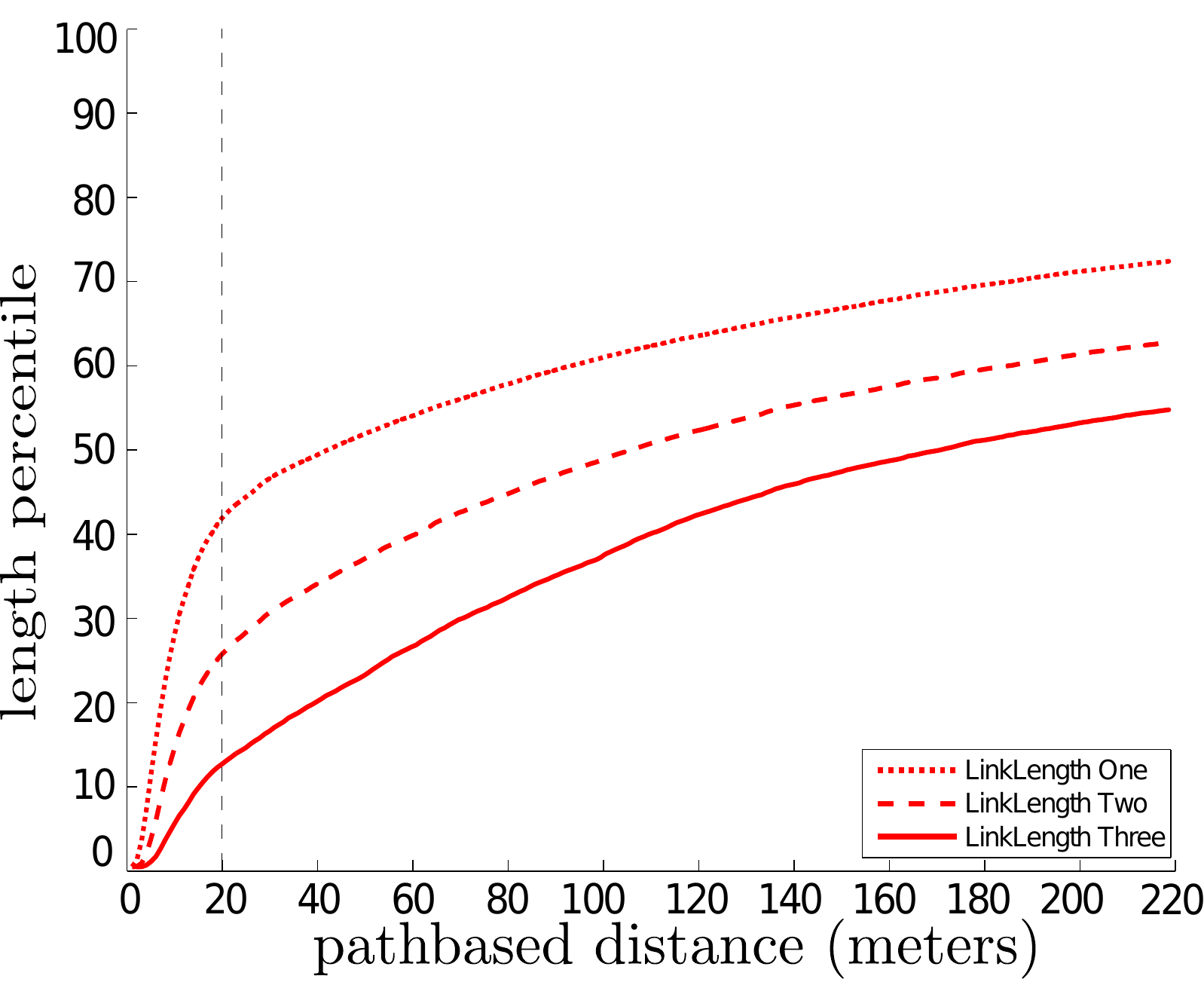}\label{fig-percentilePBberlinl_osm}}

\caption{For each point $p$ in a graph $G$, we assign to it the value
$\Delta_{k,e}$, where $e$ is the edge containing
  $p$.  Above, we plot the cumulative distribution of these assigned values.
  In particular, we take note of when the path-based distance is at most $20$
  meters (indicated by the vertical gray lines),
  since the path-based distance exceeding $20$ meters indicates that the graphs
  are not similar near the edge $e$.}
\label{fig-percentilePB}
\end{figure}

\paragraph{Cumulative Distribution of Local Signature Values}
For a given edge $e\in E_G$ and a given integer $k\geq 1$, let
$\Delta_{k,e}=\distance{\Pi_{e}^k}{\Pi_H}$ be the
path-based signature at $e$. Similar to the distribution of $\Delta_{k,v}$
discussed above, the distribution of $\Delta_{k,e}$ provides insight
into the similarity between $G$ and $H$.  Therefore, we investigate this
distribution in more detail.
%
Given a distance threshold $x$ and a fixed link-length $k$, we define the
weighted cumulative distribution of
$\Delta_{k,e}$ as follows:
\begin{equation*}
 CDF(x; G,H,k) = \frac{\sum_{e \in E'} \text{length}(e)}{\text{length}(G)},
\end{equation*}
where $E' = \{e ~:~ \Delta_{k,e} \leq x\}$.  In \figref{percentilePB}, we plot
$CDF(x;G,H,k)$ and $CDF(x;H,G,k)$ for
three different data sets.  We interpret these plots as follows: assume
$y=CDF(x;G,H,k)$, then $y\%$ of the points in
$G$ observe a path-based distance of at most $x$ meters.

\addtocounter{footnote}{-1}
\footnotetext{
The $x$
and $y$-coordinates are the offset (in meters)
from an arbitrary location, given in UTM
coordinates.  
That location is UTM Zone 33U $391,200$ meters east,
$5,819,400$ meters north in (a)-(b), and UTM Zone 33U $39,0800$ meters east,
$5,817,900$ meters north in (c)-(d).}

\figref{percentilePBberlin_osm} shows results for computing the distance from
the 2013 OSM map to the 2007 TA map for \bsmall.  Here, we can see $68\%$ of the
streets (more precisely, $181.56$ km out of
$267$ km) have very close correspondence (less than or equal
 to $20$ meters) to
the map of TA based on \length\ one.
That means that for $68\%$ of  locations chosen in the OSM map, there exists a
path in $H$ that has Fr\'echet distance 
at most $20$ meters to the edge containing the chosen location.  Due to the 
large
Fr\'echet distance 
to any path in the TA
dataset, we can conclude that about $32\%$ of the streets in this area of Berlin
were either omitted in the TA map or
may have been new constructions between $2007$ and $2013$.
This result is
consistent with the observation in Section~\ref{subsec-goodVertices} that
$54\%$ 
of the vertices are $\Delta_1$-separated.
On the other
hand,  $85\%$  of the streets that existed in the 2007 map have a
corresponding street in the 2012 map.  New
roads being built is a common 
occurrence, but removing roads is not.  Since these
street maps were taken from different sources, one explanation is that
different types of roads can be ignored by OSM
but would be recorded by TA.

Contrasting the \bsmall\ dataset is the \asmall\ dataset.  In \asmall, we see
that $85\%$ of the streets from the 2007
TA map have corresponding link-length one paths in the 2010 OSM map, and $95\%$
of the OSM map
have a corresponding path in the TA map. Perhaps some of this discrepancy
can be explained by time.  The Berlin
road network had six years to change; whereas, the Athens dataset had only three
years to change.

Looking at $CDF(x;H,G,k)$ for $k>1$ provides further insights.  The cumulative
distribution $CDF(x;H,G,1)$ gave
information akin to Hausdorff distance.  The distribution of $CDF(x;H,G,2)$
describes how well short distances (link-length two) are preserved between the 
graphs.  The larger $k$ gets, the longer
the paths are that need to be mapped from
$G$ to~$H$.  In the \blarge\ dataset, only $81\%$, $57\%$ and $34\%$ of the TA 
can find paths close to all
link-length one, two and three paths respectively. That also means $19\%$ of streets in TA
are missing (or have dissimilar correspondence) in OSM, $81-57=24\%$ streets in TA are dissimilar 
to OSM  because an adjacent turn and/or street is missing (similar to \figref{linktwobsmall2}).

\subsection{Experiments with Perturbed Data}
\label{subsec-controlled}

\begin{figure}[hb]
\centering
\subfloat[$p=0.1$]
{\label{fig-pGraph-p10}\includegraphics[width=.25\textwidth]
{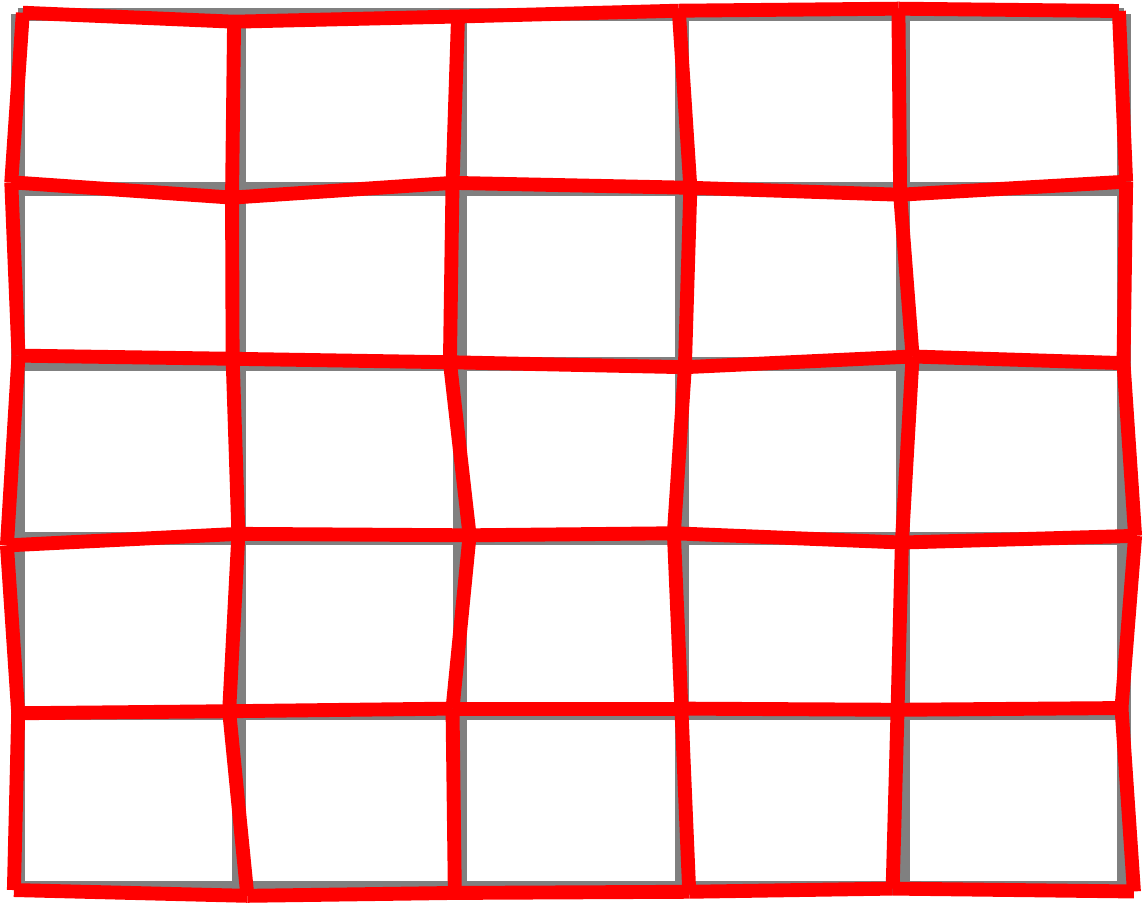}}
\hspace{1cm}
\subfloat[$p=0.5$]{\label{fig-pGraph-p50}\includegraphics[width=.25\textwidth]
{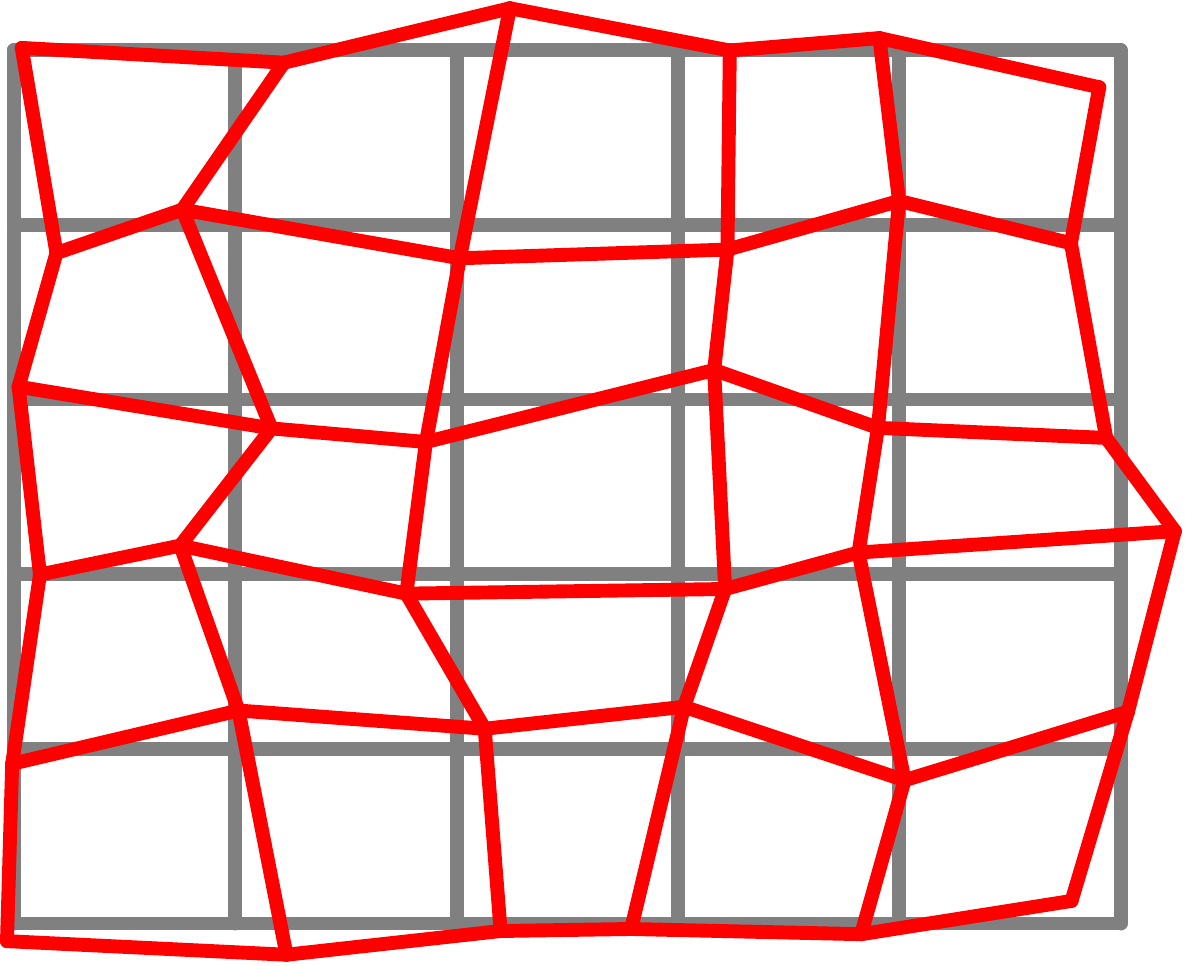}}
\hspace{1cm}
\subfloat[$p=0.9$]{\label{fig-pGraph-p90}\includegraphics[width=.25\textwidth]
{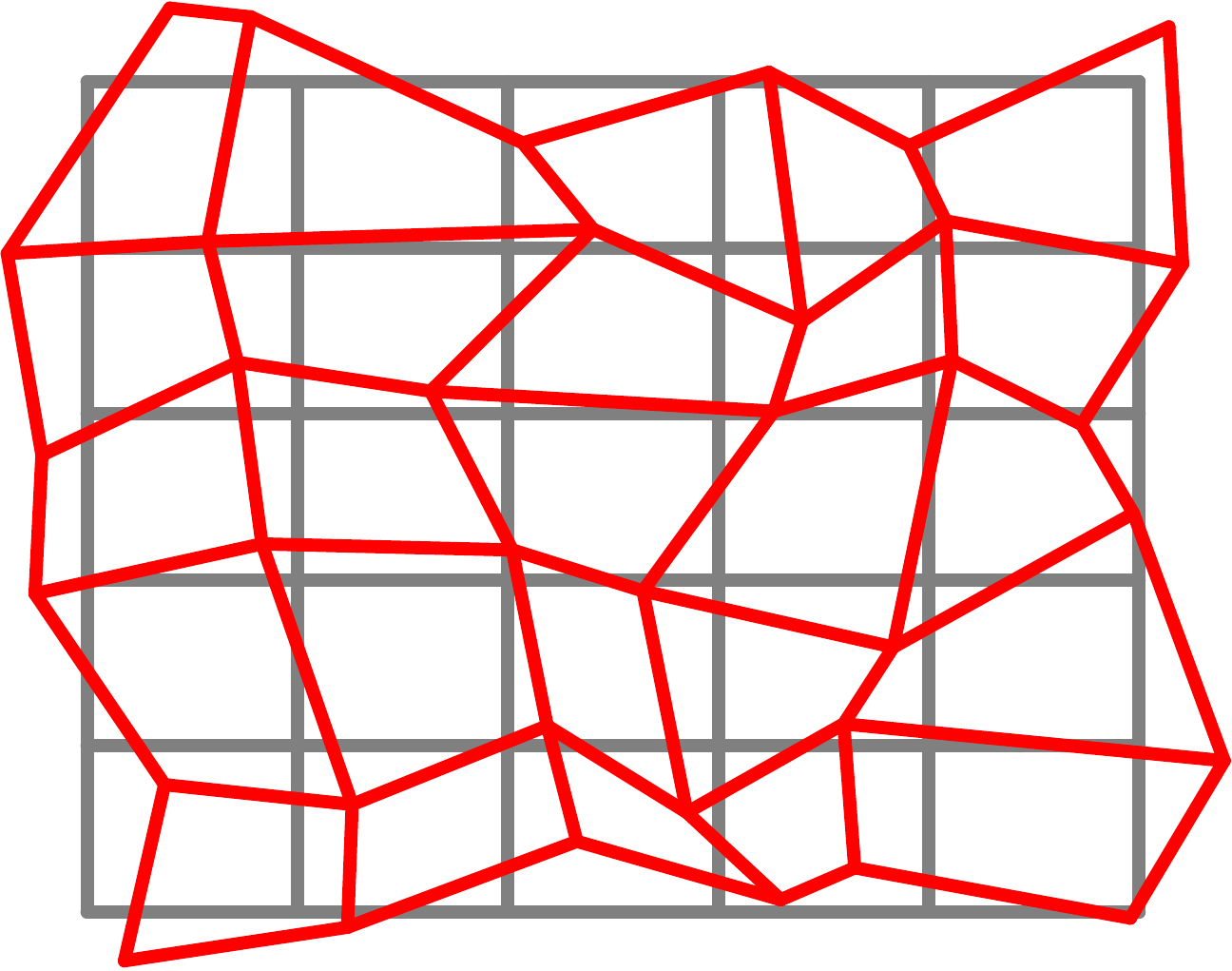}}
\caption{The original $G$ is the grey graph, a
regular grid over $\left[0, 10\right] \times \left[0, 10\right]$. As the perturbation
index $p$ increases, the perturbation of vertices in a graph increases.
The red graphs represent perturbed graphs for $p = 0.1$, $0.5$ and $0.9$, 
from left to right.}
\label{fig-pGraphs}
\end{figure}

In order to assess the ability of our distance measure to quantify dissimilarity 
between 
maps, we created a map $G$ and nine sets of estimations of that map with
increasing allowable deviations from $G$.  The map $G$ is a regular grid over
$[0,10] \times [0,10]$ using the coordinates with even integers as the
vertices; see the grey graph in \figref{pGraphs}.
The perturbation parameter is $p$, which is allowed to be between zero and one.  
We perturbed
the vertex at
$(i,j)$ in $G$ by choosing two numbers $\alpha$ and $\beta$ uniformly at random in the
interval $[-p,p]$ and moving the vertex at $(i,j)$ to $(i+\alpha,j+\beta)$.
Thus, as $p$ increases, the distance of the perturbed graph~$G_p$ to the 
original graph $G$ should increase as well, in expectation.  For each
value~$p$, 
we generate $100$ perturbed graphs: $G_p^1, \ldots, G_p^{100}$. For example,
see 
the red graphs in \figref{pGraphs} illustrating~$G_{0.1}^1,
G_{0.5}^1$, and $G_{0.9}^1$.

\begin{table}[bht]
\tbl{Three sample maps.\label{table-theta}}
{
\begin{tabular}{|r|l|c|r|}
\hline
\multicolumn{1}{|c|}{$p$} & \multicolumn{1}{c|}{$\theta$} & 
\multicolumn{1}{c|}{$\distance{\Pi^3_{G_p}}{\Pi_G}$} &
\multicolumn{1}{c|}{$\sqrt{2}p$}
\\\hline\hline
$0.1$ & $83.00^\circ$ & $0.125$ & $0.14$ \\\hline
$0.5$ & $55.43^\circ$ & $0.695$ & $0.71$ \\\hline
$0.9$ & $29.83^\circ$ & $1.195$ & $1.27$ \\
\hline
\end{tabular}
}
\end{table}
We first take a look at the three graphs in \figref{pGraphs}.
The values for $\theta$ and the distance $\distance{\Pi^3_{G_p^1}}{\Pi_G}$
 are listed in \tableref{theta}. 
We observe that when increasing the perturbation  parameter $p$, the angle 
$\theta$ decreases.  Hence, the upper bound for
$\distance{\Pi^3_{G_p^1}}{\Pi_G}$ from \corref{runtime} becomes
quite large (for $G_{p}^1$ it is $0.628$ and for $G_{0.9}^1$ it is $11.675$).
However, we notice that due to the perturbation scheme,
$\distance{\Pi^3_{G_p^1}}{\Pi_G} \leq \sqrt{2}p$ since each coordinate can be
perturbed by at most $p$. And, in fact, the link-three based distance is quite
close to that bound.

\begin{figure}[ht]
\centering
\includegraphics[width=.5\textwidth]{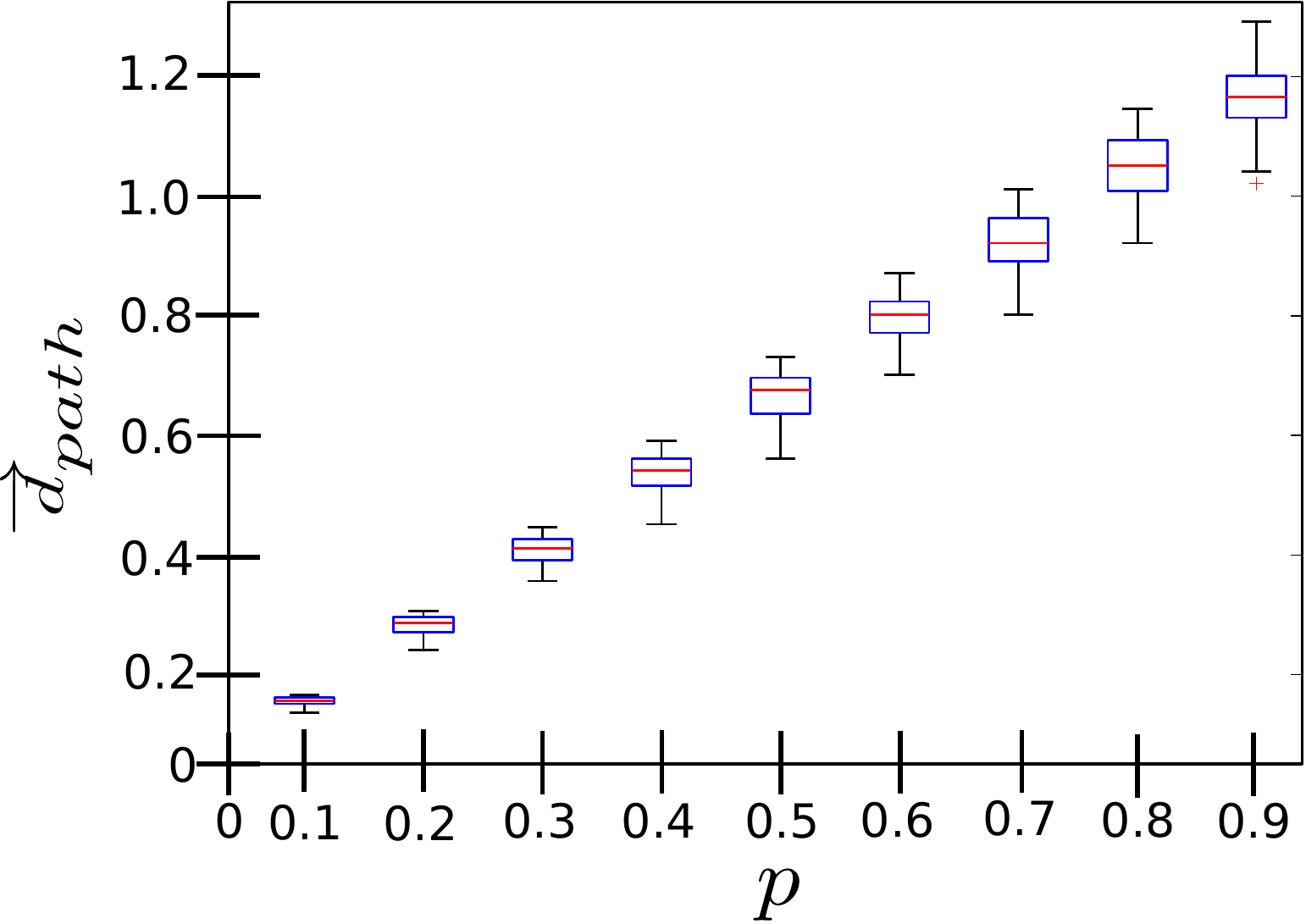}
\caption{For fixed $p$, we show the boxplot for the
distribution of observed distances. We can see as $p$ increases the path based
distances increase as well. The path based distance captures dissimilarities at
different levels.
\label{fig-controlled_result}}
\end{figure}
We compute $\distance{{\Pi^3}_{G_p^i}}{\Pi_G}$ for each $G_p^i$ and  summarize
our results in \figref{controlled_result}, using boxplots to illustrate the
distribution of path-based distances for each $p$. We observe that
the path-based 
distance   increases as the perturbation parameter $p$ increases, indicating 
that our distance measure can capture dissimilarities of varying
levels. Graphs that are more similar (lower values of $p$) have the smallest distances, and graphs that are more dissimilar (higher values of $P$) have the largest distances. 
%


\subsection{Comparison with {Biagioni and Eriksson 2012}}
\label{subsec-comparejames}
In this section, we compare our distance measure with the sampling-based
distance measure presented in
\subsecref{comparingStreetMaps}.  We used the code provided
by \citeN{Biagioni:2012:MIF:2424321.2424333}, making
modifications to allow for a different input format as well as to make the
output comparable to the path-based signature presented in this
paper.  In particular, we take vertices (including degree
two) from a graph as the seed locations
instead of the random sampling used in \cite{Biagioni:2012:MIF:2424321.2424333}.
The resulting marbles and holes distance ($F$-score) has three
parameters:
\begin{enumerate}
\item{ Sampling density: how densely the map should be sampled (marbles for
generated map and holes for ground-truth
    map); we use one sample every five meters.}
\item{ Matched distance: the maximum distance between a matched marble-hole
pair; we vary from $10$ to $220$ meters.}
\item{ Maximum path length: from seed, the maximum distance from start location
one will explore; we use $300$ meters.}
\end{enumerate}

\begin{figure}[tbph]
\centering
\includegraphics[width=0.45\textwidth]{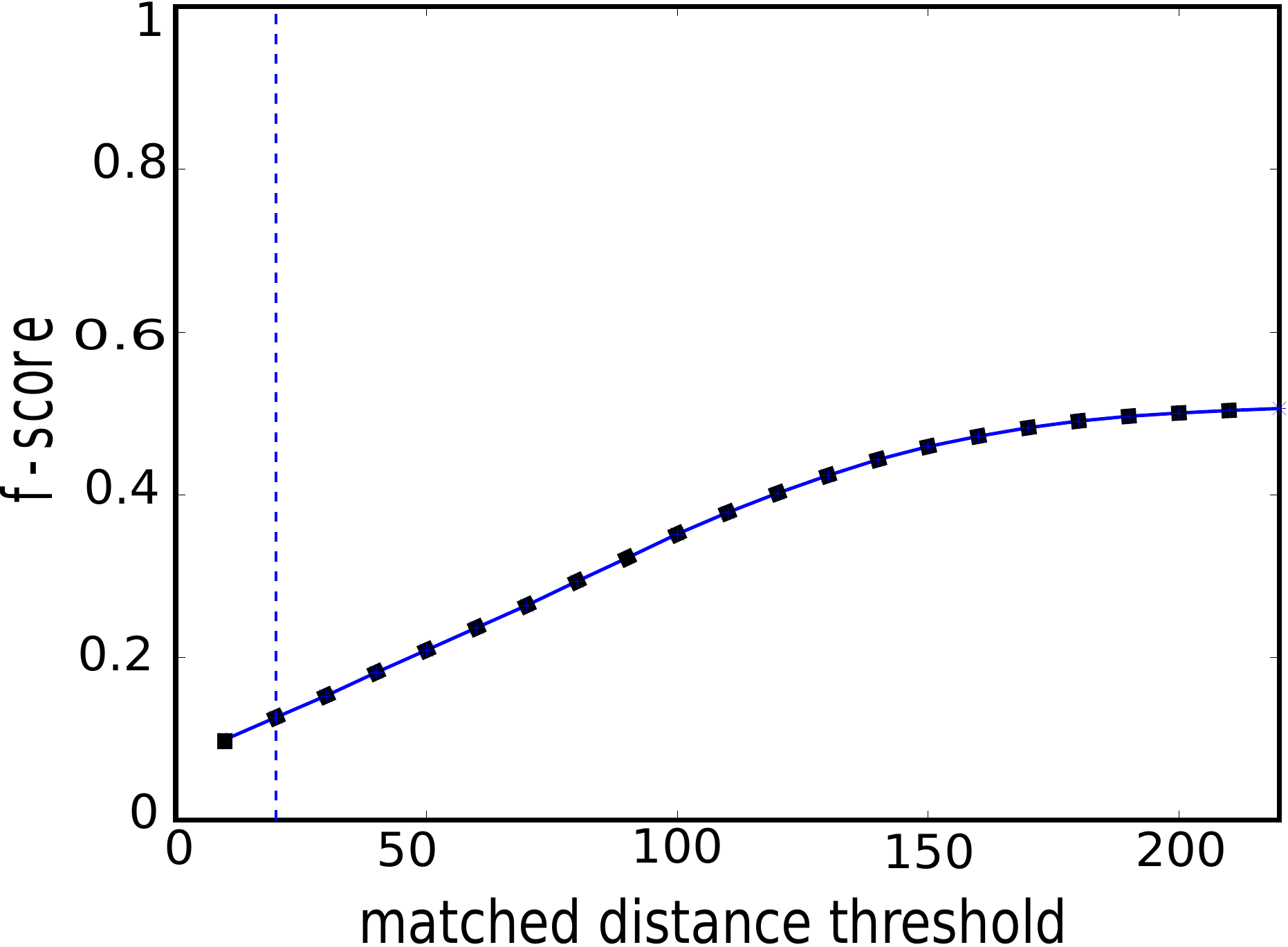}
\caption{By increasing the matched distance threshold, the $F$-score also
increases as it is easier for marbles and holes to be matched.  This
threshold is just one of the three parameters to the approach of
\citeN{Biagioni:2012:MIF:2424321.2424333}.}
\label{fig-fscore}
\end{figure}

Before commenting on the differences between the signatures, we first compare
and contrast the $F$-score to the path-based
distance.
We compute the $F$-scores for the Berlin-small dataset by varying the {\em
matched distance} from $10$ meters to $220$ meters, summarizing the results in
\figref{fscore}.
Here, we can see the $F$-score is quite low and this finding is
consistent with the observation we can draw from
\figref{percentilePBberlin_osm} that $31\%$ of the roads in OSM Berlin-small
map are new construction.
Although the two Berlin-small maps do not look very dissimilar,
the addition of more roads means that the topology of the maps has
changed.  Even if these changes are localized to a small area, the addition of
topological features punishes the whole graph for being dissimilar in a
tiny portion.  Choosing the matched distance to be $20$ meters,
the $F$-score is only  $0.1265$.
This computation took $17$
minutes, which is on the same order of magnitude as the
computation of our distance measure.

\begin{figure}[tbph]
\centering
\subfloat[$F$-score-signature]
{\label{fig-james_signature}\includegraphics[width=.45\textwidth]
{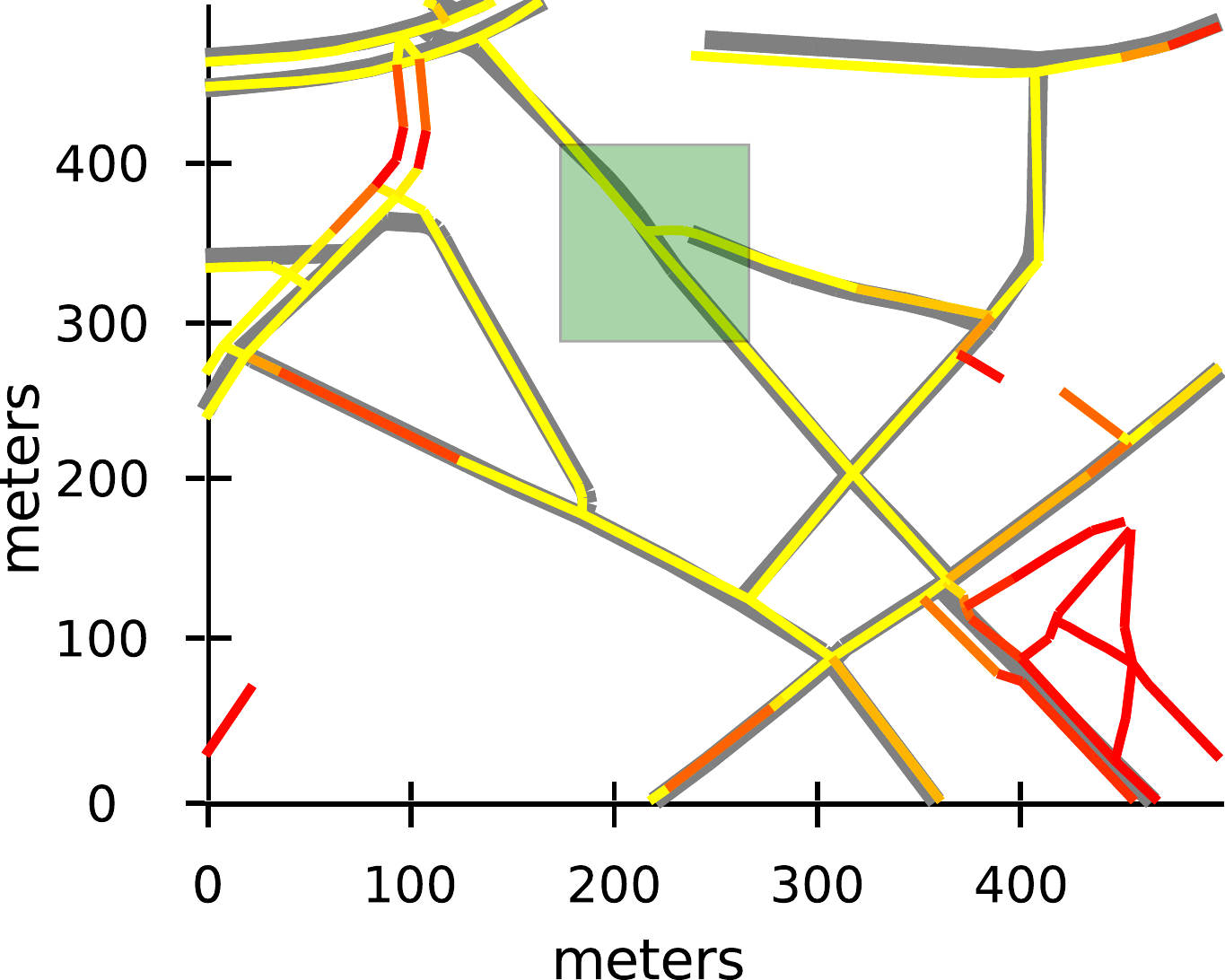}}\ \ \
\subfloat[$\Delta_{2}$-signature]
{\label{fig-signature_linktwo}\includegraphics[width=.45\textwidth]
{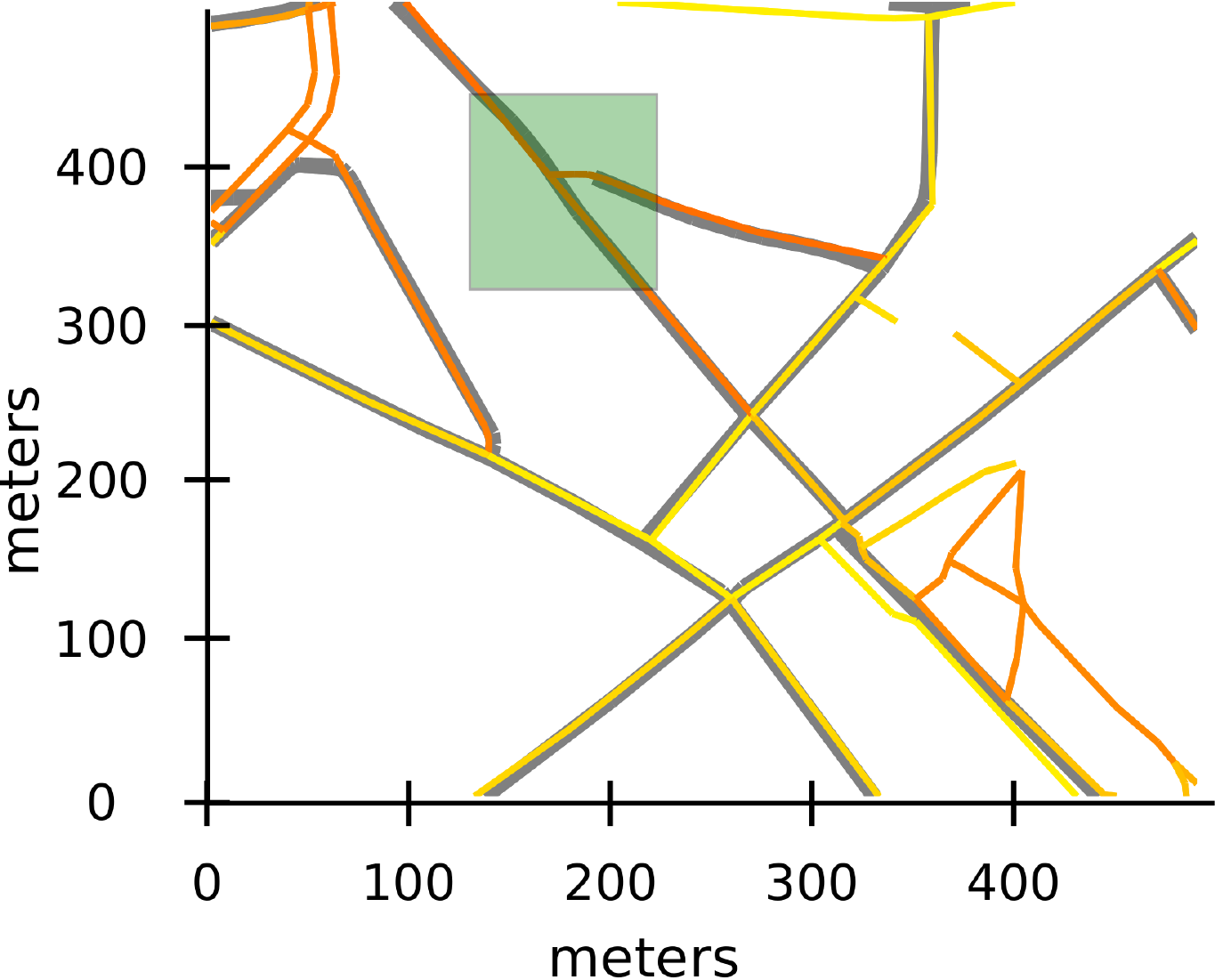}}
\caption{OSM map overlayed on the TA map (in gray) for the \bsmall\ dataset\footnotemark.
The matched distance is $20$ meters.
Note that the gray TA map has an intersection, while the colored OSM map does not. The $\Delta_2$ signature captures this as indicated by the darker orange color.
}

\label{fig-comparejames}
\end{figure}
\footnotetext{The $x$
and $y$-coordinates are the offset (in meters)
from an arbitrary location, given in UTM
coordinates.  That location is UTM Zone 33U $391,450$ meters east,
$5,818,600$ meters north.}

In order to compare two distance measures in a finer level, we compute
$F$-scores
for each start location individually
and compute edge signatures by averaging the $F$-scores at the two endpoint
vertices.   We compute this $F$-score signature and plot its heat-map in an
analogous fashion as we do for the path-based
local signatures: we color an edge
yellow if it observes similar behavior in both graphs (i.e., high $F$-score) and
red if the distance indicates that the
graphs are dissimilar (i.e., low $F$-score).

Balancing the tuning parameters above is difficult. \figref{comparejames} shows
a case where the graph sampling based
distance measure fails to capture the difference between the two road networks
due to the fact that the maximum path
length was set too high.  In gray, we plot the TA map, and we overlay
that with the OSM map colored
according to the adapted $F$-score in \figref{james_signature} and our local
signature in \figref{signature_linktwo}.
In the green box, we notice that there is an intersection in the OSM map that is
missing in the TA map, since one of the
streets ends before it meets the other street.  The $F$-score measure fails to
identify the missing intersection since
there is a detour available to reach the other road within $300$ meters of the
seed (the intersection in the green box).

    \begin{figure}[tbph]
    \centering
    \includegraphics[height=1in]{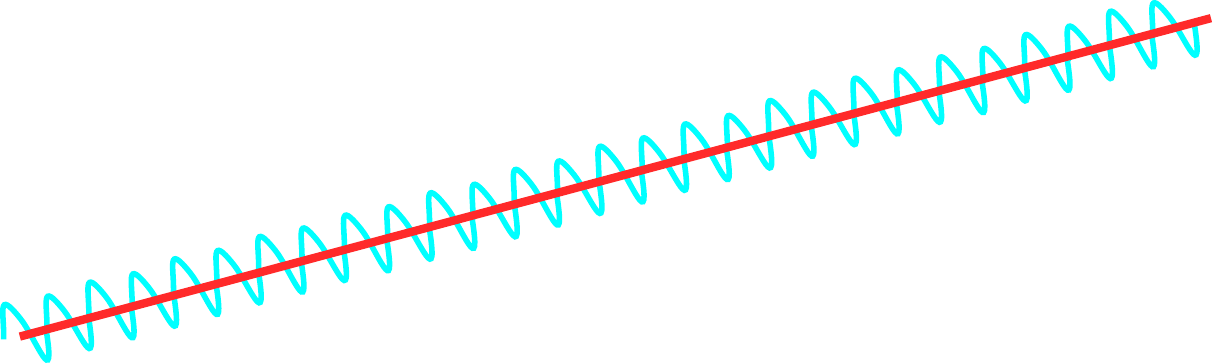}
    \caption{We consider evaluating the two distance measures on these two
    paths (taken to be graphs).  In this
      example, one graph is straight and the second graph oscillates frequently,
  but
    the deviation from the straight line
      path is relatively small.  Hence, the path-based distance is small
  indicating
    that the graphs are very close, but the
      $F$-score will be close to zero indicating that the graphs are very
    dissimilar.}
    \label{fig-james-toyexample}
    \end{figure}

Again, as this distance measure is based on one-to-one correspondence, it picks
up streets on the OSM map very
clearly (in red), which are missing in the TA map;  whereas our
measure picks them based on their proximity to the nearest street (darker yellow to
orange); see lower right corner of \figref{comparejames}.
There are examples for which our distance measure would find the graphs to be
similar, but the $F$-score would be
very low indicating that the graphs are not similar.  For example, consider
\figref{james-toyexample}. The first graph (in orange)
consists of exactly one straight line path.
The second graph (in cyan) consists of one path close to the orange one, but
oscilates frequently.  Depending on the circumstances, one distance measure
would be preferred over the other.  If the exact distance of the path traveled
matters (e.g., in computing the cost of transporting goods), then
using the $F$-score may be preferred; whereas, if the topology of the map is
important (e.g., when deciding if roads have been closed or new roads
added), then the path-based distance would be preferred.

The above illustrates one of the strengths of our distance measure: no tuning
parameters must be adjusted in order to
get a meaningful distance measure. Moreover, we have theoretical guarantees that
capture the difference in navigation
patterns between the two~graphs.

\section{Conclusion and Future Work}
In this paper, we formally defined a path-based distance measure for
the comparison of street map graphs. We provided a polynomial-time
algorithm to approximate this distance, by approximating the maximum
Fr\'echet distance over an infinite collection of pairs of paths.
This is the first distance measure for comparing street maps
that gives theoretical quality guarantees to compare travel paths
between the street maps, and which can be approximated in
polynomial~time.

Summarizing the differences between the maps with a single number
gives a global view of the differences, which may not provide enough
detail about those differences. In general, finding correspondences
between regions (or paths) of the street maps is a challenging task in
map comparison.  In this paper, we
defined a vertex-based as well as an edge-based local signature, which
allows for a natural visualization of the path-based distance.
These local signatures provide the means to distinguish similar regions, as well
as dissimilar regions, between two street maps. 

We have made the code for computing our path-based distance available
on {\tt www.mapconstruction.org}, a website recently established for
benchmarking map construction algorithms.  The largest current
comprehensive comparison of map construction algorithms using
different distance measures is provided in \cite{akpwComparison13}.
Among the distance measures used is the path-based distance defined in
this paper.

The work on the path-based distance measure has exposed several ideas
warranting further investigation. 
The major constraint on the theoretical side was the exclusion of degree-three
vertices.
However, we have noticed that in practice the link-length three distance measure appears 
to accurately capture similarities and differences between maps.
To close the gap between theory and practice, we ask what can be proven about
path-based distances between networks that include degree-three vertices? 

As mentioned in \secref{background}, finding the closest path in $H$ given a
path $p$ in $G$ is called map-matching.  Using the Fr\'echet distance to define
\emph{closest} is just one of the ways that this can be done.  One of the
limiting factors in this framework is that the Fr\'echet distance
captures the worst-case behavior.  In the future, we will investigate
different map-matching techniques. We further plan on studying
the use of alternative input models for the graphs, including directed
graphs and non-planar graphs that can model bridges and tunnels.  In
addition, instead of using link-length three paths, we wonder if we
can find a different set of paths that allows us to prove tighter
approximation bounds.

To date, there are only a few approaches for comparing planar embedded
graphs, and the definition of such distance measures varies depending
on the context.  Although this paper provides a new means of comparing
road networks, there is still a need to develop more techniques for
road network comparison.





\begin{acks}
 This work has been supported by the  National Science Foundation grants 
CCF-0643597 and
  CCF-1216602. We thank Dieter Pfoser  for providing the TeleAtlas maps, Sophia 
Karagiorgou for helping with data conversion, James Biagioni for providing 
his code, and the anonymous referees for providing thoughtful feedback.
\end{acks}

\bibliographystyle{ACM-Reference-Format-Journals}
\bibliography{gs_tsas}


\begin{thebibliography}{00}


\ifx \showCODEN    \undefined \def \showCODEN     #1{\unskip}     \fi
\ifx \showDOI      \undefined \def \showDOI       #1{{\tt DOI:}\penalty0{#1}\ }
  \fi
\ifx \showISBNx    \undefined \def \showISBNx     #1{\unskip}     \fi
\ifx \showISBNxiii \undefined \def \showISBNxiii  #1{\unskip}     \fi
\ifx \showISSN     \undefined \def \showISSN      #1{\unskip}     \fi
\ifx \showLCCN     \undefined \def \showLCCN      #1{\unskip}     \fi
\ifx \shownote     \undefined \def \shownote      #1{#1}          \fi
\ifx \showarticletitle \undefined \def \showarticletitle #1{#1}   \fi
\ifx \showURL      \undefined \def \showURL       #1{#1}          \fi

\bibitem[\protect\citeauthoryear{A. and Madhvanath}{A. and Madhvanath}{2014}]%
        {A.:2014:AMO:2676410.2629622}
{Bharath A.} {and} {Sriganesh Madhvanath}. 2014.
\newblock \showarticletitle{Allograph Modeling for Online Handwritten
  Characters in Devanagari Using Constrained Stroke Clustering}.
\newblock {\em ACM Transactions on Asian Language Information Processing\/}
  {13}, 3 (2014), 12:1--12:21.
\newblock


\bibitem[\protect\citeauthoryear{Aanjaneya, Chazal, Chen, Glisse, Guibas, and
  Morozov}{Aanjaneya et~al\mbox{.}}{2011}]%
        {Aanjaneya:2011:MGR:1998196.1998203}
{Mridul Aanjaneya}, {Frederic Chazal}, {Daniel Chen}, {Marc Glisse},
  {Leonidas~J.\ Guibas}, {and} {Dmitriy Morozov}. 2011.
\newblock \showarticletitle{Metric Graph Reconstruction from Noisy Data}. In
  {\em Proc.\ ACM SoCG}. 37--46.
\newblock
\showISBNx{978-1-4503-0682-9}
\showDOI{%
\url{http://dx.doi.org/10.1145/1998196.1998203}}


\bibitem[\protect\citeauthoryear{Ahmed, Fasy, and Wenk}{Ahmed
  et~al\mbox{.}}{2014a}]%
        {afw2014lhdist}
{Mahmuda Ahmed}, {Brittany~Terese Fasy}, {and} {Carola Wenk}. 2014a.
\newblock \showarticletitle{Local Persistent Homology Based Distance Between
  Maps}. In {\em SIGSPATIAL}. ACM.
\newblock


\bibitem[\protect\citeauthoryear{Ahmed, Karagiorgou, Pfoser, and Wenk}{Ahmed
  et~al\mbox{.}}{2014b}]%
        {akpwComparison13}
{Mahmuda Ahmed}, {Sophia Karagiorgou}, {Dieter Pfoser}, {and} {Carola Wenk}.
  2014b.
\newblock A Comparison and Evaluation of Map Construction Algorithms.
\newblock   (2014).
\newblock
\newblock
\shownote{ArXiv preprint 1402.5138.}


\bibitem[\protect\citeauthoryear{Ahmed and Wenk}{Ahmed and Wenk}{2012}]%
        {csm_esa2012}
{Mahmuda Ahmed} {and} {Carola Wenk}. 2012.
\newblock \showarticletitle{Constructing Street Networks from {GPS}
  Trajectories}. In {\em Proc.\ European Symp.\ Algorithms}. 60--71.
\newblock


\bibitem[\protect\citeauthoryear{Alt, Efrat, Rote, and Wenk}{Alt
  et~al\mbox{.}}{2003}]%
        {aerwMPM03}
{Helmut Alt}, {Alon Efrat}, {G\"unter Rote}, {and} {Carola Wenk}. 2003.
\newblock \showarticletitle{Matching Planar Maps}.
\newblock {\em J.\ Algorithms\/} (2003), 262--283.
\newblock


\bibitem[\protect\citeauthoryear{Alt and Godau}{Alt and Godau}{1995}]%
        {altgodau}
{Helmut Alt} {and} {Michael Godau}. 1995.
\newblock \showarticletitle{Computing the {Fr\'echet} Distance between Two
  Polygonal Curves}.
\newblock {\em Int.\ J.\ Comput.\ Geom.\ and Applications\/}  {5} (1995),
  75--91.
\newblock


\bibitem[\protect\citeauthoryear{Biagioni and Eriksson}{Biagioni and
  Eriksson}{2012}]%
        {Biagioni:2012:MIF:2424321.2424333}
{James Biagioni} {and} {Jakob Eriksson}. 2012.
\newblock \showarticletitle{Map Inference in the Face of Noise and Disparity}.
  In {\em Proc.\ 20th ACM SIGSPATIAL}. 79--88.
\newblock
\showISBNx{978-1-4503-1691-0}
\showDOI{%
\url{http://dx.doi.org/10.1145/2424321.2424333}}


\bibitem[\protect\citeauthoryear{Chen, Guibas, Hershberger, and Sun}{Chen
  et~al\mbox{.}}{2010}]%
        {cghsRnrop10}
{Daniel Chen}, {Leonidas Guibas}, {John Hershberger}, {and} {Jian Sun}. 2010.
\newblock \showarticletitle{Road Network Reconstruction for Organizing Paths}.
  In {\em Proc.\ ACM-SIAM Symp.\ on Discrete Alg.} 1309--1320.
\newblock


\bibitem[\protect\citeauthoryear{Cheong, Gudmundsson, Kim, Schymura, and
  Stehn}{Cheong et~al\mbox{.}}{2009}]%
        {DBLP:conf/wea/CheongGKSS09}
{Otfried Cheong}, {Joachim Gudmundsson}, {Hyo-Sil Kim}, {Daria Schymura}, {and}
  {Fabian Stehn}. 2009.
\newblock \showarticletitle{Measuring the Similarity of Geometric Graphs}. In
  {\em SEA}. 101--112.
\newblock


\bibitem[\protect\citeauthoryear{Conte, Foggia, Sansone, and Vento}{Conte
  et~al\mbox{.}}{2004}]%
        {cfsvTYGM04}
{Donatello Conte}, {Pasquale Foggia}, {Carlo Sansone}, {and} {Mario Vento}.
  2004.
\newblock \showarticletitle{Thirty Years Of Graph Matching In Pattern
  Recognition}.
\newblock {\em Int.\ J.\ Pattern Recognit.\ Artificial Intell.\/} {18}, 3
  (2004), 265--298.
\newblock


\bibitem[\protect\citeauthoryear{Eppstein}{Eppstein}{1995}]%
        {Eppstein:1995:SIP:313651.313830}
{David Eppstein}. 1995.
\newblock \showarticletitle{Subgraph Isomorphism in Planar Graphs and Related
  Problems} {\em (SODA)}. SIAM, Philadelphia, PA, USA, 632--640.
\newblock
\showISBNx{0-89871-349-8}
\showURL{%
\url{http://dl.acm.org/citation.cfm?id=313651.313830}}


\bibitem[\protect\citeauthoryear{Ge, Safa, Belkin, and Wang}{Ge
  et~al\mbox{.}}{2011}]%
        {DBLP:conf/nips/GeSBW11}
{Xiaoyin Ge}, {Issam Safa}, {Mikhail Belkin}, {and} {Yusu Wang}. 2011.
\newblock \showarticletitle{Data Skeletonization via {R}eeb Graphs}. In {\em
  25th Annual Conf.\ Neural Info.\ Proc.\ Sys.} 837--845.
\newblock


\bibitem[\protect\citeauthoryear{Karagiorgou and Pfoser}{Karagiorgou and
  Pfoser}{2012}]%
        {Karagiorgou:2012:VTD:2424321.2424334}
{Sophia Karagiorgou} {and} {Dieter Pfoser}. 2012.
\newblock \showarticletitle{On Vehicle Tracking Data-Based Road Network
  Generation} {\em (SIGSPATIAL '12)}. ACM, New York, NY, USA, 89--98.
\newblock
\showISBNx{978-1-4503-1691-0}
\showDOI{%
\url{http://dx.doi.org/10.1145/2424321.2424334}}


\bibitem[\protect\citeauthoryear{Kim, Jung, and Kim}{Kim et~al\mbox{.}}{1996}]%
        {Kim:1996:OCC:244032.244050}
{Hang~Joon Kim}, {Jong~Wha Jung}, {and} {Sang~Kyoon Kim}. 1996.
\newblock \showarticletitle{On-line Chinese Character Recognition Using
  ART-based Stroke Classification}.
\newblock {\em Pattern Recogn. Lett.\/} {17}, 12 (Oct. 1996), 1311--1322.
\newblock
\showISSN{0167-8655}
\showDOI{%
\url{http://dx.doi.org/10.1016/0167-8655(96)00078-5}}


\bibitem[\protect\citeauthoryear{Liu, Biagioni, Eriksson, Wang, Forman, and
  Zhu}{Liu et~al\mbox{.}}{2012}]%
        {Liu:2012:MLS:2339530.2339637}
{Xuemei Liu}, {James Biagioni}, {Jakob Eriksson}, {Yin Wang}, {George Forman},
  {and} {Yanmin Zhu}. 2012.
\newblock \showarticletitle{Mining Large-Scale, Sparse {GPS} Traces for Map
  Inference: Comparison of Approaches} {\em (KDD)}. ACM, New York, NY, USA,
  669--677.
\newblock
\showISBNx{978-1-4503-1462-6}
\showDOI{%
\url{http://dx.doi.org/10.1145/2339530.2339637}}


\bibitem[\protect\citeauthoryear{Mondzech and Sester}{Mondzech and
  Sester}{2011}]%
        {MondzechS11}
{Juliane Mondzech} {and} {Monika Sester}. 2011.
\newblock \showarticletitle{Quality Analysis of OpenStreetMap Data Based on
  Application Needs}.
\newblock {\em Cartographica\/}  {46} (2011), 115--125.
\newblock


\bibitem[\protect\citeauthoryear{Shi, Damper, and Gunn}{Shi
  et~al\mbox{.}}{2003}]%
        {Shi:2003:OHC:964161.964163}
{Daming Shi}, {Robert~I. Damper}, {and} {Steve~R. Gunn}. 2003.
\newblock \showarticletitle{Offline Handwritten Chinese Character Recognition
  by Radical Decomposition}.
\newblock  {2}, 1 (March 2003), 27--48.
\newblock
\showISSN{1530-0226}
\showDOI{%
\url{http://dx.doi.org/10.1145/964161.964163}}


\end{thebibliography}

\received{May 2014}{October 2014}{February 2015}

\end{document}